\documentclass[a4paper,aps,preprintnumbers,showpacs,twocolumn,superscriptaddress,nofootinbib]{revtex4-1}

\usepackage{amsmath}
\usepackage{amssymb}
\usepackage{graphicx}
\usepackage{epsfig}
\usepackage{color}
\usepackage{url}
\usepackage{times}
\usepackage{bm}         

\newcommand{\vk}{v_{\rm k}}
\newcommand{\ks}{\,{\rm km/s}}
\newcommand{\cf}{cf.~}
\newcommand{\ie}{i.e.,~}
\newcommand{\eg}{e.g.~}
\newcommand{\be}{\begin{equation}}
\newcommand{\ee}{\end{equation}}
\newcommand{\bea}{\begin{eqnarray}}
\newcommand{\eea}{\end{eqnarray}}
\newcommand{\nn}{\nonumber}

\newcommand{\bwt}{\begin{widetext}}
\newcommand{\ewt}{\end{widetext}}

\font\tenscr=rsfs10 scaled1100
\font\sevenscr=rsfs7 
\font\fivescr=rsfs5 
\skewchar\tenscr='177
\skewchar\sevenscr='177
\skewchar\fivescr='177
\newfam\scrfam
\textfont\scrfam=\tenscr
\scriptfont\scrfam=\sevenscr
\scriptscriptfont\scrfam=\fivescr

\def\scri{{\fam\scrfam I}}
\def\scre{{\fam\scrfam E}}

\begin{document}

\title{Black-hole horizons as probes of black-hole dynamics I:
  post-merger recoil in head-on collisions}

\author{Jos\'e Luis Jaramillo}
\affiliation{
Max-Planck-Institut f{\"u}r Gravitationsphysik, Albert Einstein
Institut, Potsdam, Germany 
}

\author{Rodrigo P. Macedo}

\affiliation{
Max-Planck-Institut f{\"u}r Gravitationsphysik, Albert Einstein
Institut, Potsdam, Germany 
}
\affiliation{
Theoretisch-Physikalisches Institut, Friedrich-Schiller-Universit{\"a}t Jena, Jena, Germany
}

\author{Philipp Moesta}
\affiliation{
Max-Planck-Institut f{\"u}r Gravitationsphysik, Albert Einstein
Institut, Potsdam, Germany 
}

\author{Luciano Rezzolla}
\affiliation{
Max-Planck-Institut f{\"u}r Gravitationsphysik, Albert Einstein
Institut, Potsdam, Germany 
}
\affiliation{
Department of Physics and Astronomy,
Louisiana State University,
Baton~Rouge, Louisiana, USA
}

\begin{abstract}
The understanding of strong-field dynamics near black-hole horizons is
a long-standing and challenging problem in general relativity. Recent
advances in numerical relativity and in the geometric characterization
of black-hole horizons open new avenues into the problem. In this
first paper in a series of two, we focus on the analysis of the recoil
occurring in the merger of binary black holes, extending the analysis
initiated in~\cite{Rezzolla:2010df} with Robinson-Trautman
spacetimes. More specifically, we probe spacetime dynamics through the
correlation of quantities defined at the black-hole horizon and at
null infinity. The geometry of these hypersurfaces responds to bulk
gravitational fields acting as test \textit{screens} in a
\textit{scattering} perspective of spacetime dynamics. Within a $3+1$
approach we build an effective-curvature vector from the intrinsic
geometry of dynamical-horizon sections and correlate its evolution
with the flux of Bondi linear momentum at large distances. We employ
this setup to study numerically the head-on collision of nonspinning
black holes and demonstrate its validity to track the qualitative
aspects of recoil dynamics at infinity. We also make contact with the
suggestion that the antikick can be described in terms of a ``slowness
parameter'' and how this can be computed from the local properties of
the horizon. In a companion paper~\cite{Jaramillo:2011rf} we will further
elaborate on the geometric aspects of this approach and on its
relation with other approaches to characterize dynamical properties of
black-hole horizons.
\end{abstract}


\maketitle

\section{Introduction}
\label{s:Intro}

Understanding the dynamics of colliding black holes (BHs) is of major
importance. Not only is this process one of the main sources of
gravitational waves (GWs), but it is also responsible for the final
recoil velocity (\ie \textit{``kick''}) of the merged object, which
could play an important role in the growth of supermassive BHs via
mergers of galaxies and on the number of galaxies containing BHs. The
recoil of BHs due to anisotropic emission of GW has been known for
decades~\cite{peres:1962,Bekenstein1973} and first estimates for the
velocity have been made using approximated and semianalytical methods
such as a particle approximation~\cite{1984MNRAS.211..933F,
  Nakamura:1983hk, Favata:2004wz}, post-Newtonian
methods~\cite{Wiseman:1992dv, Kidder:1995zr, Blanchet:2005rj,
  Damour-Gopakumar-2006} and the close-limit
approximation~\cite{Andrade:1997pc, Sopuerta:2006wj}. However, it is
only thanks to the recent progress in numerical relativity that
accurate values for the recoil velocity have been
computed~\cite{Baker:2006nr, Gonzalez:2006md, Campanelli:2007cg,
  Herrmann:2007ac, Koppitz-etal-2007aa:shortal, Lousto:2007db,
  Pollney:2007ss:shortal, Healy:2008js}.

Indeed, simulations of BHs inspiralling on quasicircular orbits have
shown, for instance, that asymmetries in the mass can lead to kick
velocities $\vk \lesssim 175\ks$~\cite{Baker:2006nr, Gonzalez:2006md},
while asymmetries in the spins can lead respectively to $\vk \lesssim
450\ks$ or $\vk \lesssim 4000\ks$ if the spins are
aligned~\cite{Herrmann:2007ac,Koppitz-etal-2007aa:shortal,
  Pollney:2007ss:shortal} or perpendicular to the orbital angular
momentum~\cite{Campanelli:2007ew, Gonzalez:2007hi,Campanelli:2007cg}
(see~\cite{Rezzolla:2008sd,Zlochower2011} for recent reviews).

In addition to a net recoil, many of the simulations show an
\textit{``antikick,"} namely, one (or more) decelerations experienced
by the recoiling BH at late times. In the case of merging BHs, such
antikicks seem to take place after a single apparent horizon (AH)
has been found~\cite{LeTiec:2009yg} (see Fig.~8 of
Ref.~\cite{Pollney:2007ss:shortal} for some examples). An active
literature has been developed over the last few years in the attempt
to provide useful interpretations to this
process~\cite{Schnittman:2007ij, Mino:2008at,
  LeTiec:2009yg,Nichols:2010, Hamerly:2010-b}. Interestingly, some of
these works do not even require the merger of the BHs. As pointed out
in~\cite{Sperhake:2010uv} when studying the scattering of BHs, in
fact, the presence of the common AH is not a necessary condition for
the antikick to occur. Furthermore, as highlighted
in~\cite{Price:2011fm}, it is also possible to describe this process
without ever discussing BHs and just using the mathematical
properties of the evolution of a damped oscillating signal\footnote{
On the other hand, if an exponentially damped oscillating signal is present,
  this is indeed a signature of the presence of a BH ringing
  down.}.

\begin{figure}
\begin{center}
\includegraphics[width=5.5cm, angle=-90]{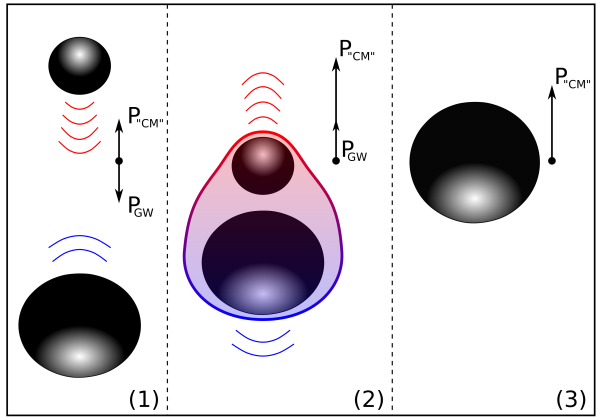}
\end{center}
\caption{Generation of the antikick in the head-on collision of two
  unequal-mass Schwarzschild BHs as described
  in~\cite{Rezzolla:2010df}. Initially the smaller BH moves faster and
  linear momentum is radiated mostly downwards, thus leading to an
  upwards recoil of the system [stage (1)]. At the merger the
  curvature is higher in the upper hemisphere of the distorted BH (\cf
  red-blue shading) and linear momentum is radiated mostly upwards
  leading to the antikick [stage (2)]. The BH decelerates till a
  uniform curvature is restored on the AH [stage (3)]. This cartoon
  should be contrasted with the results of numerical simulations in
  Fig.~\ref{fig:nocartoon}.}
\label{fig:cartoon}
\end{figure}

Although the presence of a common AH is not a necessary condition for
the appearance of an antikick (which could indeed be produced also by
the scattering of a system involving one or two neutron stars), when a
common AH \textit{is present} through the merger of BH binary, we can
use information on the latter to gain insight in the physical
mechanisms behind the antikick\footnote{Here and in the companion paper we will
  show that even when a horizon is not present, the considerations
  made here can be extended on a suitably defined 2-surface.}. We
believe that constructing an intuitive picture of the dynamics of
general relativity in a region of very strong field is not only
of general interest but also of practical use to explain this process. In
Ref.~\cite{Rezzolla:2010df}, in fact, a new conjecture was suggested
in which the antikick produced in the head-on collision of two BHs
with unequal masses was understood in terms of the dissipation of the
AH intrinsic deformation. As shown in the schematic cartoon in
Fig.~\ref{fig:cartoon} (\cf Fig. 1 of~\cite{Rezzolla:2010df} and also
Fig.~\ref{fig:nocartoon} for a comparison with numerical data), the
kick and antikick can be easily interpreted in terms of simple
dynamical concepts. Initially the smaller BH moves faster and linear
momentum is radiated mostly downwards, thus leading to an upwards
recoil of the system [stage (1)]. When a single AH is formed at the
merger, the curvature is higher in the upper hemisphere of the
distorted BH and linear momentum is radiated mostly upwards leading to
the antikick [stage (2)]. The BH decelerates till a uniform curvature
is restored on the AH [stage (3)]. The qualitative picture shown in
the cartoon was then investigated by exploiting the analogy between
this process and the evolution of Robinson-Trautman (RT)
spacetimes~\cite{Robinson:1962zz,Kramer80} and by showing that a
one-to-one correlation could be found between the properties of the AH
perturbation and the size of the recoil
velocity~\cite{Rezzolla:2010df}.

In this paper and in its companion~\cite{Jaramillo:2011rf} paper (hereafter
paper I and II, respectively), we provide further support to the
conjecture proposed in~\cite{Rezzolla:2010df} by extending our
considerations in~\cite{Rezzolla:2010df} to more generic initial data
and, more importantly, by investigating in detail numerical spacetimes
describing the head-on collision of two BHs with unequal masses. To do
this we introduce a {\em cross-correlation} picture in which the
dynamics of the spacetime can be read off from two ``screens''
provided naturally by the black-hole event horizon $\scre$ and by future null
infinity $\scri^+$. In practice, using the standard $3+1$ approach in
general relativity, we replace these screens with effective ones
represented, respectively, by a dynamical horizon ${\mathcal H}^+$ and
by a timelike tube ${\cal B}$ at large spatial distances. We then
define a phenomenological curvature vector
$\tilde{K}^{\mathrm{eff}}_i(t)$ in terms of the (mass multipoles of
the) Ricci scalar curvature ${}^2\!R$ at ${\mathcal H}^+$ and show
that this is closely correlated with a {\em geometric quantity}
$(dP_i^{\cal B}/dt)(t)$, representing the variation of the Bondi
linear momentum in time on $\scri^+$. This construction, which is free
of fitting coefficients and valid beyond the axisymmetric scenario
considered here, correlates quantities on the AH with quantities at
large distance, thus providing us with two important tools.  Besides
confirming the association of recoil dynamics with the dissipation of
anisotropic distribution of curvature on the AH, it opens a new route
to the analysis of strong-field effects in terms of purely local
quantities evaluated either on the AH or on other suitable surfaces.

This first article is organized as follows. Section~\ref{s:Exe_summ}
introduces an executive summary, where the main concepts presented in
these two papers are summarized for those not wishing to enter into
the mathematical details. Section~\ref{s:RT}, on the other hand,
extends the analysis carried out in~\cite{Rezzolla:2010df} for RT
spacetimes by considering more general initial data and by analyzing
aspects of the evolution of the AH curvature. Section~\ref{s:BBH}
extends the methodology and diagnostic tools to BH spacetimes
representing the head-on collision of unequal-mass BHs. In particular,
we develop the mathematical tools necessary to measure the relevant
quantities on the two screens and we show how they closely
correlate. Finally, the conclusions are discussed in
Sec.~\ref{s:conclusions}, while the Appendix is used to provide
details on our definitions of the correlation and matching of
time series.

This paper also builds on the material presented in its companion
paper II, where we present a more detailed discussion of the
mathematical aspects of our framework. In particular, we revisit there
the evolution of relevant geometric objects on the AH and introduce
preferred null normals on a dynamical horizon. In our discussion of a
newslike function on the dynamical horizon and its relation to the
problem of quasilocal linear momentum, we also stress the importance
of the inner horizon when evaluating fluxes across the horizon.

We use a spacetime signature $(-,+,+,+)$, with abstract index notation
(first letters, $a$, $b$, $c$..., in Latin alphabet) and Latin
midalphabet indices, $i,j,k...$, when making explicit the spacelike
character of a tensor. Greek indices are used for expressions in
particular coordinates systems.  We also employ the standard
convention for the summation over repeated indices. Finally, all the
quantities are expressed in a system of units in which
$c=G=M_\odot=1$, unless otherwise stated.

\section{The cross-correlation approach: an executive summary}
\label{s:Exe_summ}

As mentioned above, this section is meant to provide a general summary
of the results and methodology of papers I and II, focusing mostly on
the conceptual aspects and leaving aside the mathematical details,
which can instead be found in the corresponding main texts.

We start by recalling that Ref.~\cite{Rezzolla:2010df} suggested an
approach to study the near-horizon nonlinear dynamics of the
gravitational fields based on the systematic analysis of the
deformations in the BH horizon geometry. In particular, it was shown
how the gravitational dynamics responsible for the antikick after a
binary merger can be understood in terms of the anisotropies in the
intrinsic curvature of the AH of the resulting merged BH. Considering
a RT spacetime, the kick velocity constructed from the Bondi momentum
(a geometric quantity at null infinity) was put in a one-to-one
correspondence with a quasilocal geometric quantity constructed on
the horizon, namely, with the effective {\em curvature parameter}
$K_{\mathrm{eff}}$. This geometric parameter $K_{\mathrm{eff}}$
encodes the part of the AH geometry whose {\em dissipation} through
gravitational radiation can be related to the final value of the
kick. Stated differently, very different binary systems, e.g., with very
different mass ratio, give rise to the same final kick velocity as
long as they share the same value of the $K_{\mathrm{eff}}$ parameter.

The following criteria were employed in Ref.~\cite{Rezzolla:2010df}
for the construction of the {\em curvature parameter}
$K_{\mathrm{eff}}$: i) $K_{\mathrm{eff}}$ should not depend on how the
AH is embedded in the spacetime; ii) $K_{\mathrm{eff}}$ should change
sign (\ie it should be an odd function) under reflection with respect
to a plane normal to a given axis. From the first requirement,
$K_{\mathrm{eff}}$ was constructed in terms of the intrinsic geometry
of the AH, namely as a functional on the Ricci scalar ${}^2\!R$
associated with the induced metric on the AH. The ansatz for
$K_{\mathrm{eff}}$ in Ref.~\cite{Rezzolla:2010df}, compatible with
requirement ii) above and within axisymmetry, had the following
structure
\begin{eqnarray}
\label{e:K_eff_Ansatz}
K_{\mathrm{eff}} = f_{\mathrm{even}}\left(M_{2\ell}\right) 
\times f_{\mathrm{odd}}\left(M_{2\ell+1} \right) \,,
\end{eqnarray}
where $M_\ell$'s are the so-called {\em isolated-horizon mass
  multipoles} associated with a spherical harmonic decomposition of
${}^2\!R$ in the axisymmetric case \cite{Ashtekar:2004gp,
  Schnetter-Krishnan-Beyer-2006}. The odd part $f_{\mathrm{odd}}$
accounts for the directionality of the kick, whereas the even part
$f_{\mathrm{even}}$ controls its intensity.

\begin{figure}[t]
\begin{center}
\vglue -2.0cm 
\hglue -2.0cm 
\includegraphics[angle=-90,width=12.0cm]{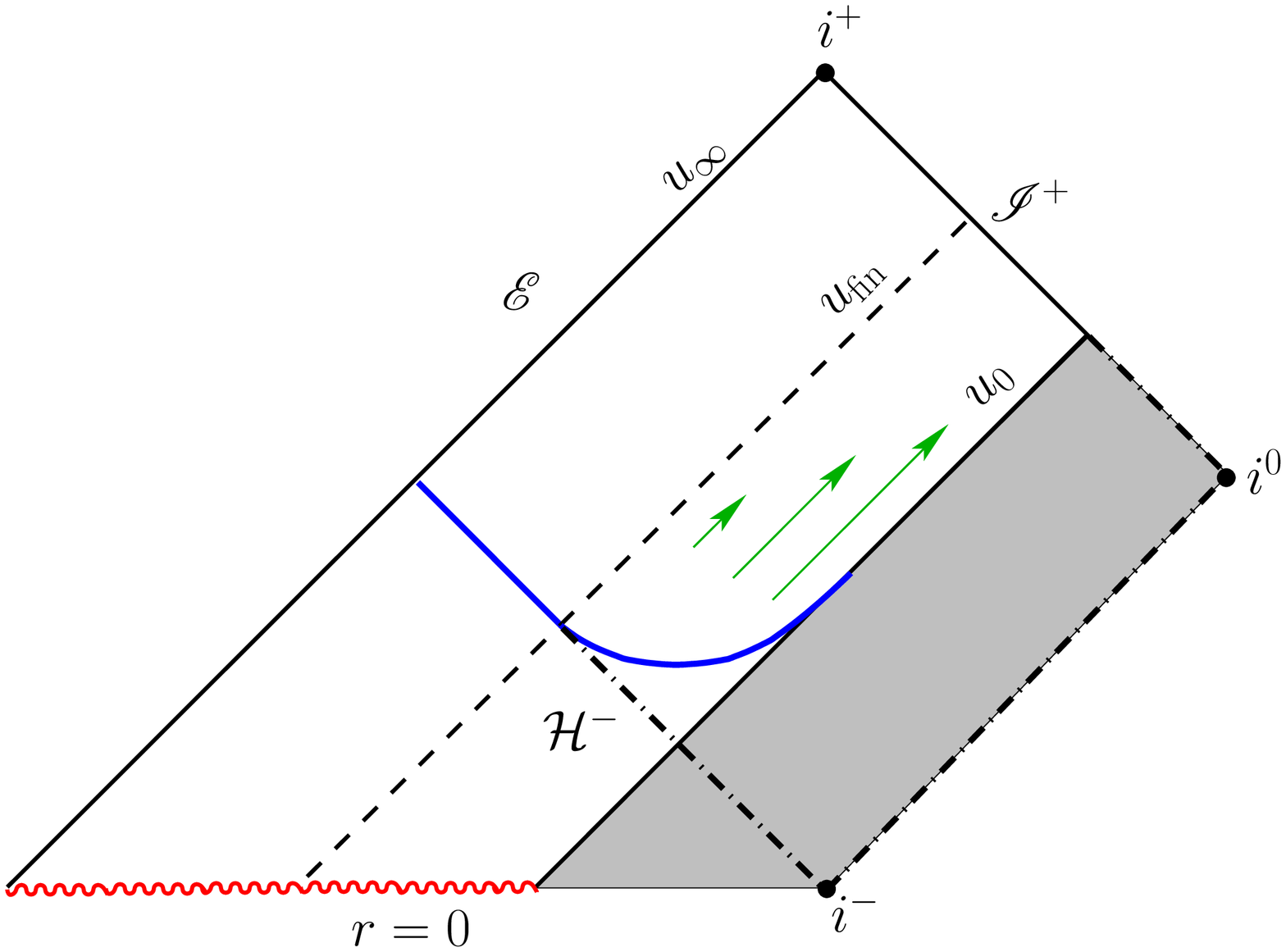}
\end{center}
\caption{Carter-Penrose diagram for the RT
  spacetime~\cite{Chrusciel:1992rv, Chow:1995va}. The solutions exist
  for $u\geqslant u_0$. The white hole emits GWs until the
  Schwarzschild spacetime is achieved, although the numerical
  simulations run until a finite $u_{\rm final}$, when an
  essentially stationary solution is found.}  %
\label{fig:RT}
\end{figure}

In order to validate this suggestion, we analyzed a family of
Robinson-Trautman (RT) spacetimes~\cite{Robinson:1962zz,Kramer80},
representing an (eternal) BH together with purely outgoing
gravitational radiation. The mathematical properties of this class of
exact solutions is already well
understood~\cite{Tod83,Singleton:90,Chrusciel:1992cj,Chrusciel:1992rv},
therefore this spacetime is a good test for numerical
schemes~\cite{Gomez97,deOliveira:2004bn} and it is an excellent toy
model for problems dealing with radiation in BH
environments~\cite{Moreschi:1996ds, Moreschi:2002, deOliveira2005,
  deOliveira:2008zc, deOliveira2008, Aranha:2008ni, Macedo:2008ia,
  Podolsky:2009an, Aranha:2010, Aranha:2010zz, deOliveira:2011pk,
  Svitek:2011}. Although the associated BH horizon is stationary,
these RT spacetimes also contain a {\em white-hole} horizon ${\cal
  H}^-$\cite{Penrose:1973, Tod89, Chrusciel:1992rv, Chow:1995va}, or,
more precisely, a {\em past outer-trapping horizon}~\cite{Hayward94a},
whose dynamics offers a particularly well-suited scenario to test our
geometric approach. This is shown in Fig.~\ref{fig:RT}, which reports
a Carter-Penrose diagram for the RT spacetime (see
also~\cite{Chrusciel:1992rv, Chow:1995va}). The solutions exist for
$u\geqslant u_0$ and the white hole emits GWs until the Schwarzschild
spacetime is achieved asymptotically. In practice, numerical
simulations run up to a finite $u_{\rm final}$ and show the
exponential convergence to a solution which is essentially stationary.

In Ref.~\cite{Rezzolla:2010df}, the functions $f_{\mathrm{even}}$ and
$f_{\mathrm{odd}}$ appearing in Eq.~(\ref{e:K_eff_Ansatz}) were
written in the simplest possible form, \ie as a linear expansion in
$M_\ell$'s
\begin{equation}
\label{e:K_eff_Ansatz_linear}
K_{\mathrm{eff}} = \left(a_2 M_2 + a_4 M_4 +\ldots \right) \times
 \left(a_3 M_3 + a_5 M_5 + \ldots \right)\,.
 \end{equation}
Then, using suitably defined initial data, a set of numerically fitted
coefficients $a_i$'s was found so that a one-to-one dependence between
the final kick velocities $\vk$ and $K_{\mathrm{eff}}$ at a given
retarded time $u$ could be found: \ie $\Delta \vk =\vk-v(u)=A\times
K_{\mathrm{eff}}(u)$, where $\vk(u)$ is the recoil velocity at time
$u$ and $A$ is a constant. This injective\footnote{Note that the relation 
is not only injective, but also linear. This is ultimately 
due to the writing of $K_{\mathrm{eff}}$ as the product of two functions 
 (of even and odd multipoles), such that each of these functions 
 is linear in the multipoles.}~relation between $K_{\mathrm{eff}}$ 
and $\vk$ permits us
to understand the degeneracy of the latter, as a function of the mass
ratio in terms of AH quantities at a given (initial) time $u$ (\cf
Fig. 3 in Ref.~\cite{Rezzolla:2010df} and Fig.~\ref{fig:Keff_dV}
below).  Moreover, the good quantitative agreement between $\vk$
calculated from full binary BH numerical simulations and from RT
models, suggested the presence of a generic behavior in this physical
process. Overall, therefore, the work in Ref.~\cite{Rezzolla:2010df}
provided an approach to understand global recoil properties in terms
of (quasi-)local quantities on the AH, and an intuitive guideline to
interpret the black-hole recoil properties in terms of the dissipation
of AH geometric quantities.

Despite the valuable insight, the treatment presented in
Ref.~\cite{Rezzolla:2010df} had obvious limitations. First, the ansatz
for $K_{\mathrm{eff}}$ in Eq.~(\ref{e:K_eff_Ansatz}) is not
straightforwardly generalizable to the nonaxisymmetric case. Second,
the phenomenological coefficients $a_\ell$'s in
Eq.~(\ref{e:K_eff_Ansatz_linear}) depend on the details of the
employed RT initial data. Finally, the white-hole horizon analysis in
RT spacetimes needs to be extended to the genuine BH horizon case. All
of these restrictions are overcome in the work reported in papers I
and II.

While the focus in Ref.~\cite{Rezzolla:2010df} was on expressing the
difference between the {\em final} kick velocity $v_{\infty}$ and the
instantaneous kick velocity $v_{\mathrm{k}}(u)$ at an (initial) given
time $u$, in terms of the geometry of the common AH at that time $u$,
we here focus on geometric quantities that are evaluated at a given
time during the evolution. More specifically, we will consider the
variation of the Bondi linear momentum vector in time
$(dP_i^{\mathrm{B}}/du)(u)$ as the relevant geometric quantity to
monitor at null infinity $\scri^+$. To this scope, we need first to
construct a vector $\tilde{K}^i_{\mathrm{eff}}(v)$ (function of an
advanced time $v$) as a counterpart on the BH horizon ${\cal
  H}^+$. Then, we need to determine how
$\tilde{K}^i_{\mathrm{eff}}(v)$ on ${\cal H}^+$ correlates to
$(dP_i^{\mathrm{B}}/du)(u)$ at $\scri^+$.

\begin{figure}[t!]
\begin{center}
\vglue-1.0cm
\includegraphics[angle=90,width=13.0cm,clip=true,angle=-90]{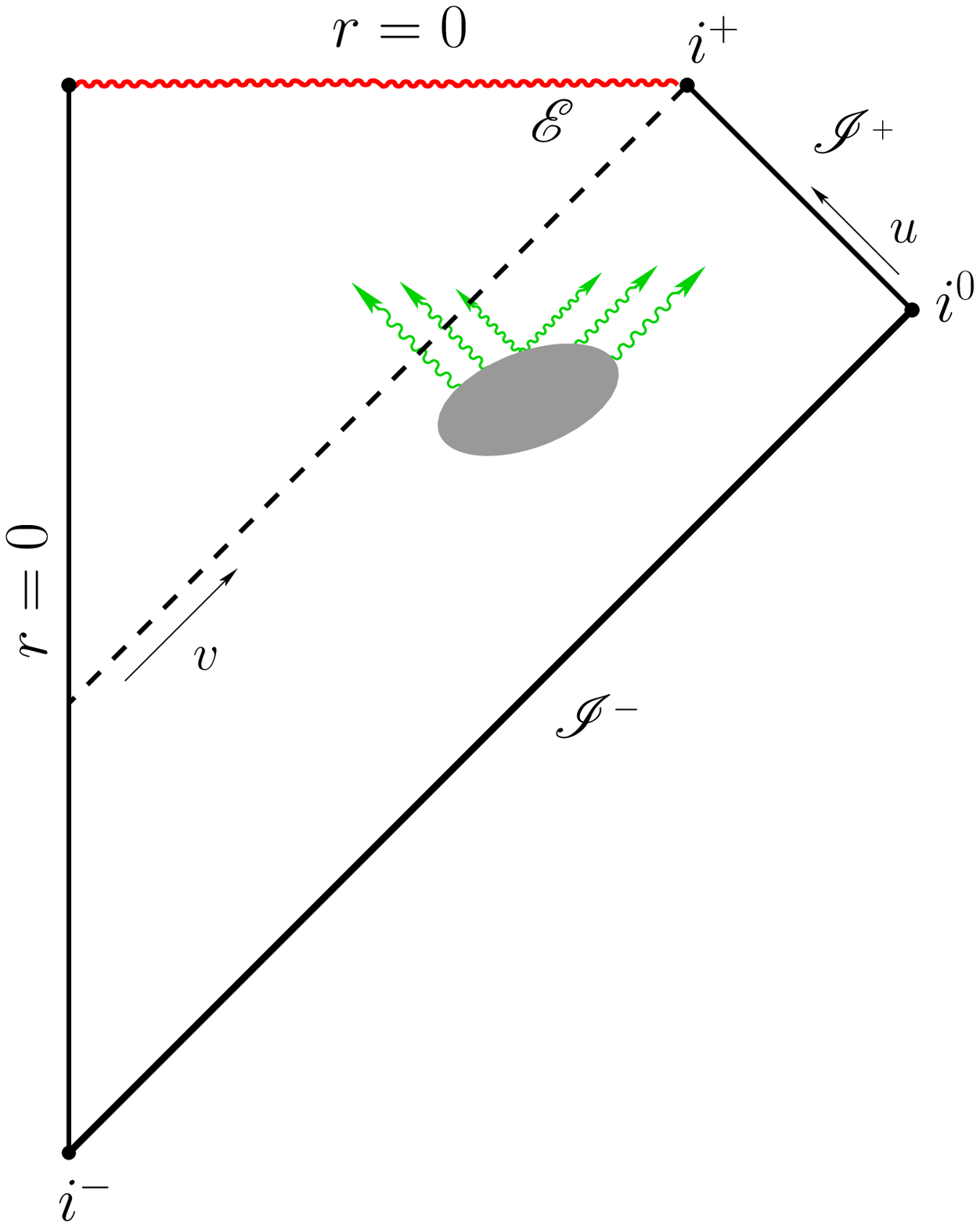}
\end{center}
\vglue-3.75cm
\caption{Carter-Penrose diagram illustrating the {\em scattering}
  approach to near-horizon gravitational dynamics in a generic
  spherically symmetric collapse. The event horizon $\scre$ and
  null infinity $\scri^+$ provide spacetime canonical screens on which
  {\em geometric quantities}, respectively accounting for horizon
  deformations and wave emission, are defined. Their cross-correlation
  encodes nontrivially information about the bulk spacetime dynamics.}
\label{fig:BHscattering}
\end{figure}

\begin{figure}[t!]
\begin{center}
\vglue-1.0cm
\includegraphics[angle=90,width=13cm,clip=true,angle=-90]{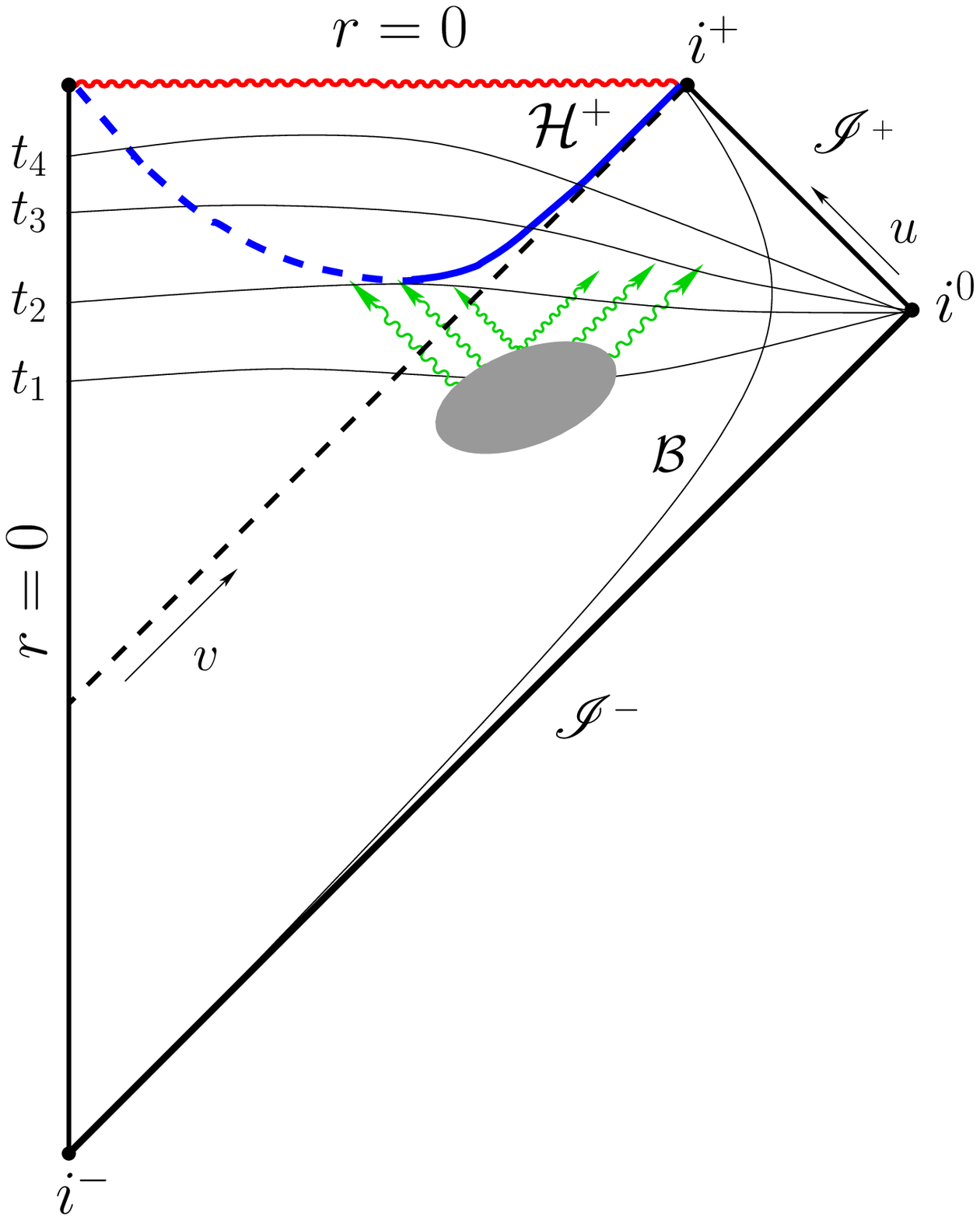}
\end{center}
\vglue-3.75cm
\caption{Carter-Penrose diagram for the {\em scattering} picture in a
  Cauchy initial value approach. The dynamical horizon ${\cal H}^+$
  and a large-distance timelike hypersurface ${\cal B}$ provide inner
  and outer screens. Note that a the dynamical horizon is split in two
  portions: outer and inner (solid and dashed blue lines,
  respectively) and that the $3+1$ slicing sets a common time $t$ for
  cross-correlations.}
\label{fig:BHscattering_3+1}
\end{figure}

In the RT case, the causal relation between the white-hole horizon
${\cal H}^-$ and null infinity $\scri^+$ made possible to establish an
explicit functional relation between ${dv_{\mathrm{k}}}/{du}$ and
$K'_{\mathrm{eff}}(u)$. In the case of generic BH horizon, however,
such a direct causal relation between the inner horizon and $\scri^+$
is lost (compare Fig.~\ref{fig:RT} with Figs.~\ref{fig:BHscattering}
and~\ref{fig:BHscattering_3+1}). However, since their respective
causal pasts partially coincide, nontrivial {\em correlations} are
still possible and expected. This can be measured through the
cross-correlations of geometric quantities $h_{\mathrm{inn}}(v)$ at
${\cal H}^+$ and $h_{\mathrm{out}}(u)$ at $\scri^+$, both considered
here as two time series\footnote{\label{f:stretching}Note that the
  meaningful definition of time series cross-correlations requires the
  introduction of a (gauge-dependent) relation between advanced and
  retarded time coordinates $v$ and $u$. In an initial value problem
  this is naturally provided by the $3+1$ spacetime slicing by time
  $t$.}. In particular, we will take $\tilde{K}^i_{\mathrm{eff}}(v)$
as $h_{\mathrm{inn}}(v)$ and $(dP_i^{\mathrm{B}}/du)(u)$ as
$h_{\mathrm{out}}(u)$.

This approach to the exploration of near-horizon gravitational
dynamics resembles therefore the methodology adopted in {\em
  scattering} experiments. Gravitational dynamics in a given spacetime
region affects the geometry of appropriately chosen {\em outer} and
{\em inner} hypersurfaces of the BH spacetime.  These hypersurfaces
are then understood as {\em test screens} on which suitable {\em
  geometric quantities} must be constructed. The correlations between
the two encode geometric information about the dynamics in the bulk,
providing information useful for an {\em inverse-scattering} approach
to the near-horizon dynamics. As a result, in asymptotically flat BH
spacetimes, null infinity $\scri^+$ and the (event) BH horizon $\scre$ provide {preferred} choices for the outer and inner
screens. This is nicely summarized in the Carter-Penrose diagram in
Fig.~\ref{fig:BHscattering}, which illustrates the cross-correlation
approach to near-horizon gravitational dynamics. The event horizon
$\scre$ and null infinity $\scri^+$ provide natural spacetime
screens on which {geometric quantities}, respectively, accounting for
horizon deformations and wave emission, are defined. Their
cross-correlation encodes information about the bulk spacetime
dynamics.

Although the picture offered by Fig.~\ref{fig:BHscattering} is quite
simple and convincing, it is not well adapted to the 3+1 approach
usually adopted in numerical studies of dynamical spacetimes. Indeed,
neither the BH event horizon nor null infinity are in general
available during the evolution\footnote{The latter would properly
  require either characteristic or a hyperboloidal evolution
  approach.}. However, we can adopt as inner and outer screens a
dynamical horizon ${\mathcal H}^+$ (future outer-trapping
horizon~\cite{Hayward94a, Ashtekar03a, Ashtekar:2004cn}) and a
timelike tube ${\cal B}$ at large spatial distances, respectively. In
this case, the time function $t$ associated with the $3+1$ spacetime
slicing provides a (gauge) mapping between the {retarded} and
{advanced} times $u$ and $v$, so that cross-correlations between
geometric quantities at ${\mathcal H}^+$ and ${\cal B}$ can be
calculated as standard time series $h_{\mathrm{inn}}(t)$ and
$h_{\mathrm{out}}(t)$. This is summarized in the Carter-Penrose
diagram in Fig.~\ref{fig:BHscattering_3+1}, which is the same as in
Fig.~\ref{fig:BHscattering}, but where the $3+1$ slicing sets an
in-built common time $t$ for cross-correlations between the dynamical
horizon ${\cal H}^+$ (\ie the inner screen) and a large-distance
timelike hypersurface ${\cal B}$ (\ie the outer screen).

Within this conceptual framework it is then possible to define a
phenomenological curvature vector $\tilde{K}^{\mathrm{eff}}_i(t)$ in
terms of the mass multipoles of the Ricci scalar curvature ${}^2\!R$
at ${\mathcal H}^+$ and show that this is closely correlated with a
quantity $(dP_i^{\cal B}/dt)(t)$ on ${\cal B}$, representing an approximation
to the variation of the Bondi linear momentum time on $\scri^+$. How to do
this in practice for a BH spacetime is the subject of the following
sections.

\section{Robinson-Trautman spacetimes: a toy model}
\label{s:RT}

We recall that the RT spacetimes are a class of solutions of the
vacuum Einstein equations admitting a congruence of null geodesics
which are hypersurface-orthogonal, shear-free but with nonvanishing
expansion. As such, it can be regarded as a white hole emitting GWs,
thus representing a valuable tool for studying the spacetime geometry
in physical conditions that are similar to the final stages on the
dynamics of BH binaries~\cite{Robinson:1962zz}. The RT metric can be
written as~\cite{Macedo:2008ia}
\begin{equation}
\label{RTM}
ds^2 = -\left(K - \frac{2M_{\infty}}{r} -
\frac{2r \partial_u Q}{Q} \right) du^2 - 2dudr + \frac{r^2}{Q^2} d\Omega^2,
\end{equation}
where $Q=Q(u,\Omega)$, $u$ is a null coordinate, $r$ is an affine
parameter of the outgoing null geodesics, and $d\Omega^2=d\theta^2 +
\sin^2\theta d\varphi^2$ is the metric of a unit sphere $S^2$. Here
$M_{\infty}$ is a constant and is related to the mass of the
asymptotic Schwarzschild BH, while the function $K(u,\Omega)$ is the
Gaussian curvature of the surface with $r = 1$ and $u =
\mathrm{const.}$, and is given by
\begin{equation}
\label{curv}
K(u,\Omega)= Q^2(1+\Delta^2_{\Omega}{\ln{Q}})\,,
\end{equation}
$\Delta^2_{\Omega}$ being the Laplacian operator on the unit sphere
$S^2$. The Einstein equations then lead to the RT evolution equation
\begin{equation}
\label{RTeq}
\partial_u Q(u,\Omega)=-{Q^3} \frac{\Delta^2_{\Omega}K(u,\Omega)}{12M_{\infty}}\,.
\end{equation}
Any regular initial data $Q=Q(u_0,\Omega)$ will smoothly evolve
according to~\eqref{RTeq} until it reaches a stationary configuration
corresponding to a Schwarzschild BH at rest or moving with a constant
speed~\cite{Chrusciel:1992cj}. Equation~\eqref{RTeq} implies the existence
of the constant of motion $\mathcal{A}\equiv \oint_{\cal
  S}{d\Omega}/{Q^2}$, which clearly represents the area of the surface
$u=\mathrm{const.}$, $r=1$ and can be used to normalize $Q$ so that
$\mathcal{A}=4\pi$.

The dynamical compact object modeled by RT spacetimes is described by
the past AH, which has a vanishing expansion of the ingoing
future-directed null geodesics. Such past AH is described by the
surface $r=R(u,\Omega)$ satisfying~\cite{Penrose:1973, Tod89,
  Chow:1995va}
\begin{equation}
\label{BHRT}
Q^2\Delta_{\Omega}^{2} \ln{R}=K-\frac{2M_{\infty}}{R}\,.
\end{equation}

The line element restricted to the AH surface $u=\mathrm{const.}$ and
$r=R(\Omega)$ is
\begin{equation}
 \label{AHmetric}
ds^2|_{_{\rm H}}=\frac{R^2}{Q^2}d\Omega^2\,,
\end{equation}
which, from Eq.~(\ref{BHRT}), has a Gaussian curvature
\begin{equation}
 \label{AHcurv}
K_{_{\rm H}}=\frac{2M_{\infty}}{R^3}\,.
\end{equation}

On the other hand, the mass and momentum are computed at future null
infinity using the Bondi 4-momentum as~\cite{Tod83, Singleton:90,
  Macedo:2008ia}
\begin{equation}
\label{BondiMoment}
P^{\alpha}(u) \equiv
\frac{M_{\infty}}{4\pi}\oint_{{\cal S}_u}\frac{\eta^{\alpha}}{Q^3}d\Omega\,,
\end{equation}
with $\left\{\eta^{\alpha}\right\}= \left\{1, \sin{\theta}\cos{\varphi},
\sin{\theta}\sin{\varphi}, \cos{\theta} \right\}$. 

From now on, we restrict our problem to axisymmetry and introduce
$x=\cos\theta$.  Clearly, all the physically relevant information is
contained in the function $Q(u,x)$, and this includes also the
gravitational radiation, which can be extracted through the radiative
part of the Riemann tensor~\cite{Robinson:1962zz,Kramer80}, which in
axisymmetry is given by
\begin{equation}
\label{Psi4RT}
 r\Psi_4=\frac{Q^2}{2}\partial_u\left[\frac{(1-x^2)\partial_x^2Q}{Q} \right]\,.
\end{equation}

The dynamics of this solution can be summarized in Fig.~\ref{fig:RT},
which shows the Carter-Penrose diagram for the RT spacetime. The final
configuration is a stationary nonradiative solution which has the
form~\cite{Macedo:2008ia}
\begin{equation}
\label{Qinfty}
S_{\pm}(\theta) \equiv Q(\infty, \theta)=\frac
{\left(1 \mp \vk x\right)}{\sqrt{1-\vk^2}}\,.
\end{equation}
Note that since the Bondi 4-momentum of the stationary solution is
\begin{equation}
\label{Bondiinfty}
P^{\alpha}(\infty)= \frac{M_{\infty}}{\sqrt{1-\vk^2}}
\left\{1,0,0,\pm \vk \right\},
\end{equation}
the parameter $\vk$ in Eq.~\eqref{Qinfty} is interpreted as the
velocity of the Schwarzschild BH in the $z$-direction.

\subsection{Mass multipoles}
\label{s:mass_multipoles}

Given a closed 2-surface ${\cal S}$, the invariant content of its
intrinsic geometry is encoded in the Ricci scalar curvature ${}^2\!R$
associated with the induced metric $q_{ab}$ on ${\cal S}$. Moreover,
if ${\cal S}$ is an axisymmetric surface, with $\phi^a$ as the axial
Killing vector, a preferred coordinate system $(\tilde{\theta},
\tilde{\varphi})$ can be constructed such that $q_{ab}$ has the
form~\cite{Ashtekar:2004gp, Schnetter-Krishnan-Beyer-2006}
\begin{eqnarray}
\label{e:axisymmetric_metric}
q_{ab} dx^adx^b= R_{_{\mathrm H}}^2
\left( f^{-1} \mathrm{\sin}^2\tilde{\theta} \;
d\tilde{\theta}^2 +
f \; d\tilde{\varphi}^2 \right) \,,
\end{eqnarray}
where $f(\tilde{\theta})=q_{ab}\phi^a\phi^b/R_{_{\mathrm H}}^2$, with
$R_{_{\mathrm H}}$ the areal radius ($A=\int_{\cal S} dA = 4\pi
R_{_{\rm H}}^2$).  The coordinate $\tilde{\theta}$ is determined by
\begin{equation}
\label{e:defzeta}
 D_{a}\tilde{\zeta} = \frac{1}{R_{_{\mathrm H}}^2} {}^2\!\epsilon_{ba} \phi^{b} ,
\end{equation}
where the coordinate $\tilde{\zeta}$ is defined by
$\tilde{\zeta}\equiv \mathrm{\cos}\tilde{\theta}$ and
${}^2\!\epsilon_{ba}$ is the alternating symbol. In addition, the
normalization condition $\oint_{\mathcal{H}} \tilde{\zeta} dA =0$ must
be imposed. We note that the Ricci scalar ${}^2\!R$ on ${\cal S}$ can
be written as~\cite{Ashtekar:2004gp}
\begin{eqnarray}
\label{e:Ricci_f}
{}^2\!R = - \frac{1}{R_{_{\mathrm H}}^2} \frac{d^2 f}{d\tilde{\zeta}^2}(\tilde{\zeta})
 \,,
\end{eqnarray}
and that regularity conditions on the metric impose
\begin{eqnarray}
\label{e:f_properties}
\lim \limits_{\tilde\zeta\to \pm1} f(\tilde\zeta)  = 0\,, \ \ 
\lim \limits_{\tilde\zeta\to \pm1}
\frac{df}{d\tilde{\zeta}}(\tilde{\zeta}) = \pm 2\,.
\end{eqnarray}
A crucial feature of this coordinate system is that the associated
expression for the area element is proportional to that of the {\em
  ``round sphere''} metric $dA=R_{_{\mathrm H}}^2
\mathrm{\sin}\tilde{\theta}\;d\tilde{\theta}d\tilde{\varphi}$. This
provides the appropriate {\em measure} on ${\cal S}$ to define the
standard spherical harmonics $Y_{\ell, m=0}(\tilde{\theta})$ with the
standard orthonormal relations
\begin{eqnarray} \oint_{\mathcal S}
Y_{\ell,0}(\tilde{\theta})Y_{\ell',0}(\tilde{\theta})dA = R_{_{\mathrm H}}^2
\delta_{\ell\ell'} \,, 
\end{eqnarray} 
so that dimensionless geometric multipoles $I_\ell$ can be introduced
as the spherical harmonics components of the Ricci scalar curvature
${}^2{\!}R$~\cite{Ashtekar:2004gp}
\be
\label{e:I_n}
I_\ell \equiv \frac{1}{4} 
\oint_{\mathcal S} {}^2\!R \; Y_{\ell,0}(\tilde{\theta}) \;dA \,, \
{}^2\!R=\frac{4}{R_{_{\mathrm H}}^2}\sum_{n=0}^\infty I_\ell Y_{\ell,0}(\tilde{\theta}) \,.
\ee
The mass multipoles $M_\ell$'s
are then defined as appropriate dimensionful rescalings of the
geometric $I_\ell$'s
\be
\label{e:IH:massmultipol}
M_\ell \equiv \sqrt{\frac{4\pi}{2 n + 1}} 
\frac{M_{_{\rm H}} (R_{_{\mathrm H}})^{\ell}}{2\pi} I_\ell \,,
\ee
where $M_{_{\rm H}}$ denotes some appropriate quasilocal mass for the
surface ${\cal S}$. Because we will consider here initial data
with zero angular momentum, $M_{_{\rm H}}$ will denote the
irreducible mass $M_{\mathrm{irr}}=\sqrt{A/(16\pi)}=
R_{\mathrm{H}}/2$. For later convenience, we introduce the rescaled
geometric multipoles $\tilde{I}_\ell$
\begin{eqnarray}
\label{e:barI_n}
\tilde{I}_\ell \equiv \frac{1}{M_{\mathrm{irr}}^2}I_\ell
=\frac{4}{(R_{_{\rm H}})^2} I_\ell,
\end{eqnarray}
with dimensions $[\tilde{I}_\ell]=[\mathrm{length}]^{-2}$. The Ricci
scalar curvature can then be written as
\begin{eqnarray}
\label{e:R_barI_n}
{}^2\!R=\sum_{\ell=0}^\infty \tilde{I}_\ell Y_{\ell0} \ \,.
\end{eqnarray}
A crucial remark for the discussion in Sec.~\ref{s:Keff} is the
vanishing of the $\ell= 1$ mode, \ie $\tilde{I}_1 = 0$, which can be
interpreted as a choice of {\em center of mass frame} of the AH
in~\cite{Ashtekar:2004gp}.  This follows by first inserting expression
(\ref{e:Ricci_f}) into the definition of $\tilde{I}_1$, so that
$\tilde{I}_1\propto \int_{-1}^{1} f''(\tilde\zeta) \tilde\zeta
d\tilde\zeta$, and then by making use of regularity conditions
(\ref{e:f_properties}) after integrating by parts.

In the particular case of a RT spacetime, the preferred axisymmetry
coordinate system $(\tilde{\theta},\tilde{\varphi})$ is related to the
RT spherical coordinates as $\tilde{\varphi}=\varphi$ and
$\tilde{\theta}=\tilde{\theta}(\theta)$ satisfying
\begin{equation}
\label{PrefCordSys}
 \partial_{\theta}\tilde{\zeta}=- \frac{\sin\theta R(\theta)^2}
{(R_{_{\rm H}})^2 Q(\theta)^2}\,.
\end{equation}
This equation is solved with the condition $\tilde{\zeta}(0)=1$ and then
one computes the mass multipoles moments through the
expression
\begin{equation}
 \label{MassMod}
M_\ell =M_{_{\rm H}} \frac{(R_{_{\mathrm H}})^{\ell+1}}{2} \oint_{\cal S} 
\frac{P_{\ell}(\tilde{\zeta})}{Q^2(\theta)R(\theta)}d\Omega\,.
\end{equation}
%

\subsection{The numerical setup}

As discussed in detail in Ref.~\cite{Macedo:2008ia}, for the numerical
solution of the Einstein equations we introduce a Galerkin
decomposition for $Q(u,x)$
\begin{equation}
\label{QGal}
Q(u,x) = \sum_{\ell=0}^N b_\ell(u)P_\ell(x),
\end{equation}
where $P_\ell(x)$ stands for the Legendre polynomial of order
$\ell$. By using standard projection techniques, Eq.~(\ref{RTeq}) can
be written as a system of ordinary differential equations
\begin{equation}
\label{RTG}
\dot{b}_\ell = -\frac{2\ell+1}{24M_{\infty}}\langle Q^3\partial_x\left[(1-x^2)\partial_xK \right],P_\ell  \rangle,\quad \ell=0,1,\dots,N,
\end{equation}
where the inner product is given by $\langle f,g \rangle = \int_{-1}^1
fg \,dx.$ In this way, the Cauchy problem for the RT
Eq.~\eqref{RTG} consists basically in choosing the initial value
of the mode functions $b_\ell(u)$ according to
\begin{equation}
\label{DATA}
b_\ell(0) = \frac{2\ell + 1}{2 }\langle Q(0,x), P_\ell \rangle,
\end{equation}
and then to solve the initial value problem given by (\ref{RTG}). Note
that, as $u\rightarrow \infty$, $b_\ell \rightarrow 0$ for $\ell>1$
and that the nonzero modes must satisfy $b_1(\infty)^2 - b_0(\infty)^2
= 1$, with the final $\vk$ parameter of Eq.~(\ref{Qinfty}) being given
simply by $\vk=-b_1(\infty)/b_0(\infty)$.

\begin{figure*}[t]
\begin{center}
\includegraphics[width=5.9cm,clip=true]{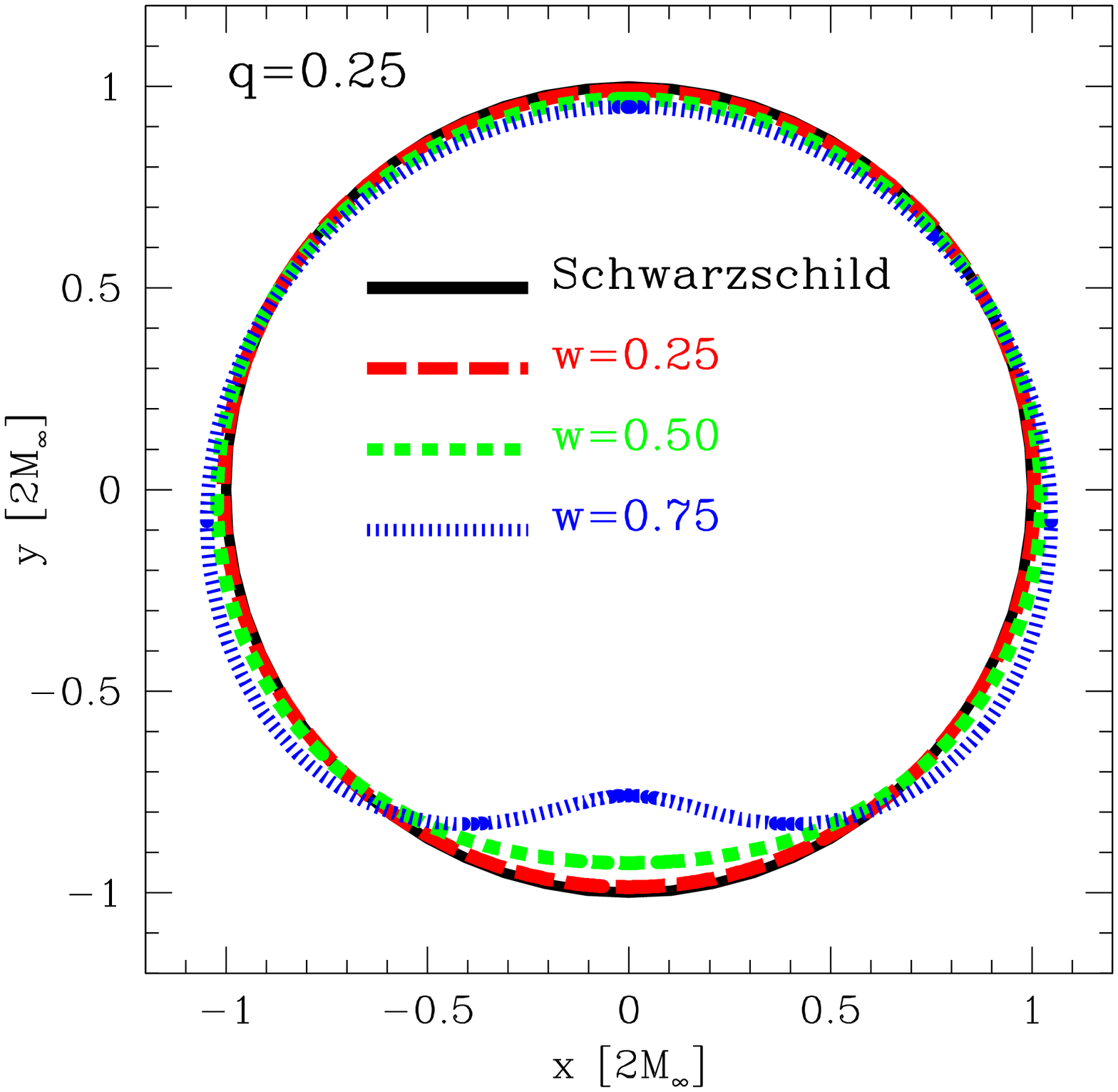}
\includegraphics[width=5.9cm,clip=true]{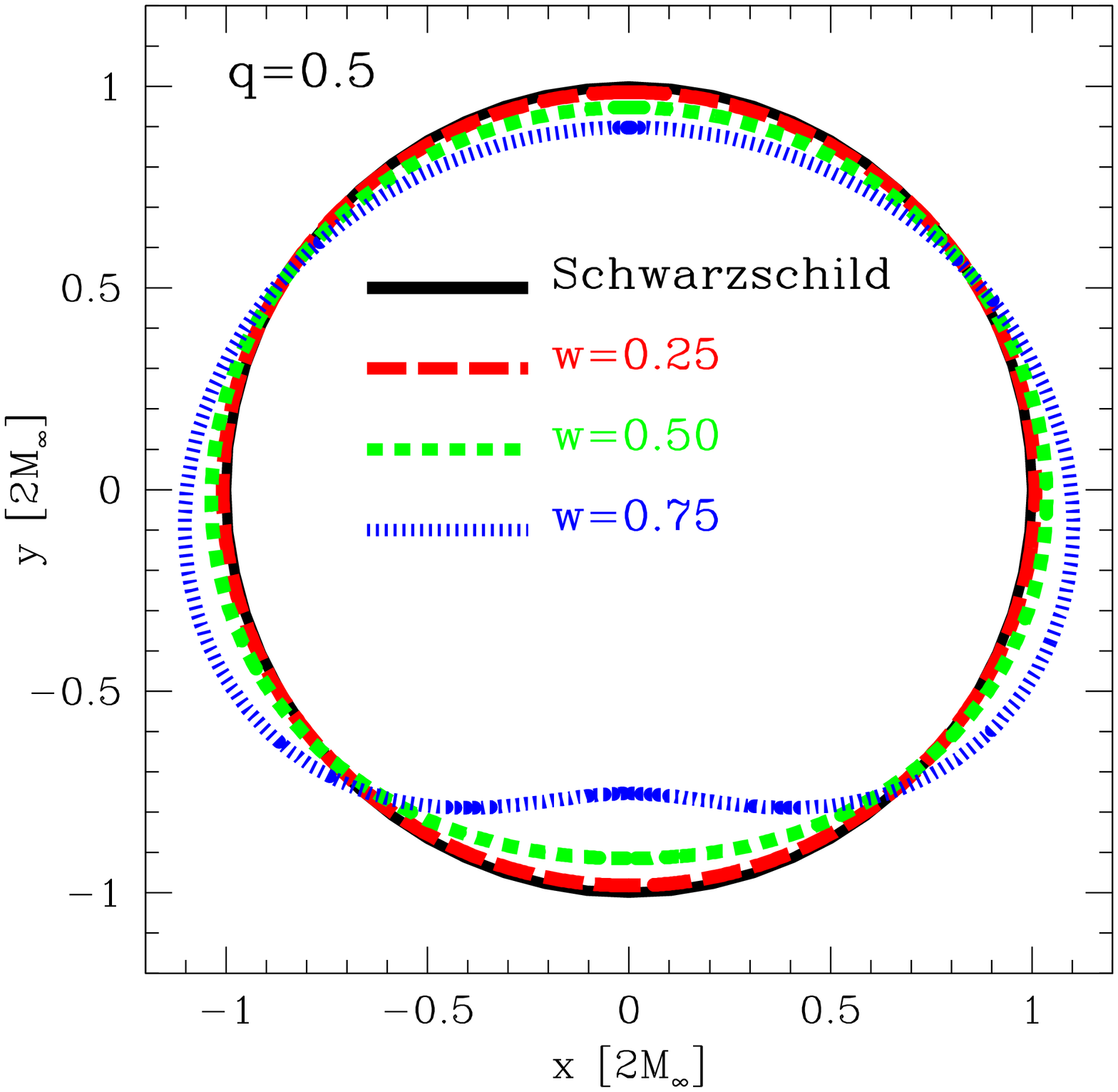}
\includegraphics[width=5.9cm,clip=true]{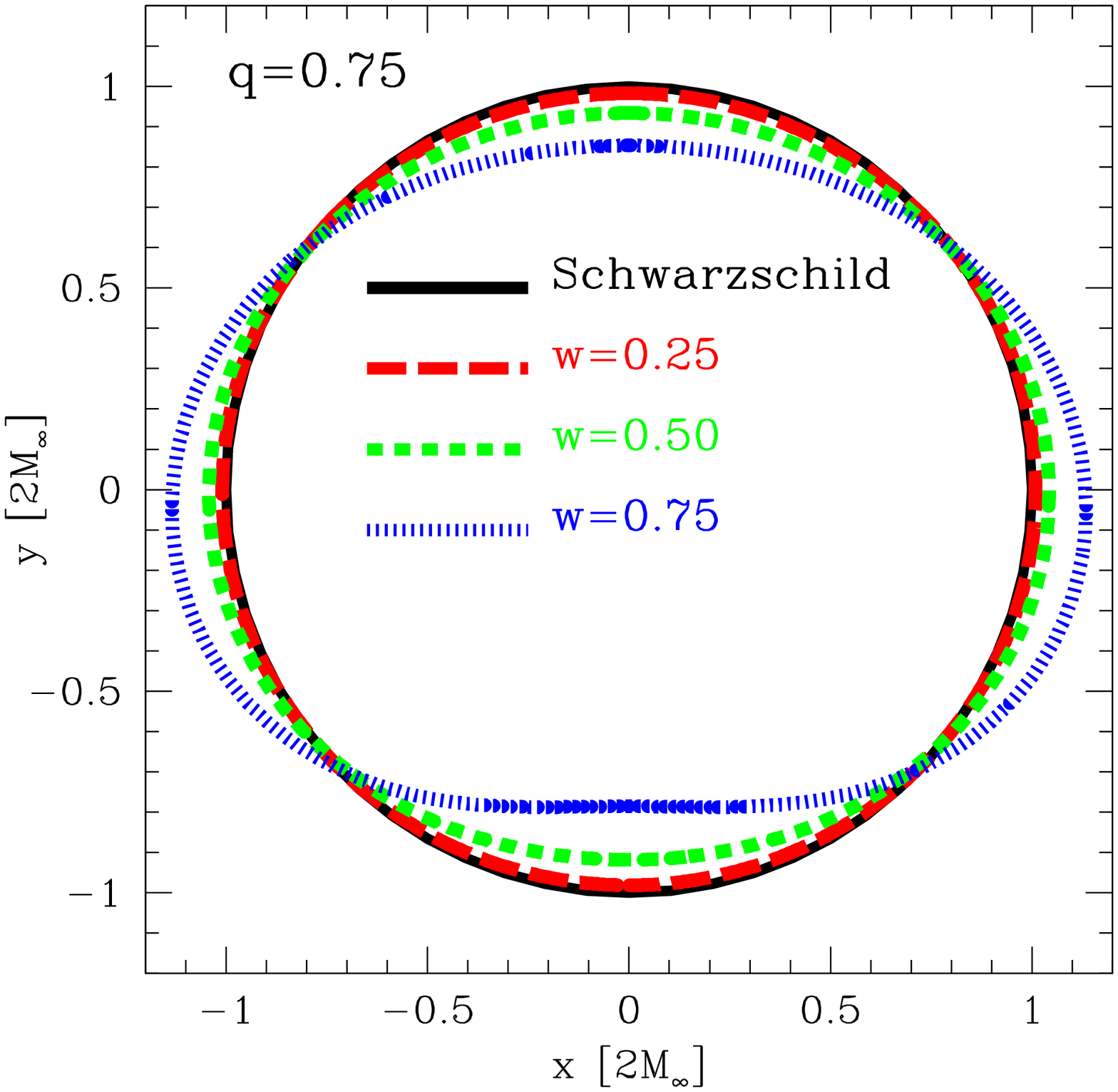}
\end{center}
\caption{AH of the head-on initial data given in
  Eq.~(\ref{Q0HeadOn}). The parameters $q$ and $w$ control the
  deformation of the surfaces. The final configuration is a
  Schwarzschild BH (continuous black line) and it is achieved after
  the deformation is dissipated with the emission of GWs. Our results
  confirm the interpretation given~\cite{Aranha:2008ni} for $q$ as the
  mass ratio of the BHs. However, we attach no physical meaning to $w$
  as done in~\cite{Aranha:2008ni}, for RT models a single deformed
  horizon, and nothing can be said about the velocities of the
  individual colliding BHs.}
\label{fig:horizonHO}
\end{figure*}

Equation (\ref{BHRT}) can be solved for the horizon $R(u,\theta)$
either by imposing regular conditions on the boundary
$\partial_{\theta}R(0)=0$ and $\partial_{\theta}R(\pi)=0$ and using an
ordinary shooting method to find $R(0)$ and $R(\pi$), or by
following the approach in~\cite{deOliveira:2009mc} introducing another
Galerkin decomposition on the horizon and truncated at the order
$N_{\text H}$
\begin{equation}
\label{BHGal}
\ln{\frac{R}{2M_\infty}}=\sum_{\ell=0}^{N_{\text H}} c_\ell(u)P_\ell(x).
\end{equation}
The projection of Eq.~(\ref{BHRT}) on the basis of the Legendre
polynomials couples the known Galerkin modes $b_{\ell}$ with the
unknown coefficients $c_{\ell}$ and the resulting algebraic nonlinear
system can be easily solved via a Newton-Raphson method.

\subsection{The initial data}

In general, any family of regular functions $Q$, \ie $0< Q(u_0,x)
<\infty, ~\forall x\in [-1,1]$ can be used as an initial data for the
RT spacetime. For any of such family, one can set a parameter $Q_0$ to
ensure the constant of motion to be $\mathcal{A}=4\pi$. Moreover, the
deformed BH will not be initially at rest in general. As a result,
given the initial velocity $v_{\rm k,0}\equiv P^3(0)/P^0(0)$, we
perform a boost ${\overline{P}}^{\alpha} = {\Lambda^{\alpha}}_{\beta} (v_{\rm
  k,0}) P^{\beta}$ with ${\Lambda^{\alpha}}_{\beta} (v_{\rm k,0})$ the
associated Lorentz transformation, so that ${\overline{P}}^{3}(0)=0$ by
construction. The kick velocity is then defined at any time as $v(u) =
\overline{P}^3(u)/\overline{P}^0(u)$.

Despite this overall simplicity, the definition of initial data that
is physically meaningful represents one of the main difficulties in
the study of RT spacetimes. Since we are here more interested in a
proof of principle than in describing a realistic configuration, we
have adopted both a prescription reminiscent of a ``head-on''
collision of two BHs~\cite{Aranha:2008ni} and a new variant of it.

\subsubsection{``Head-on'' initial data}

As a first set of initial data we consider the one developed in
Ref.~\cite{Aranha:2008ni}
\begin{equation}
\label{Q0HeadOn}
Q(0,\theta)=Q_0\left[\frac{1}{\sqrt{1 - w x}}+
\frac{q}{\sqrt{1 + w x}} \right]^{-2},
\end{equation}
which was interpreted to represent the final stages (\ie after a
common AH is formed) of a head-on collision of two boosted BHs with
opposite velocities $w \in [-1,1]$ and mass ratio $q \in
[0,1]$~\cite{Aranha:2008ni}. Figure~\ref{fig:horizonHO} shows the
shape of the surface $R(u_0,\tilde{\theta})$ for different values of
those parameters. It is worth remarking that despite the name, this
initial data does not represent a binary system but is, strictly
speaking, only a distorted horizon. In a more conservative approach,
one can regard $q$ and $w$ just as free parameters that control the
deformation on the horizon, and this is the view we will adopt
hereafter. However, a number of interesting analogies with the head-on
collision of two BHs have been suggested~\cite{Aranha:2008ni,
  Aranha:2010zz}, and will be further discussed below.

The interpretation of $q$ as a mass-ratio parameter is not totally
unreasonable. For instance, Fig.~3 of~\cite{Rezzolla:2010df} showed
the final value of the velocity in RT spacetime evolved from the
head-on initial data against the reduced mass ratio $\nu \equiv
q/(1+q)^2$. The curve obeys the distribution
\begin{equation}
\label{Vkick_mass}
\vk=A\nu^2\sqrt{1-4\nu}(1+B\nu),
\end{equation} 
as found in all numerical simulations~\cite{Rezzolla:2008sd}, with the
value of $A$ and $B$ depending on the particular choice of
$w$. 

Since the solution does not exist for $u<u_0$, it is impossible to
assign a value for $w$ that could account for any previous stages on
the evolution of the binary. With the original interpretation of $w$
as the velocities of the BHs, a first trivial estimate, as proposed
by~\cite{Moreschi:1996ds}, is to assume a Newtonian evolution of two
particles with masses $M_1$ and $M_2$, which start at rest at an
initial distance $L_0$. At a given distance $L$, in the frame where
$v_1=-v_2=w$ one has
\begin{equation}
\label{Westimative}
w=\sqrt{\frac{1}{2}\left(\frac{M}{L}-\frac{M}{L_0} \right)},
\end{equation}
with $M=M_1+M_2$. Choosing $L_0 \simeq 6\,M$ and $L \simeq 2\,M$, one
obtains $w \simeq 0.41$.  Furthermore, still in
Ref.~\cite{Rezzolla:2010df} it was shown that $w=0.425$ presents a
surprisingly good match with some results found in the close-limit
approximation, where the initial data for the ringdown phase was given
by a previous plunge with the BHs inspiralling towards each other from
the innermost circular orbit until $\thicksim
2M$~\cite{LeTiec:2009yg}.

\begin{figure*}[t]
\begin{center}
\resizebox{0.6\linewidth}{!}{ \includegraphics{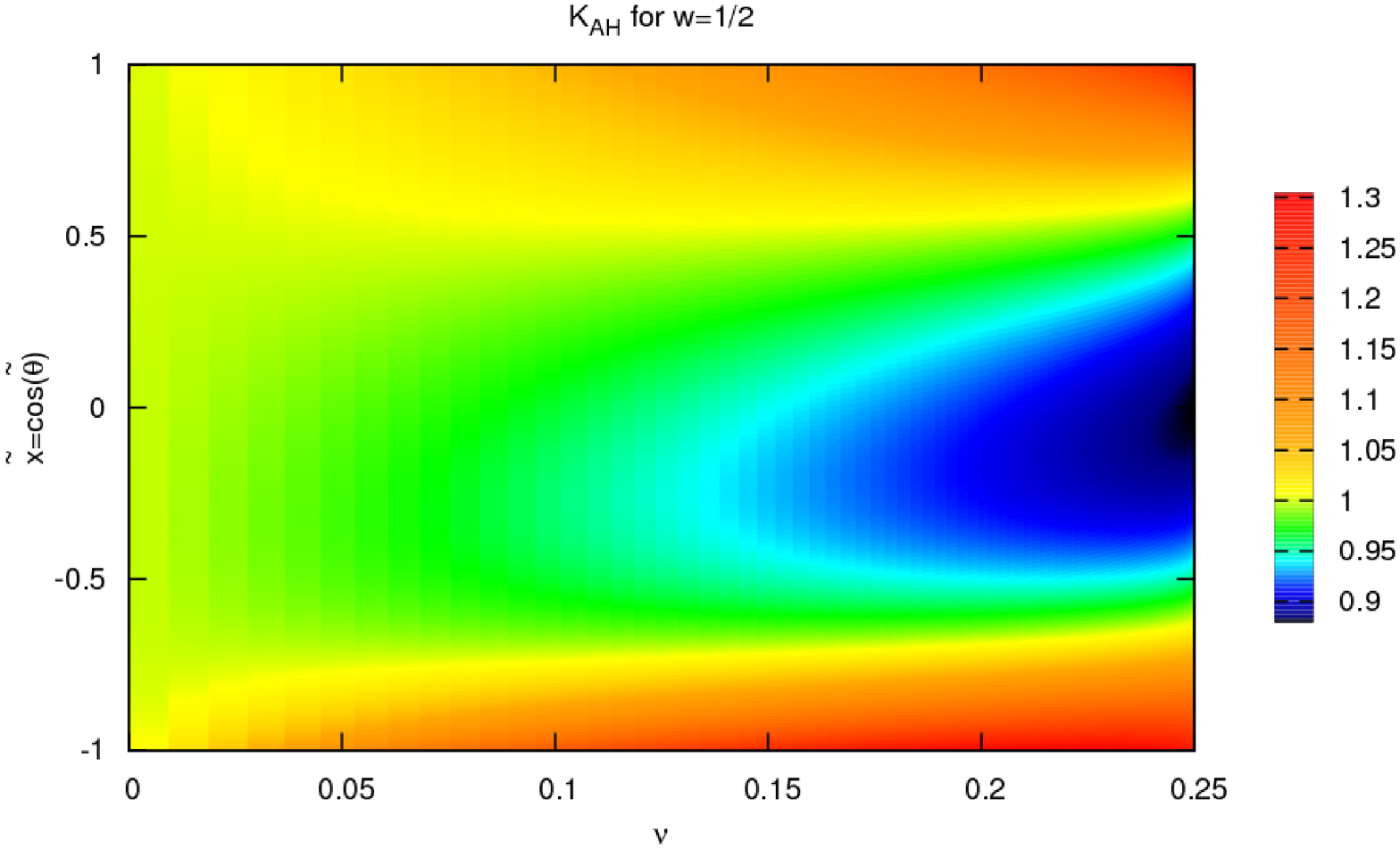}}
\hglue 0.5cm
\resizebox{0.35\linewidth}{!}{ \includegraphics{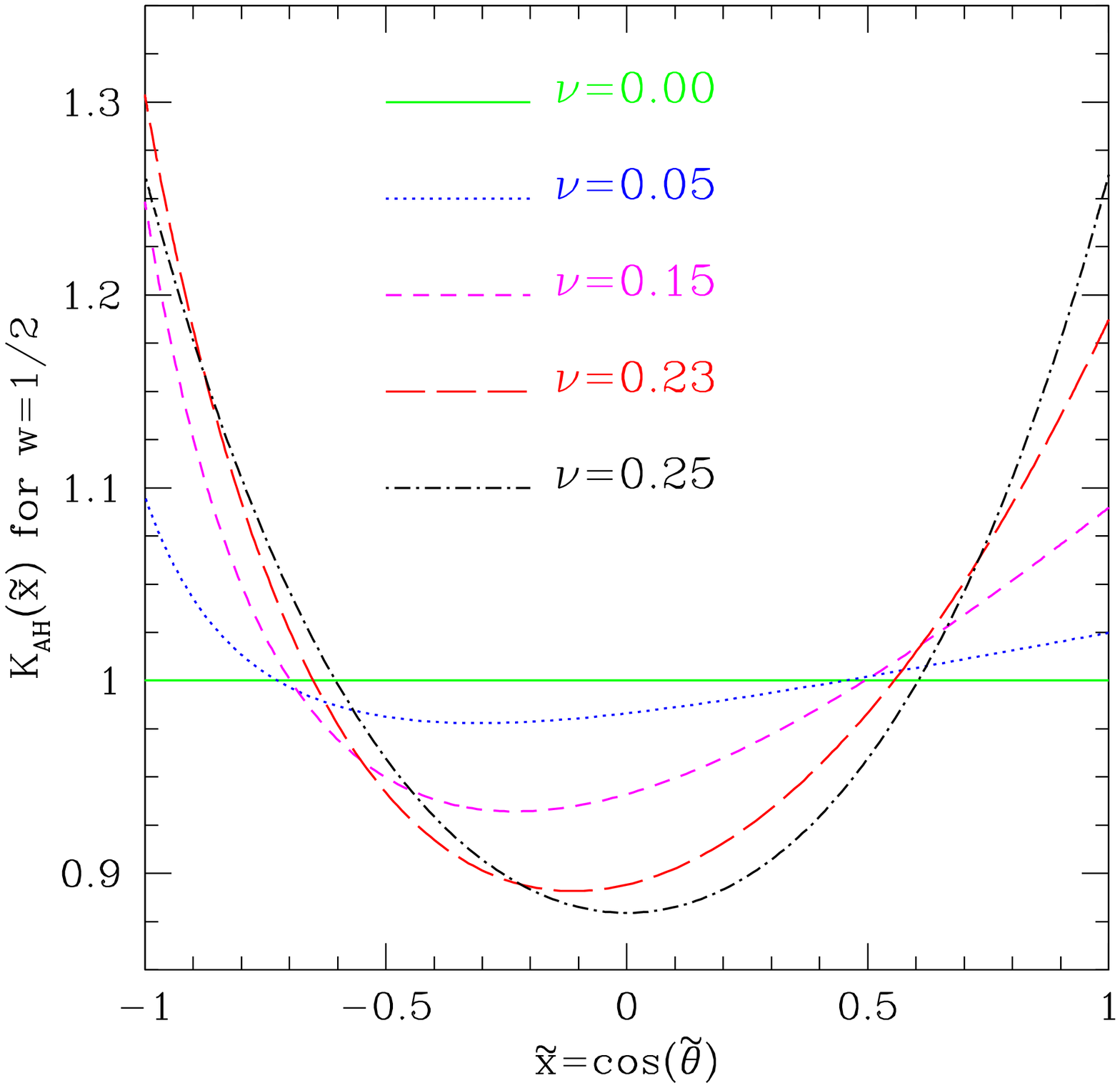}}
\end{center}
\caption{\textit{Left panel:} Horizon mean curvature $K_{_{\rm H}}$ for
  $w=1/2$ shown as a function of reduced mass ratio $\nu$ and of the
  polar angle $x=\cos\theta$. Note that the distribution is symmetric
  with respect to the equatorial plane for $\nu=0$ and
  $\nu=1/4$. \textit{Right panel:} Horizon curvature for some
  representative values of the reduced mass ratio. The low-$\nu$
  branch is characterized by large curvature gradients across the AH
  but small values of the curvature, while the high-$\nu$ branch small
  curvature gradients and large values of the
  curvature~\cite{Rezzolla:2010df}.}
\label{fig:K_vs_nu}
\end{figure*}

It is important to remark that one should not expect a complete
agreement between the values of $A$ and $B$ from the head-on collision
in RT spacetimes with the ones found in numerical-relativity
calculations of binary BHs in quasicircular
orbits~\cite{Gonzalez:2006md}. The first ones, in fact, (and modulo
the interpretative issues discussed above) can only account for the
post-merger phase, while the second ones account for the whole
recoil. A complete discussion on the dependence of $A$ and $B$ with
respect to $w$ can be found in~\cite{Aranha:2010zz}. For the sake of
convenience we will fix $w=0.5$, but our results do not depend upon
this choice. The substitution $w \rightarrow -w$ just changes the sign
of the recoil velocity.

\begin{figure}[b]
\begin{center}
\includegraphics[width=8.0cm,clip=true]{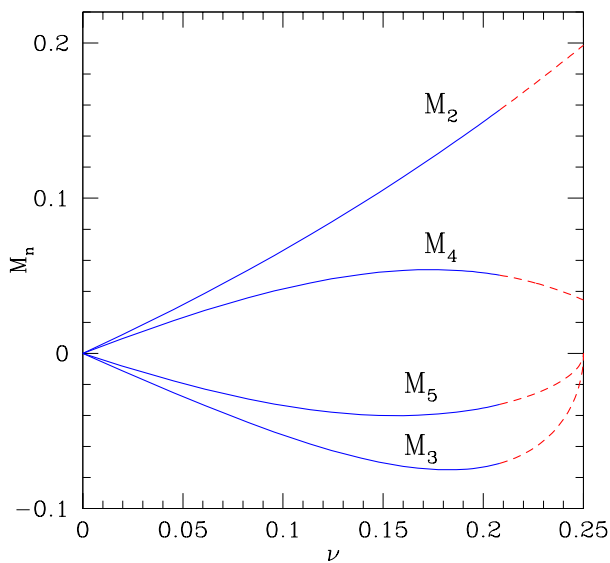}
\end{center}
\caption{Lower-order mass moments for $w=1/2$. Note that they all
  vanish at $\nu=0$, while only the odd ones are zero for
  $\nu=1/4$. As in Fig.~3 of~\cite{Rezzolla:2010df}, the colors
  represent the two different branches composing the curve $\vk$
  versus $\nu$. The color changes at the mass ratio for which the
  recoil velocity is at a maximum.}  
\label{fig:fig6}
\end{figure}

The relation between the kick velocity and the reduced mass ratio
expressed by Eq.~(\ref{Vkick_mass}) has a peak $v_{\rm k}^{\rm max}$
for $\nu^{\rm max} \approx 0.195$~\cite{Gonzalez:2006md}. As the
recoil vanishes for $\nu = 0$ and $\nu = 1/4$, there will always be
two values of the mass parameter leading to the same recoil when
$v<v_{\rm k}^{\rm max}$. It is natural to wonder whether two
completely different systems (namely systems with different mass
ratios) share some common physical property which could lead to the
same recoil. Thinking in terms of the horizon's intrinsic deformation
provides a simple way to explain this degeneracy. System with $\nu <
\nu^{\rm max}$, in fact, are characterized by large curvature
gradients across the AH but small values of the curvature, while
system with $\nu >\nu^{\rm max}$ are characterized by small curvature gradients and
large values of the curvature. This intuitive picture can be best
appreciated in Fig.~\ref{fig:K_vs_nu}, which reports the horizon mean
curvature $K_{_{\rm H}}$ for $w=1/2$ as a function of reduced mass
ratio $\nu$ and of the polar angle $\tilde{x}=\cos\tilde{\theta}$
(left panel) or for some representative values of the reduced mass
ratio (right panel). Note that the low-$\nu$ branch is characterized
by large curvature gradients between the north and south poles of the
AH, but small values of the curvature, while the high-$\nu$ branch is characterized by
small curvature gradients and large values of the curvature.  As
discussed in Ref.~\cite{Rezzolla:2010df}, it is the product of the
deformations on the horizon with the gradients across the equator that
yields the same recoil for two apparently different systems.

\begin{figure*}[t!]
\begin{center}
\resizebox{0.55\linewidth}{!}{ \includegraphics{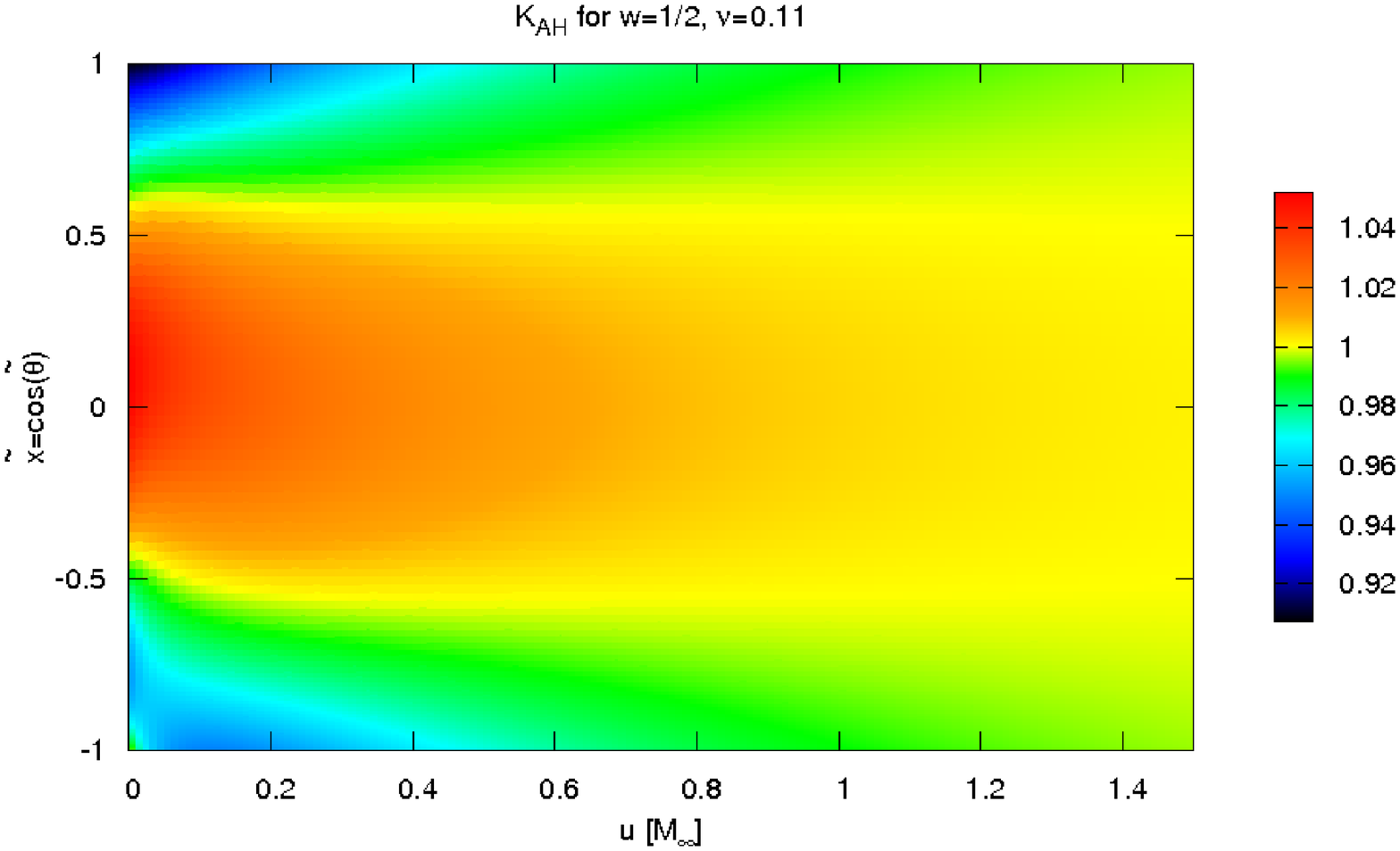}}
\hglue 0.5cm
\resizebox{0.35\linewidth}{!}{\includegraphics{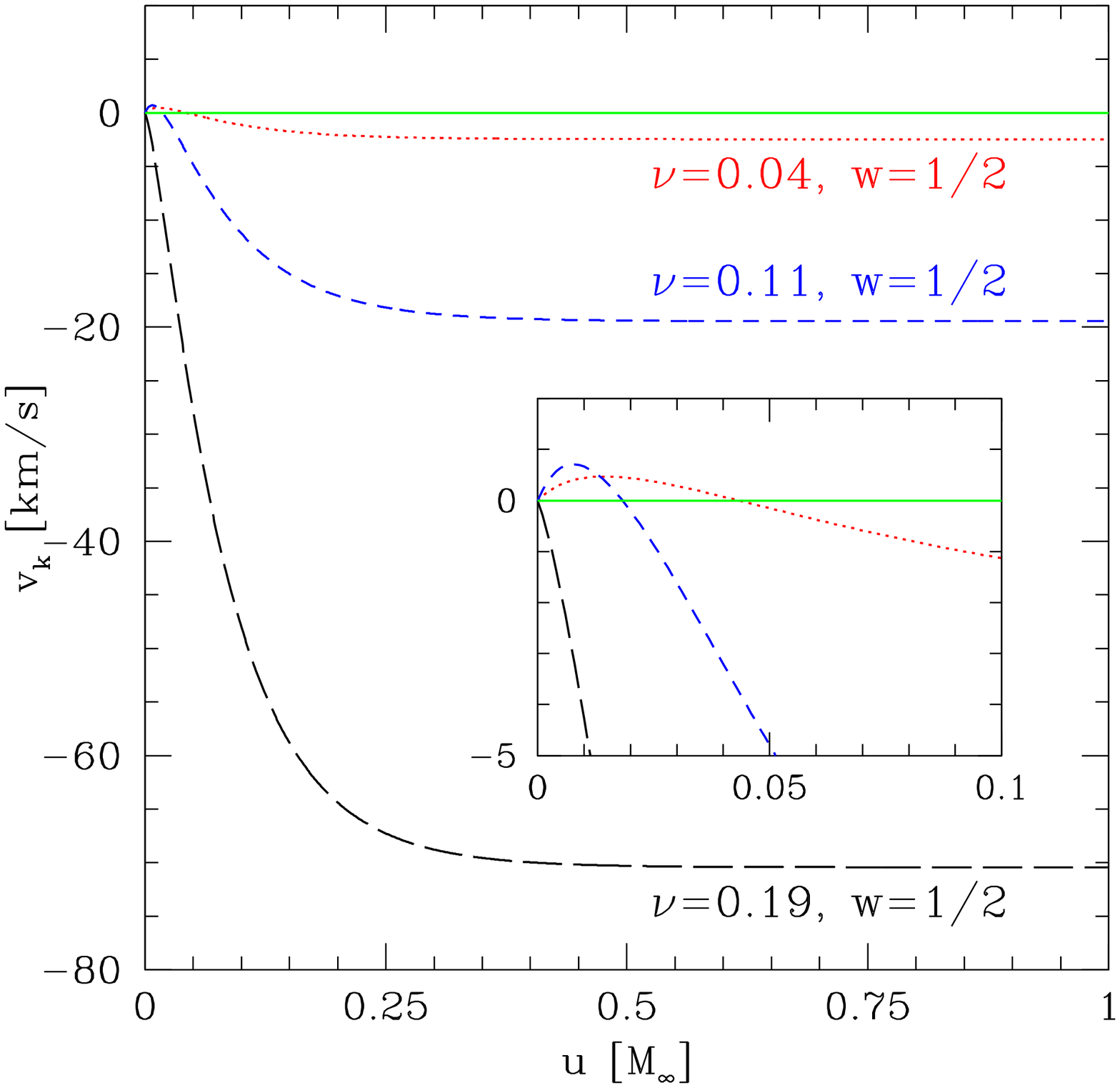}}
\end{center}
\caption{\textit{Left panel:} Curvature evolution for the distorted-AH
  initial data with $w=1/2, \nu=0.11$. Note the small curvature excess
  at the south pole, which is rapidly radiated away to yield an almost
  uniform distribution after $u/M_{\infty} \simeq 1.5$. \textit{Right
    Panel:} evolution of the recoil velocity for representative values
  of the parameter $q$, some of which lead to no-monotonic changes in
  the velocity.}
\label{fig:fig8}
\end{figure*}

Complementary information to the one in the left panel of
Fig.~\ref{fig:K_vs_nu} is depicted in Fig.~\ref{fig:fig6}, which shows
the typical behavior of the lower-order mass moments as a function of
the reduced mass ratio. Notice they all vanish for $\nu=0$, since this
configuration represents an undistorted BH. For $\nu=1/4$, on the
other hand, only the odd modes are zero, indicating that the
configuration is symmetric with respect to the equatorial plane and
the emission of GW will not give rise to a recoil. Also note that the
maximum of the odd modes does not correspond to the mass ratio at
which the recoil velocity reaches its highest value. Even though a net
emission of momentum will only take place when there is an asymmetry
on the horizon across the equator (\ie $M_{\rm odd} \ne 0$), the
intensity of the emission will also depend on how deformed the BH is
(\ie $M_{\rm even} \ne 0$).

Without loss of generality, we can use the even modes to measure
overall distortions on the horizon, while the odd ones measure the
asymmetries between the north and south hemispheres. To account for
both contributions, we constructed in~\cite{Rezzolla:2010df} an
effective-curvature parameter as the product of two functions
depending solely on the even or the odd modes, \ie $K_{\mathrm{eff}} =
f_{\mathrm{even}} \left(M_{2n}\right) \times f_{\mathrm{odd}}
\left(M_{2n+1} \right)$. This quantity represents a measure of the
global curvature properties of the initial data, from which the recoil
depends in an injective way. Indeed, Fig.~4 in~\cite{Rezzolla:2010df}
showed that with a suitable choice of coefficients, \ie $K_{\rm
  eff}=M_2|\sum_{n=1} M_{2n+1}/3^{n-1}|$, the correlation between
$K_{\rm eff}$ measured at the initial time against the final velocity
is actually linear.

\subsubsection{Distorted-AH initial data}
\label{secDistAH}

The evolution produced by the head-on initial data~\eqref{Q0HeadOn}
leads to a monotonic increase/decrease of the recoil velocity once the
initial data is specified on the white hole. However, a more complex
(\ie nonmonotonic) dynamics can be easily produced through a simple
variation of the head-on initial data. We refer to this new family of
initial conditions as to the ``distorted AHs'' and we express them as
\begin{equation} \label{DistortedID}
Q(u_0,x) = Q_{\text{HO}}(x; q,w) + qx^2 Q_{\text{HO}}(x; q,-w),
\end{equation}
where $Q_{\text{HO}}(x; q,w)$ corresponds to the head-on initial data
(\ref{Q0HeadOn}). Clearly, this is just a mathematical choice and no
physical significance can be associated to this initial data.

Note that~\eqref{DistortedID} maintains the symmetries provided by the
mass-ratio parameter $q$: for $q=0$ one recovers the nondeformed
Schwarzschild BH and $q=1$ gives an even initial data, \ie
$Q(u_0,x)=Q(u_0,-x)$, which leads to a zero final recoil. Furthermore,
note that the resulting recoil velocity does not lead to the scaling
expressed by Eq.~(\ref{Vkick_mass}), thus giving strength to the idea
that the head-on initial data is closely related to the merger of a
binary system as proposed in Refs.~\cite{Aranha:2008ni,
  Aranha:2010zz}.

As anticipated, the use of this initial data leads to a more
interesting dynamics and this is shown in Fig.~\ref{fig:fig8}, whose
left panel reports the curvature evolution for $w=1/2$ and $\nu=0.11$,
while the right panel reports the evolution of the recoil velocity for
representative values of the parameter $q$, some of which lead to
nonmonotonic changes in the velocity\footnote{This evolution is
  related to the Bondi momentum as defined by~\cite{Tod83,
    Singleton:90} and given by Eq.~(\ref{BondiMoment}). Recently, a
  different approach has been proposed in~\cite{Aranha:2010zz}, where
  they showed a slightly different profile for the velocity time
  evolution. However, the difference is not important in our argument,
  since we are mainly concerned with values of the curvature at the
  initial time, when we boost $P^a$ to its rest frame, and its
  correlation with the asymptotic final velocity, when the momentum is
  unambiguously defined.}. Note the small curvature excess at the
south pole, which is rapidly dissipated as GWs are emitted so as to
yield an almost uniform distribution. As pointed out
in~\cite{Rezzolla:2010df}, the velocity reaches its final value when
there is no asymmetry in the deformation between the north ($\tilde{x}
> 0$) and south ($\tilde{x} < 0$) hemispheres, \ie after $u/M_{\infty}
\simeq 0.5$.

\begin{figure}[t]
\begin{center}
\includegraphics[width=8.0cm,clip=true]{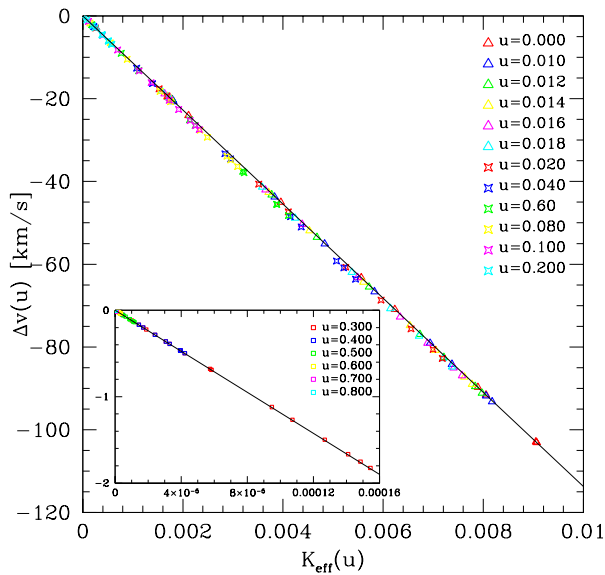}
\end{center}
\caption{Difference $\Delta \vk(u) \equiv \vk(u_\infty)-\vk(u)$
  against the effective-curvature parameter for several values of the
  ``initial'' time $u$. The linear relation between the effective
  curvature (\ref{e:Keff_DistorAH}) and the kick velocity is preserved
  along the evolution as indicated by the different symbols, each
  representing a specific time in the evolution. This remains the case
  also at late times as shown in the inset.}
\label{fig:Keff_dV}
\end{figure}

To prove that the approach discussed in the previous subsection is
indeed generic, we define an effective curvature $K_{\mathrm{eff}}$
also for this family of initial data, again in terms of the product of
odd and even mass moments
\begin{widetext}
\begin{eqnarray} 
\label{e:Keff_DistorAH}
K_{\rm eff}&=&
(M_2 + a_4\,M_4 + a_6\,M_6 + a_8\,M_8 + a_{10}\,M_{10}) 
\times 
(M_3 + a_5\,M_5 + a_7\,M_7 + a_9\,M_9)\,,
\label{KeffDistID} 
\end{eqnarray}
\end{widetext}
to find that the set of coefficients \mbox{$a_4=0.304, a_6=0.178$},
\mbox{$a_8=0.086, a_{10}=-0.186$} and \mbox{$a_5=0.076, a_7=-0.090$},
$a_9=-0.183$ leads to the expected injective, and actually linear, behavior. 
Clearly, and as it is also natural to expect, the coefficients are different 
from those found for the head-on initial data, and they will always depend 
upon the specific family of initial data considered. The remarkable feature
though, is that they remain constant in time. This is illustrated in
Fig.~\ref{fig:Keff_dV}, which shows that the effective curvature
(\ref{KeffDistID}) is still linear with respect to the relative
velocity $\Delta \vk(u) \equiv v(u_\infty)-v(u)$ at any time during
the evolution (this is shown by the different coloured symbols, each
of which refers to a specific time) and also at late times (see
inset). This time-independent property is general, and not limited to
this particular family of initial data. In particular, it is also
found, for instance, in the head-on case. This result reflects the
fact that the deformations of the horizon evolve in time in a
self-similar manner, so that although the ranges for $K_{\rm eff}$
change in time (becoming smaller as the deformations are radiated
away), the corresponding recoil velocities maintain the same
proportionality (\cf Fig.~\ref{fig:Keff_dV}).

\begin{figure}[t]
\begin{center}
\includegraphics[angle=0, width=8.0cm,clip=true]{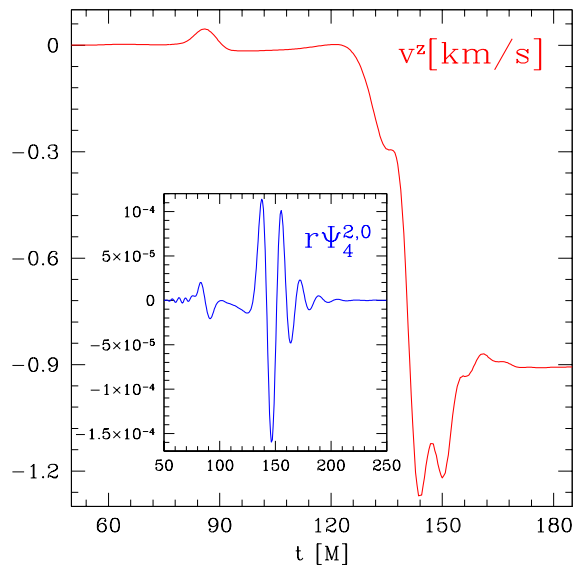}
\end{center}
\caption{Evolution of the velocity (red curve) measured with the flux
  of momentum carried out by the GWs. Note the antikick at about $t/M
  \approx 145$ that decelerates the system before the final kick
  velocity is reached. The GW signal is instead shown in the inset, namely 
  the dominant $\Psi_4^{2,0}$ multipole (blue curve).}
\label{fig:BBH_v_kick}
\end{figure}

As a concluding remark, we can summarize as follows the insight gained
through the study of RT spacetimes: the construction of the effective-curvature 
parameter depends quantitatively on the family of initial
data considered, but that for any choice of data, it is possible to
find an explicit expression that relates the effective-curvature
parameter to the (final) recoil velocity through a one-to-one
mapping. What however will not depend on the initial data is the
functional dependence of the effective-curvature parameter as
expressed by~\eqref{e:K_eff_Ansatz}. Indeed, in the next section we
will generalize the idea and functional form of the effective-curvature
 parameter to account for the dynamics in binary BH
spacetimes.

\begin{figure*}
\begin{center}
\includegraphics[width=5.5cm]{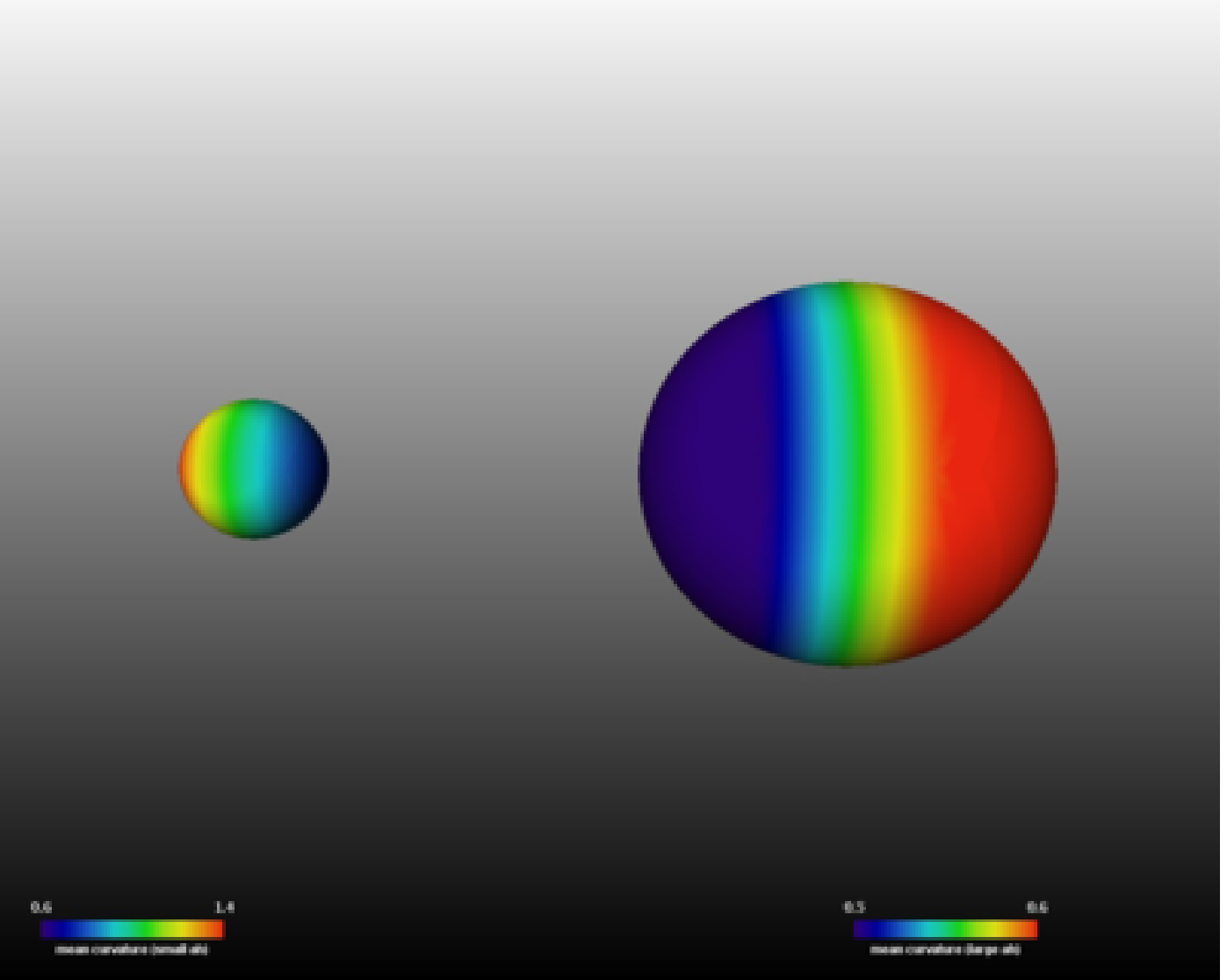}
\hskip 0.1cm
\includegraphics[width=5.5cm]{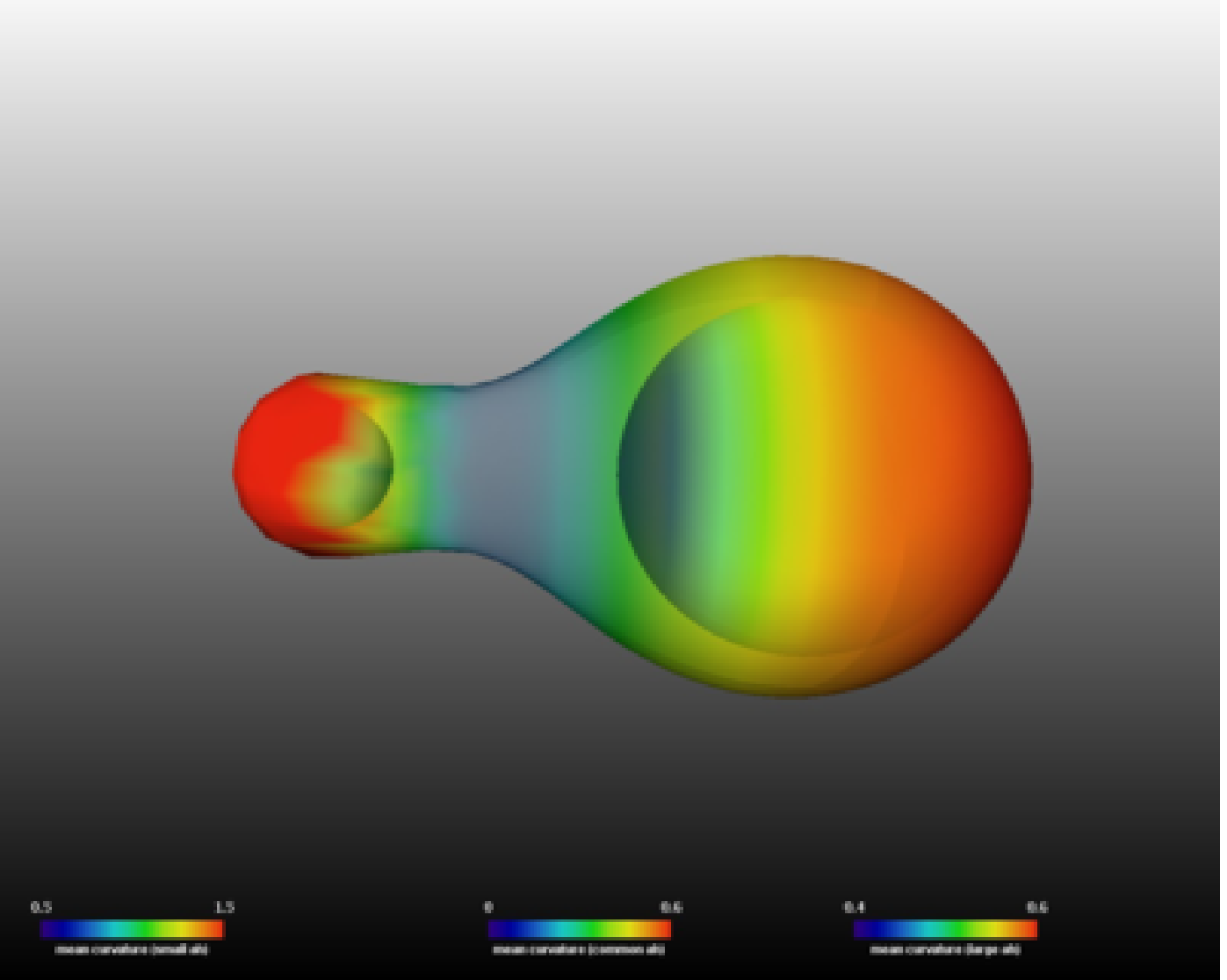}
\hskip 0.1cm
\includegraphics[width=5.5cm]{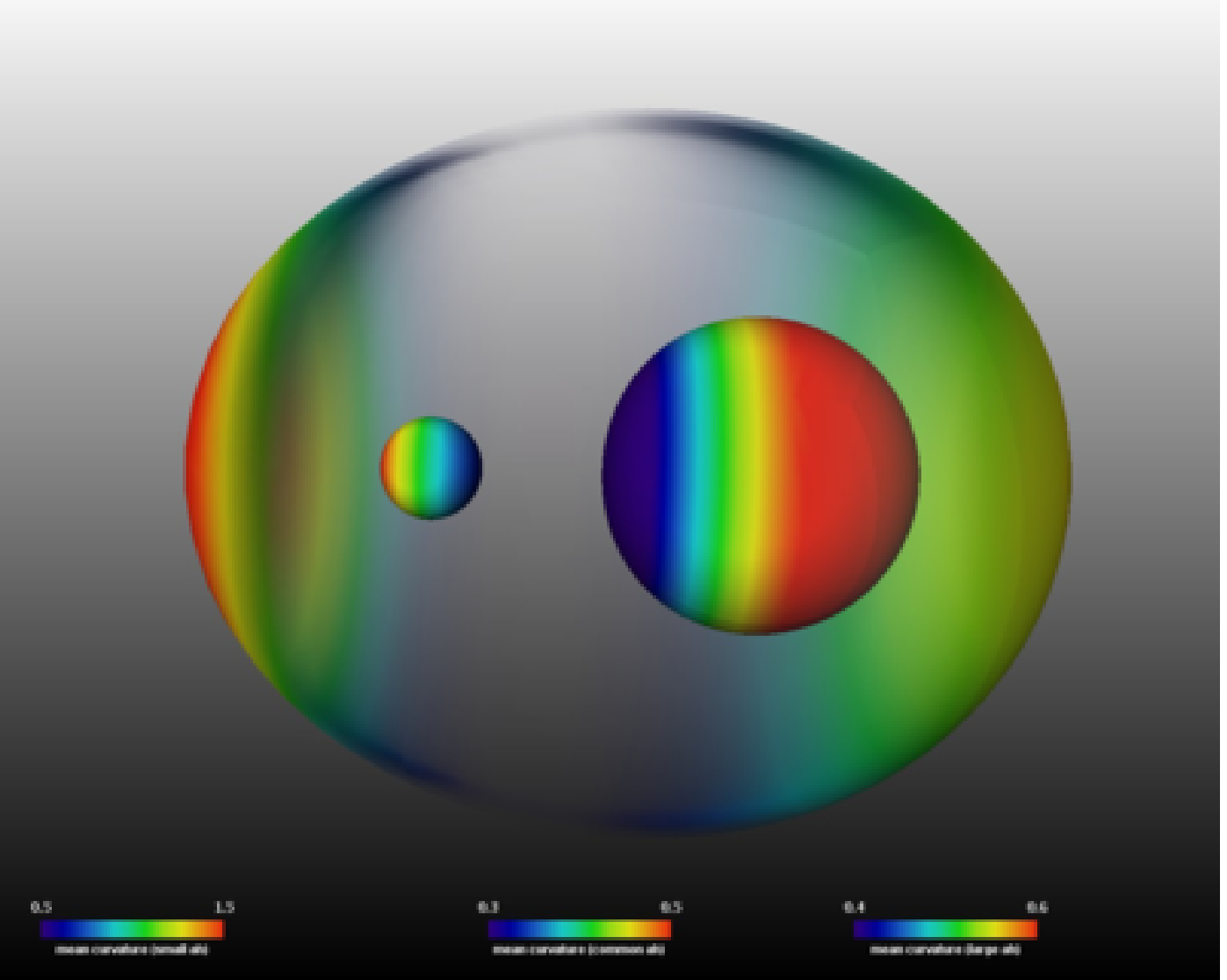}
\end{center}
\caption{Realization of the cartoon in Fig.~\ref{fig:cartoon} with
  numerical data from a simulation of head-on collision with mass
  ratio $q=1/2$. The color code shows the mean curvature on the
  apparent horizons, which has the same anisotropic behavior of the
  intrinsic curvature. As described in Fig.\ref{fig:cartoon}, once the
  common horizon is formed, the curvature is stronger in the region of
  the smaller BH and is dissipated during the evolution.}
\label{fig:nocartoon}
\end{figure*}

\section{Black-Hole spacetimes: Head-on collisions}
\label{s:BBH}
\subsection{Numerical Setup and Results}
\label{s:num_methods}

The numerical solution of the Einstein equations has been performed
using a three-dimensional finite-differencing code solving a
conformal-traceless ``$3+1$'' BSSNOK formulation of the Einstein
equations (see~\cite{Pollney:2009yz1} for the full expressions) using
the \texttt{Einstein Toolkit}~\citep{einsteintoolkitweb}, the
\texttt{Carpet}~\citep{Schnetter-etal-03b} adaptive mesh-refinement
driver, \texttt{AHFinderDirect}~\cite{Thornburg2003:AH-finding} to
track the AHs, and \texttt{QuasiLocalMeasures}~\cite{Dreyer02a} to
evaluate the mass multipoles associated with them. Recent
developments, such as the use of $8$th-order finite-difference
operators or the adoption of a multiblock structure to extend the size
of the wave zone have been recently presented
in~\cite{Pollney:2009ut,Pollney:2009yz1}. Here, however, to limit the
computational costs and because a very high accuracy in the waveforms
is not needed, the multiblock structure was not used. Also, for
compactness we will not report here the details of the formulation of
the Einstein equations solved for the form of the gauge conditions
adopted. All of these aspects are discussed in detail
in~\cite{Pollney:2009yz1}, to which we refer the interested reader. 

Our initial data consists of head-on (\ie zero angular momentum)
Brill-Lindquist initial data with a mass-ratio of $q = 1/2$. The
initial separation of both BHs is $10M$ and they are initially located
at $(0,0,6.6666$) and $(0,0,-3.3333)$ to reflect their center-of-mass
offset. Both BHs have no angular nor linear momentum initially. We use
a 3D Cartesian numerical grid with seven levels of mesh-refinement for
the higher mass and eight levels of mesh-refinement for the lower mass
BH. The resolution of our finest grid is $h=M/64$, while the angular
grid used to find the AHs and evaluate any property on these
2-surfaces has a resolution of $65$ points in $\varphi$-direction and
$128$ points in $\theta$-direction. The extraction of GWs is performed
calculating $\Psi_4$ at finite-radius detection spheres with radii of
$r_1=60\,M\,,r_2=85\,M$ and $r_3=145\,M$ and then extrapolating to
infinity.

Some of the most salient results of the numerical simulations are
summarized in Fig.~\ref{fig:BBH_v_kick}, which reports the evolution
of the recoil velocity (red curve) measured with the flux of momentum
carried out by the GWs. Note the development of the antikick at about
$t/M \approx 145$ (followed by several smaller oscillations) that
decelerates the BH before the final kick velocity is reached. Also
shown in the inset is GW signal in its larger multipolar component
$\Psi_4^{2,0}$ (blue curve). Similarly, Fig.~\ref{fig:nocartoon}
provides a realization of the cartoon in Fig.~\ref{fig:cartoon} with
numerical data from a simulation of head-on collision with mass ratio
$q=1/2$. Shown with a color code is the mean curvature on the apparent
horizons, which shares the same qualitative properties, and in
particular the anisotropic behavior, of the intrinsic curvature. As
intuitively described in Fig.\ref{fig:cartoon}, once the common
horizon is formed, the curvature is stronger in the region of the
smaller BH and is dissipated as the evolution proceeds. Note that the
curvature distribution is anisotropic already at the beginning, as the
BHs are tidally distorting each other.

\subsection{Geometric quantities at the BH horizon: 
$\tilde{K}^{\mathrm{eff}}_i(t)$}
\label{s:Keff}

As discussed in Secs.~\ref{s:Intro} and \ref{s:Exe_summ}, the
analysis of the recoil dynamics in generic BH spacetimes requires a
shift with respect to the methodology used in RT spacetimes. When considering
standard $3+1$ numerical solutions of BH spacetimes, in fact, we 
study the near-horizon dynamics responsible for the BH recoil in
terms of the time cross-correlations between a vector $(dP_i^{\cal
  B}/dt)(t)$ at a large-radius hypersurface ${\cal B}$ and an
effective-curvature vector $\tilde{K}^{\mathrm{eff}}_i(t)$ constructed
from the intrinsic geometry on the dynamical BH horizon ${\cal
  H}^+$. The vector $(dP_i^{\cal B}/dt)(t)$ on ${\cal B}$ approximates
the Bondi linear momentum flux $(dP_i^{\mathrm{B}}/dt)(t)$ at
$\scri^+$. From now on we will systematically refer to
$(dP_i^{\mathrm{B}}/dt)(t)$ (and to $\scri^+$ instead of ${\cal B}$),
understanding that we are actually using an approximation.

The construction of $\tilde{K}^{\mathrm{eff}}_i(t)$ at ${\cal H}^+$ is
based on the following two guidelines: a)
$\tilde{K}^{\mathrm{eff}}_i(t)$ is built out of the intrinsic geometry
Ricci scalar curvature ${}^2\!R$ on ${\cal H}^+$ sections; b) the
functional form of $(dP_i^{\mathrm{B}}/dt)(t)$ in terms of the
geometry at $\scri^+$ guides the choice of the functional dependence
of $\tilde{K}^{\mathrm{eff}}_i(t)$ on ${}^2\!R$. The first requirement
is motivated by the success in the RT case, whereas the second one
aims at preserving those basic structural features of the specific
function to be cross-correlated.

Following these guidelines, we start by expressing the flux of Bondi
linear momentum at null infinity. In terms of a retarded time $u$
parameterizing $\scri^+$, its Cartesian components can be written as
\begin{eqnarray}
\label{e:flux_Bondi_momentum_u}
\frac{dP_i^{\mathrm{B}}}{du}(u) =  \lim 
\limits_{(u, r \to \infty)}
\frac{r^2}{8 \pi} \oint_{{\cal S}_{u,r}} s_i \; |{\cal N}(u,\Omega)|^2 d\Omega \,,
\end{eqnarray}
where $r$ parameterizes the large-radius spheres ${\cal S}_{u,r}$ along
a $u=\mathrm{const.}$ hypersurface, $r^2d\Omega$ is the area element
on ${\cal S}_{u,r}$, $s^i$ is its normal unit vector with Cartesian
components $s^i=\left(\mathrm{\sin}\theta\mathrm{\cos}\varphi,
\mathrm{\sin}\theta\mathrm{\sin}\varphi,
\mathrm{\cos}\theta\right)$, and the {\em news functions} ${\cal
  N}(u)$ can be expressed in terms of the Weyl scalar $\Psi_4$ as
\begin{eqnarray}
\label{e:News_scri_u}
{\cal N}(u, \Omega) = 
\int_{-\infty}^u \Psi_4(u',\Omega) du' \,.
\end{eqnarray}

In our $3+1$ setting with an outer boundary at a finite spatial
distance we need to express the flux with respect to the time function
$t$ parameterizing the spatial slices $\Sigma_t$, so that we
replace ${\cal S}_{u,r}$ with ${\cal S}_{t,r}$ 
\begin{equation}
\label{e:flux_Bondi_momentum}
\frac{dP_i^{\mathrm{B}}}{dt}(t) = \lim \limits_{r \to \infty}
\frac{r^2}{16 \pi} \oint_{{\cal S}_{t,r}} s_i \left| \int_{-\infty}^t
\Psi_4(t',\Omega) dt'\right|^2 d\Omega \,,
\end{equation}
and where we can think of $t$ (related to $u$ by $u=t-r$ near
$\scri^+$) as parameterizing the cuts of $\scri^+$ by hyperboloidal
slices or, alternatively, the cuts of the timelike
hypersurface ${\cal B}$ approximating $\scri^+$ at large $r$. We can
now rewrite expression (\ref{e:flux_Bondi_momentum}) in terms of a
generic vector $\xi^i$ {transverse} to ${\cal S}_{u,r}$ (\ie with a
generically nonvanishing component along the normal to ${\cal
  S}_{u,r}$), so that the component of the flux of Bondi linear
momentum along $\xi^i$ is
\begin{eqnarray}
\label{e:flux_Bondi_momentum_xi}
\frac{dP^{\mathrm{B}}[\xi]}{dt}(t) =  \lim \limits_{r \to \infty}  
\frac{r^2}{16 \pi} \oint_{{\cal S}_{t,r}} (\xi^i s_i) 
\left| \int_{-\infty}^t \Psi_4(t',\Omega) dt'\right|^2 d\Omega \,. \nn\\
\end{eqnarray}

We take this expression as the starting point for the construction of
$\tilde{K}^\mathrm{eff}_i$. It provides the functional form of the
Bondi linear momentum flux in terms of the relevant component of the
Riemann tensor at $\scri^+$, namely $\Psi_4$. Then, the two
above-mentioned guidelines for the construction of
$\tilde{K}^\mathrm{eff}_i$ can be met by considering a heuristic
substitution of $\Psi_4$ by ${}^2\!R$ in
expression~(\ref{e:flux_Bondi_momentum_xi}).

It is important to note that in the same way in which the outgoing
null coordinate $u$ parameterizes naturally $\scri^+$, the ingoing null
coordinate $v$, which runs along $\scri^-$, is a natural label
to parameterize the horizon ${\cal H}^+$. However, within our $3+1$
setting, we use Eq.~(\ref{e:flux_Bondi_momentum_xi}) as the ansatz leading
to the following proposal for the component
$\tilde{K}_\mathrm{eff}[\xi](t)$ of $\tilde{K}^\mathrm{eff}_i(t)$
along a vector $\xi^i$ (tangent to the slice $\Sigma_t$) transverse to
the section ${\cal S}_t$ of ${\cal H}^+$
\begin{eqnarray}
\label{e:3D_Keff}
\tilde{K}^\mathrm{eff}[\xi](t) \equiv  - \frac{1}{16\pi}
\oint_{{\cal S}_t} (\xi^is_i) \left|\tilde{\cal N}(t,\Omega)\right|^2 dA \,,
\end{eqnarray}
with
\begin{eqnarray}
\label{e:news_Rt}
\tilde{\cal N}(t,\Omega) \equiv 
\int_{t_c}^t {}^2\!R(t',\Omega) dt' + \tilde{\cal N}^{t_c}(\Omega) \,.
\end{eqnarray}

In the equations above, $dA$ is the area element of ${\cal S}_t$, the
global negative sign accounts for the relative change of the
orientation of outgoing vector normal to inner and outer boundary
spheres, $s^i$ are the components of the unit normal vector to ${\cal
  S}_t$ tangent to $\Sigma_t$, and $\tilde{\cal N}^{t_c}(\Omega)$ is a
generic function on the surface to be fixed.

Some remarks are in order concerning expressions (\ref{e:3D_Keff}) and
(\ref{e:news_Rt}). First, there is a clear asymmetry between
expressions (\ref{e:flux_Bondi_momentum_xi}) and (\ref{e:3D_Keff})
when substituting the complex quantity $\Psi_4$ at $\scri^+$ (encoding
two independent modes corresponding to the GW polarizations) by the
real quantity ${}^2\!R$ on the inner horizon (a single dynamical
mode). Inspection of Eq.  (\ref{e:flux_Bondi_momentum_xi}) immediately
suggests an alternative to ${}^2\!R$ by the natural {inner boundary}
analogue of $\Psi_4$, \ie $\Psi_0$. However, this strategy must face
the issue of identifying an appropriate null tetrad at ${\cal H}^+$
for the very construction of $\Psi_0$. Second, the lower limit in the
time integration, $t \to -\infty$, appearing in
Eq.~(\ref{e:flux_Bondi_momentum_xi}) must be replaced by the time
$t_c$ of first appearance of the common horizon, when quantities as
${}^2\!R(t,\Omega)$ start to be well-defined. However, there is still
a deeper difference between ${\cal N}(t,\Omega)$ and $\tilde{\cal
  N}(t,\Omega)$. Even though one can construct the former as in
Eq.~(\ref{e:News_scri_u}), \ie as the time integral of $\Psi_4$, the
definition of the news function is local in time depending only on
quantities on ${\cal S}_t$ and  not requiring
the knowledge of the past history of $\scri^+$. The latter though is
here {defined} as the time integral of ${}^2\!R$ and there is no
reason to expect the same local-in-time behavior, specially as $t
\rightarrow \infty$. In particular, we fix the function
$\tilde{\cal N}^{t_c}(\Omega)$ by imposing $\displaystyle
\lim_{t\rightarrow \infty} \tilde{\cal N}(t,\Omega) = 0$. 

All the points raised above are addressed in detail in the
accompanying paper II and we adopt here a purely effective approach to
$\tilde{K}_\mathrm{eff}^i(t)$, since ${}^2\!R$ represents an
unambiguous geometric object that captures the (possibly many, if matter 
is included)
relevant dynamical degrees of freedom in a single effective mode.
Ultimately, this {heuristic} proposal for the effective curvature is
acceptable only as long as it can be correlated with
$dP_i^{\mathrm{B}}/dt$, and this is what we will show in the
following.

\subsubsection{Axisymmetric BH spacetimes}

As a first application of the ansatz~\eqref{e:3D_Keff}, we consider
the axisymmetric case of the head-on collision of two BHs with unequal
masses. We adopt therefore a coordinate system $(r, \theta, \varphi)$
adapted to the horizon ${\cal H}^+$ so that $r=\mathrm{const.}$
characterizes sections ${\cal S}_t$ and we can write $s_i=M D_ir$,
with $M^{-2}=D_i r D^i r$ (\ie $M^{-2}=\gamma^{rr}$).  Then, taking
advantage of the axisymmetry, we adopt on ${\cal S}_t$ the preferred
coordinated system $(\tilde{\theta}, \tilde{\varphi)}$ discussed in
Sec.~\ref{s:mass_multipoles} and consider the Cartesian-like
coordinates constructed from $(r, \tilde{\theta}, \tilde{\varphi})$ by
standard spherical coordinates relations: $x=r
\,\mathrm{\sin}\tilde{\theta}\, \mathrm{\cos}\tilde{\varphi}, y=r \,
\mathrm{\sin}\tilde{\theta}\, \mathrm{\sin}\tilde{\varphi}, z=r \,
\mathrm{\cos}\tilde{\theta}$. In these coordinates we have $s_i=M
(\mathrm{\sin}\tilde{\theta}\, \mathrm{\cos}\tilde{\varphi},
\mathrm{\sin}\tilde{\theta}\, \mathrm{\sin}\tilde{\varphi},
\mathrm{\cos}\tilde{\theta})$. Assuming the $z$-axis to be adapted 
to the axisymmetry, we choose $\xi^i$ in Eq.~(\ref{e:3D_Keff}) as
$(\xi_z)^i=M^{-1} (\partial_z)^i$, so that $(\xi_z)^i s_i
=\mathrm{\cos}\tilde{\theta}$. Inserting expression (\ref{e:R_barI_n})
in Eqs.~(\ref{e:3D_Keff}) and (\ref{e:news_Rt}) we obtain
\begin{eqnarray}
\label{e:Keff_z}
\tilde{K}_{z}^\mathrm{eff}(t) &\equiv&
\tilde{K}^\mathrm{eff}[\xi_z](t) =  \\
& = & - \frac{1}{16 \pi}
\oint_{{\cal S}_t} (\mathrm{\cos}\tilde{\theta}) 
\left( 
\sum_{\ell=0}^\infty 
\tilde{\cal N}_\ell(t') Y_{\ell,0}(\tilde{\theta}) 
dt'\right)^2 dA, \nn
\end{eqnarray}
with
\begin{eqnarray}
\label{e:news_R}
\tilde{\cal N}_\ell(t)  \equiv
\int_{t_c}^t dt'\tilde{I}_\ell(t') + \tilde{\cal N}_\ell^{t_c} \ ,
\end{eqnarray}
being the coefficients of a multipolar expansion of
Eq.~(\ref{e:news_Rt}). Inserting the form $dA = R^2_\mathrm{H}
\mathrm{\sin}\tilde{\theta} d\tilde{\theta}d\tilde{\varphi}$ of the area
element on ${\cal S}_t$ and performing the angular integration we
finally find
\begin{equation}
\label{e:Keff_m=0}
\tilde{K}^{\mathrm{eff}}_z(t) = - \frac{R_{\mathrm{H}}^2}{16\pi} 
\sum_{\ell=2} \tilde{\cal N}_\ell  \left(D^{(0)}_{\ell, 0}  
\tilde{\cal N}_{\ell-1}
+  D^{(0)}_{\ell+1, 0} \tilde{\cal N}_{\ell+1}\right)   \,,
\end{equation}
with
\begin{eqnarray}
\label{e:D0m=0}
D^{(0)}_{\ell,0} &\equiv& \frac{\ell}{\sqrt{(2\ell+1)(2\ell-1)}} \,.
\end{eqnarray}
As for the definition of $K_{\mathrm{eff}}$ in the RT case,
Eq.~(\ref{e:Keff_m=0}) is quadratic in the (geometric) mass
multipoles, \ie the spherical harmonic components of the intrinsic
curvature Ricci scalar curvature ${}^2\!R$, although it involves a
time integration, absent in (\ref{e:K_eff_Ansatz_linear}). Also, it is
an odd function under reflection with respect to $z=\mathrm{const.}$
planes and it involves only products of {odd} and {even} multipoles,
precisely one of the criteria for the construction of
$K_{\mathrm{eff}}$ leading to the ansatz in
Eq.~(\ref{e:K_eff_Ansatz})\footnote{Note that expression
  (\ref{e:Keff_m=0}) cannot be factorized as a product of even and odd
  functions, as proposed in (\ref{e:K_eff_Ansatz}).}. In essence,
expression (\ref{e:Keff_m=0}) for $\tilde{K}^{\mathrm{eff}}_z$
fulfills the two basic requirements for the curvature parameter
$K_{\mathrm{eff}}$ with the added value that it is fully general and,
in contrast with (\ref{e:K_eff_Ansatz_linear}), no phenomenological
parameters need to be fitted. An additional and crucial feature is
that terms involving $\ell=0, 1$ are absent, due to the 
vanishing\footnote{The function $\tilde{\cal
    N}^{t_c}(\Omega)$ in Eq.~(\ref{e:news_Rt}) does  not introduce
  $\ell = 1$ modes either.}  of
$\tilde{I}_1$ as discussed after 
Eq.~\eqref{e:R_barI_n}.

The quantity $\tilde{K}^{\mathrm{eff}}_z$ at the horizon ${\cal H}^+$
is to be correlated with the component $(dP_z^{\mathrm{B}}/dt)(t)$ of the
flux of Bondi linear momentum at $\scri^+$, which is useful to express
in its multipolar expansion. First, we decompose $\Psi_4$ into its
multipoles
\begin{eqnarray}
\label{e:Psi_4_multipole}
\Psi_4 = \sum_{\ell\geq 2 ,m\leq|\ell|} 
\Psi_4^{\ell,m}{}_{-2}\!Y^{\ell,m}(\theta,\varphi) \,,
\end{eqnarray}
where ${}_{-2}\!Y^{\ell,m}(\theta,\varphi)$ are the spin-weighted
$s=-2$ spherical harmonics. The explicit expression for the component
of $(dP_i^{\mathrm{B}}/dt)(t)$ along the $z$-axis (\eg
Ref.~\cite{Alcubierre:2008}) then becomes
\bwt
\begin{eqnarray}
\label{e:Bondi_momentum_time_derivative}
\frac{dP^z_{\mathrm{B}}}{dt}(t) 
&=& \lim \limits_{r \to \infty}\frac{r^2}{16 \pi}\sum_{\ell\geq 2 ,m\leq|\ell|} 
 {\cal N}^{\ell, m} \times \left(C^{(-2)}_{\ell, m}  \bar{\cal N}^{\ell, m}  \right.
+ \left. D^{(-2)}_{\ell, m}  \bar{\cal N}^{\ell-1, m}  
+  D^{(-2)}_{\ell+1, m} \bar{\cal N}^{\ell+1, m}\right)\,,
\end{eqnarray}
\ewt
with 
\begin{equation}
\label{e:N_psi4_lm} 
{\cal N}^{\ell, m} \equiv \int_{-\infty}^t dt'
\Psi_4^{\ell, m}
\end{equation} 
being the corresponding multipolar components of the {news functions}
introduced in (\ref{e:News_scri_u}), with the coefficients
$C^{(-2)}_{\ell, m}$ and $ D^{(-2)}_{\ell, m}$ given by
\begin{eqnarray}
\label{e:coefficients_News_cd}
C^{(-2)}_{\ell, m} &\equiv& \frac{2m}{\ell(\ell+1)}\,, 
\end{eqnarray}
\begin{eqnarray}
D^{(-2)}_{\ell, m} &\equiv& \frac{1}{\ell}
\sqrt{\frac{(\ell -2)(\ell + 2)(\ell -m)(\ell + m)} 
{(2\ell - 1)(2\ell + 1)} } \,.
\end{eqnarray}
The axisymmetric reduction of
expression~\eqref{e:Bondi_momentum_time_derivative} is obtained by
setting $m=0$ in the expressions above. Note that
$\Psi_4$ is purely real in this case\footnote{For instance, the GW
  cross polarization $h_\times$ vanishes: $\Psi_4=-\ddot{h}_+
  +i\ddot{h}_\times$.}. The resulting coefficients are therefore
\begin{eqnarray}
\label{e:coefficients_News_m=0}
C^{(-2)}_{\ell, 0} &=& 0\,, \\
D^{(-2)}_{\ell, 0} &=& \sqrt{\frac{(\ell -2)(\ell + 2)}
{(2\ell - 1)(2\ell + 1)} } \,,
\end{eqnarray}
and we can write
Eq.~(\ref{e:Bondi_momentum_time_derivative}) as
\bwt
\begin{eqnarray}
\label{e:Bondi_momentum_time_derivative_m=0}
\frac{dP_z^{\mathrm{B}}}{dt}(t) = \lim \limits_{r \to \infty}\frac{r^2}{16 \pi}\sum_{\ell\geq2} 
\int_{-\infty}^t dt' \Psi_4^{\ell, 0} \int_{-\infty}^t dt''   
 \left(D^{(-2)}_{\ell, 0}  \Psi_4^{\ell-1, 0} 
+  D^{(-2)}_{\ell+1, 0} \Psi_4^{\ell+1, 0}\right) .
\end{eqnarray}
\ewt 

Expression~\eqref{e:Bondi_momentum_time_derivative_m=0} has obvious
similarities with Eq.~(\ref{e:Keff_m=0}) for
$\tilde{K}^{\mathrm{eff}}_z(t)$.  First, the (real) modes
$\Psi_4^{\ell, 0}$ play a role analogous to those of the mass
multipoles $\tilde{I}_\ell$. The common geometric nature of the
underlying quantities $\Psi_4$ and ${}^2\!R$ as curvatures, in
particular, their dimensions as second derivatives of the metric, is
indeed at the heart of the definition of the geometric multipoles
$\tilde{I}_\ell$'s by Eqs. (\ref{e:barI_n}) and (\ref{e:R_barI_n}) as
the correct analogues of $\Psi_4^{\ell, 0}$. Second, modes $\ell=0,1$
are absent in both expressions. This is nontrivial since the reasons
underlying each case are different: the $s=2$ spin weight of $\Psi_4$
in Eq.~(\ref{e:Bondi_momentum_time_derivative_m=0}) and the vanishing
of $\tilde{I}_1$ in (\ref{e:Keff_m=0}), respectively. This is a
crucial feature for it directly impacts the determination of the mode
dominating the dynamical behavior and, therefore, singles out the
Ricci scalar ${}^2\!R$ as a preferred quantity to be monitored instead
of any other (spin-weighted $s\ne2$) function that could measure in
some way the deformations of the horizon (for instance, the mean
curvature).  Besides the similarities, there are also differences
between expressions (\ref{e:Bondi_momentum_time_derivative_m=0}) and
(\ref{e:Keff_m=0}). First, the coefficients $D^{(s)}_{\ell, 0}$ in
(\ref{e:D0m=0}) and (\ref{e:coefficients_News_m=0}) differ due to the
different spin weight of ${}^2\!R$ and $\Psi_4$. Therefore, the
correlation between $(dP_z^{\mathrm{B}}/dt)(t)$ and
$\tilde{K}_i^{\mathrm{eff}}$ encodes information about the relative
weight of the different couplings. Second, the lower time-integration
bound ($t\to-\infty$) is well-defined for $(dP_z^{\mathrm{B}}/dt)(t)$,
whereas $\tilde{K}_z^{\mathrm{eff}}(t)$ can be measured only after the
formation of the common horizon. Finally, due to the absence in the
general case of a preferred coordinate system on ${\cal S}_t$ and
their associated spherical harmonics, there is no natural multipolar
expression for $\tilde{K}_i^{\mathrm{eff}}$ in the nonaxisymmetric
case and one must resort to the full expression (\ref{e:3D_Keff}).

\subsection{Correlation between the screens}
\label{s:correlation}

The effective-curvature vector $\tilde{K}^{\mathrm{eff}}_i$ introduced
in the previous section can now be used as a probe of the degree of
correlation between the geometry at the horizon and the geometry far
from the BH. More specifically, we aim at assessing the correlation
between $h_{\mathrm{inn}}(t) = \tilde{K}^{\mathrm{eff}}_z(t)$ at the
horizon and $h_{\mathrm{out}}(t)=(dP_z^{\cal B}/dt)(t)$ at large
distances, considering these two quantities as time series. As
discussed in Sec.~\ref{s:Exe_summ}, the use of a common time variable
$t$ for functions $h_{\mathrm{inn}}$ and $h_{\mathrm{out}}$ assumes a
(gauge) mapping (\cf footnote $2$ in Sec.~\ref{s:Exe_summ}) between
the advanced $v$ and retarded $u$ times, parameterizing ${\cal H}^+$
and $\scri^+$, respectively. The $3+1$ slicing by hypersurfaces $\{
\Sigma_t \}$ provides such a mapping, though an intrinsic
time-stretching ambiguity between the signals at the two screens is
present, due to the gauge nature of the slicing.  This will be
discussed in more detail later in this section.

To quantify the similarities in the time series we employ the
correlation function between time series $h_1(t)$ and $h_2(t)$, ${\cal
  C}(h_1,h_2;\tau)$, defined as
\begin{equation}
\label{e:correl_function}
{\cal C}(h_1,h_2;\tau)=\int_{-\infty}^\infty h_1(t+\tau) h_2(t) dt \,.
\end{equation} 
The structure of ${\cal C}(h_1,h_2;\tau)$ encodes a quantitative
comparison between the two time series as a function of the time shift
$\tau$ (referred to as ``lag'') between them. This correlation
function encodes the frequency components held in common between $h_1$
and $h_2$ and provides crucial information about their relative
phases. Because the time series are intrinsically different by a time
lag, we measure the correlation between $h_1$ and $h_2$ as
\begin{equation}
\label{e:correl_estimator}
{\cal M}(h_1,h_2)=
\max\limits_{\tau} 
\left(\frac{{\cal C}(h_1,h_2;\tau)}
{\left[{\cal C}(h_1,h_1;0)\,{\cal C}(h_2,h_2;0)\right]^{\frac{1}{2}}}
\right) \ \,.
\end{equation} 
This number is confined between $0$ and $1$ (where $1$ indicates
perfect correlation, and $0$ no correlation at all) and provides the
maximum {matching} between the time series $h_1$ and $h_2$ obtained by
shifting one with respect to the other in time, and then normalized in
frequency space. Besides providing a measure of the correlation,
expression Eq.~\eqref{e:correl_estimator} also gives a quantitative
estimate of the coordinate time delay $\tau_\mathrm{max}$ between the two
signals.

\begin{figure*}[t]
\begin{center}
\includegraphics[angle=0, width=5.9cm,clip=true]{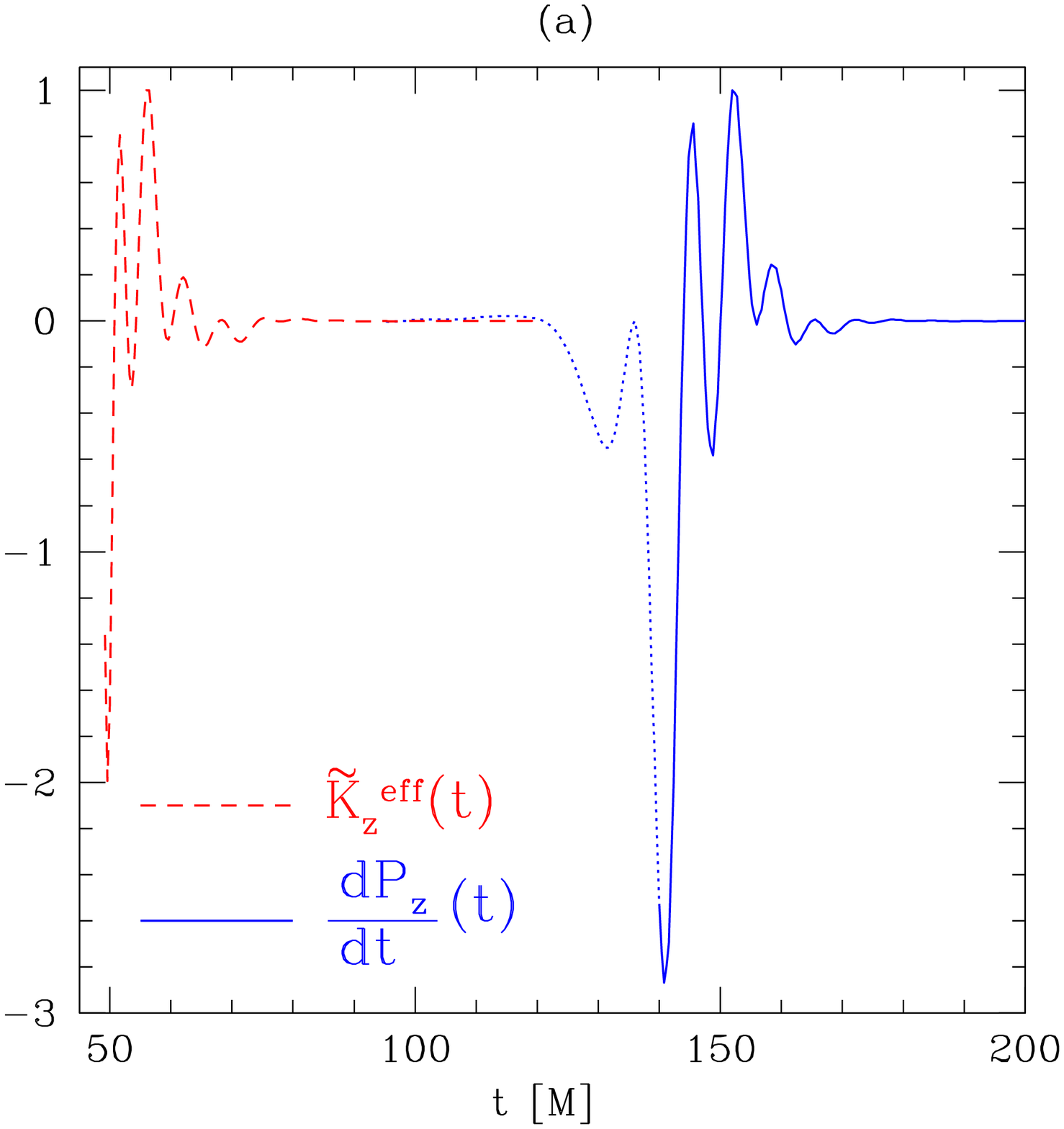}
\includegraphics[angle=0, width=5.9cm,clip=true]{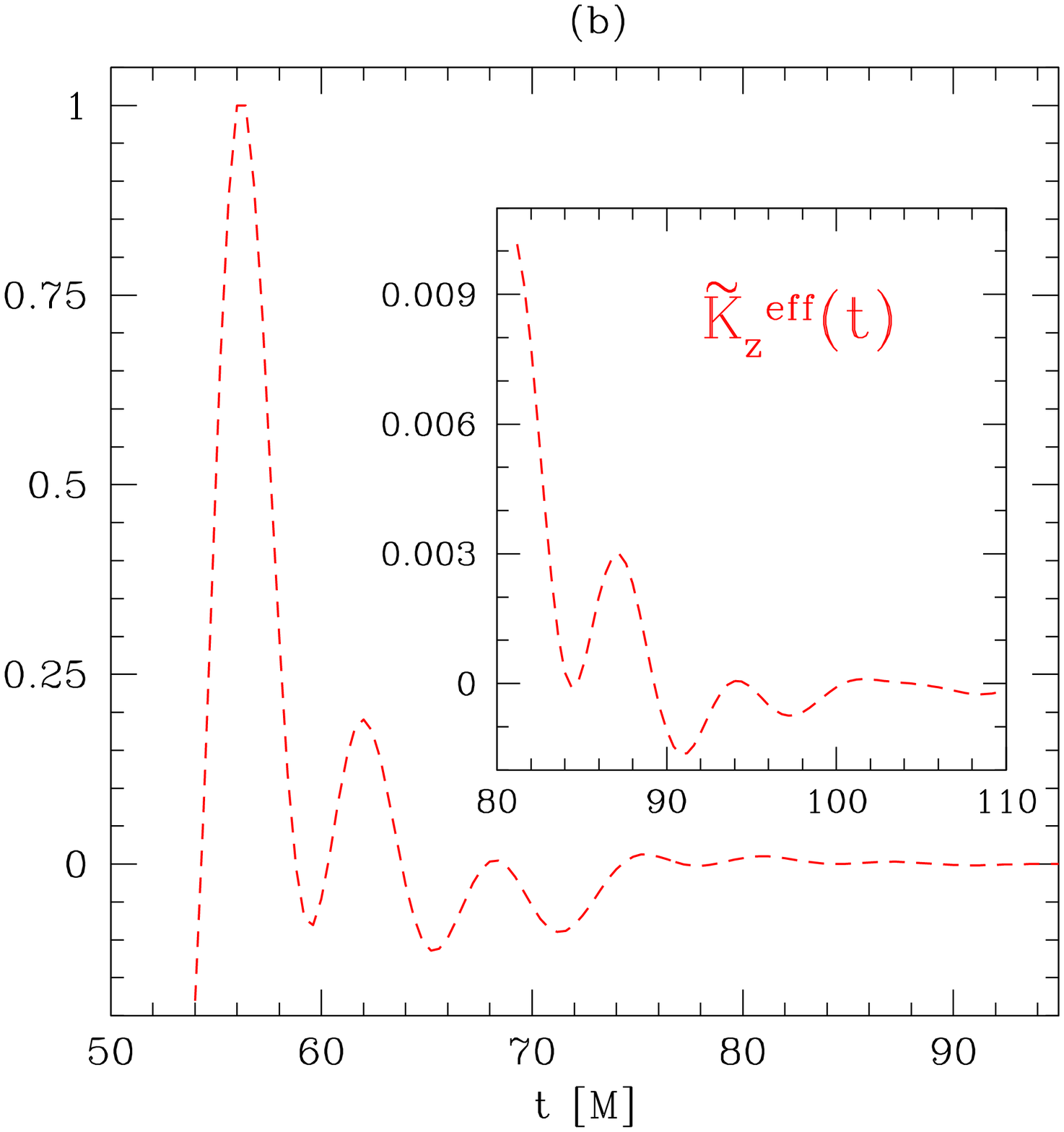}
\includegraphics[angle=0, width=5.9cm,clip=true]{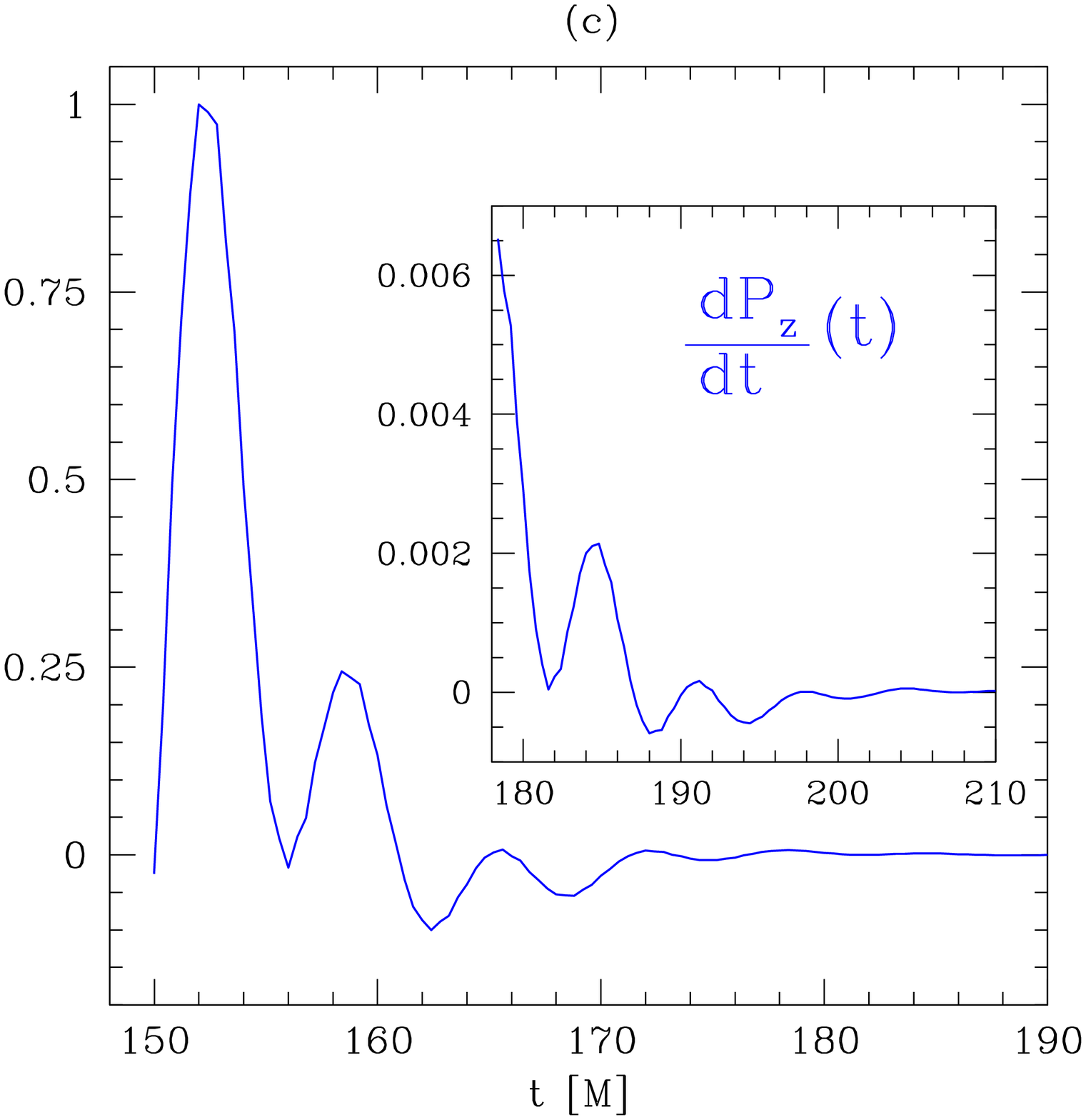}
\end{center}
\caption{Effective curvature $\tilde{K}^{\mathrm{eff}}_z$ defined at
  the horizon via Eq.~(\ref{e:3D_Keff}) (red dashed curve) and flux of
  momentum $(dP_z^{\cal B}/dt)$ evaluated at an approximation of
  $\scri^+$ with Eq.~(\ref{e:flux_Bondi_momentum_u}) (blue dotted and
  solid curves). These quantities encode, respectively, the information
  of the common horizon deformation and the flux of momentum carried
  away by GWs in the head-on collision of BHs with mass ratio
  1/2. Note, at panel (a), the qualitative agreement between those
  curves, which allows us to distinguish the momentum radiated before
  (blue dotted curve) and after (blue solid curve) the merge. Panel
  (b) and (c) compares the same quantities for latter times, where one
  can still see the good agreement.}  %
\label{fig:flux_time}
\end{figure*}

Note that one should not expect a perfect match between $(dP_z^{\cal
  B}/dt)(t)$ at $\scri^+$ and $\tilde{K}^{\mathrm{eff}}_z(t)$ at
${\cal H}^+$, even if the latter results to be a good estimator of the
former. Indeed, (nonlinear) gravitational dynamics in the bulk
spacetime affect and distort the possible relation between both
quantities. Furthermore, given the related but different nature of
${}^2\!R$ and $\Psi_4$ it is not obvious that a correlation should be
found at all.

In order to assess the validity of the approach, we construct
$\tilde{K}^{\mathrm{eff}}_z(t)$ and $(dP_z^{\cal B}/dt)(t)$ from the
numerical simulations described in Sec.~\ref{s:num_methods}. Note that
because $\tilde{I}_1$ vanishes identically, the contributions
$\tilde{\cal N}_0 \tilde{\cal N}_1$ and $\tilde{\cal N}_1 \tilde{\cal
  N}_2$ are absent in the expression for
$\tilde{K}^{\mathrm{eff}}_z(t)$. Furthermore, since higher-order
multipoles $\tilde{I}_{\ell}$ become increasingly difficult to calculate,
we truncate expression (\ref{e:Keff_m=0}) at $\ell=6$; in our case, 
this has little influence on the overall results as we will show 
that the lowest
(\ie $\ell=2, 3$) modes are by large the dominant ones. 

The values for $(dP_z^{\cal B}/dt)$ and $\tilde{K}^{\mathrm{eff}}_z$
as functions of the time $t$, and corresponding to the numerical
simulations described in the previous Sec.~\ref{s:num_methods}, are
presented in Fig.~\ref{fig:flux_time}. The signals have been
normalized with respect to their maximum value [a global rescaling
does not affect the cross-correlations properties of two functions
$h_1(t)$ and $h_2(t)$, \cf Appendix~\ref{a:correlation}]. In
Fig.~\ref{fig:flux_time} (a), the quantity
$\tilde{K}^{\mathrm{eff}}_z$ is shown from the time $t_c\approx 49.2
M$ of first appearance of the common horizon (red dashed line). After
time $t_{\mathrm{max}}\approx 120 M$ the error in the calculation of
the $\tilde{I}_\ell$ multipoles becomes comparable with the value of
the multipoles, spoiling the evaluation of the integrals in
(\ref{e:news_R}). Hence, we set $\tilde{K}^{\mathrm{eff}}_z$ to zero
for $t>t_{\mathrm{max}}$. Similarly, the flux of Bondi linear momentum
$(dP_z^{\cal B}/dt)$ as computed by an observer at $100\,M$ from the
origin, is split in a part before the appearance of the common AH
(blue dotted line) and in one which is to be compared with
$\tilde{K}^{\mathrm{eff}}_z$ (blue dashed line). In panels (b) and (c)
of Fig.~\ref{fig:flux_time} we show instead
$\tilde{K}^{\mathrm{eff}}_z$ and $(dP_z^{\cal B}/dt)$ separately and
in different time intervals for a better emphasis of the similarities.

Some interesting remarks on Fig.~\ref{fig:flux_time} can be made
already at a qualitative level. In particular, it is clear that
$\tilde{K}^{\mathrm{eff}}_z$ succeeds in tracking key features of
$(dP_z^{\cal B}/dt)$. This is apparent in the relative magnitude of
the three first positive peaks in the two signals and the qualitative
agreement is maintained in time. As expected, some specific features
of $(dP_z^{\cal B}/dt)$ are not faithfully captured in
$\tilde{K}^{\mathrm{eff}}_z$, such as the magnitude of the negative
peak around $t\approx 148M$ relative to the neighboring
peaks. However given the heuristic character of
$\tilde{K}^{\mathrm{eff}}_z$ and the fact that its geometric
definition does not leave room for any tuning, the overall qualitative
agreement with $(dP_z^{\cal B}/dt)$ at $\scri^+$ already represents a
remarkable result, shedding light on the near-horizon dynamics. This
agreement between $(dP_z^{\cal B}/dt)$ and
$\tilde{K}^{\mathrm{eff}}_z$ is indeed the main result of this section
and the ultimate justification for the introduction of
$\tilde{K}^{\mathrm{eff}}_z$. It is also worth stressing that attempts
employing other quantities (\eg a blind application of the methods
used for RT spacetimes) would not lead to such a clear matching.

From a quantitative point of view, the correlation analysis for the
time intervals shown in Fig.~\ref{fig:flux_time} (b) and (c) indicates
that the two signals yield a typical correlation ${\cal M} \approx
0.93$ and a time lag $\tau=97M$ (we recall that the observer is at
$100\,M$ and that the common AH has the size of a couple of
$M$). However, as one tries to extend the analysis to the very first
time of the formation of the apparent horizon, the correlation drops
significantly. The reason for this drop is related to the stretching
of the time coordinate between the two screens. In addition to the
obvious time delay between the $(dP_z^{\cal B}/dt)(t)$ and
$\tilde{K}^{\mathrm{eff}}_z(t)$ due to the finite (coordinate) speed
of light, in fact, the dependence of the two signals in coordinate
time $t$ is not the same and is stretched between the two screens.
This effect is the result of the in-built gauge mapping between
sections of $\scri^+$ and the horizon ${\cal H}^+$ defined by the
spacetime slicing, but also of the physical blueshift (redshift) of
signals at the inner (outer) boundaries in the BH spacetime.

Although approaches to disentangle the physical and gauge
contributions can be derived, for instance by introducing proper times
of suitably defined observers, this goes beyond the scope of this
paper. Rather, we opt here for a more straightforward approach in
which only comparisons based on sequences of (absolute values and
signs of the) maxima and minima in the signals $h_{\mathrm{inn}}(t)$
and $h_{\mathrm{out}}(t)$ are considered significant, since the
relative shape of $h_{\mathrm{inn}}(t)$ and $h_{\mathrm{out}}(t)$ can
be subject to a time reparametrization. This association is possible
when the quantities which are compared are scalars, so that the values
of maxima and minima are well-defined and independent of coordinates.
This is possible in the case of axisymmetry as it gives a privileged
direction along which to contract the effective-curvature vector
$\tilde{K}^{\mathrm{eff}}_i$. In a more generic configuration one
would need to build an appropriate frame to produce scalars by
contraction with tensorial quantities.  Once the correspondence
between maxima and minima in the two signals $h_{\mathrm{inn}}(t)$ and
$h_{\mathrm{out}}(t)$ is established, a mapping $t_{\mathrm{out}} =
t_{\mathrm{out}}(t_{\mathrm{inn}})$ can be easily constructed. With
this matching, the calculation of the correlation parameter gives
typically values ${\cal M}\geqslant0.9$ for any chosen time
interval~\footnote{ \label{foot:mapping}Interesting information can also be gained by
  studying in more detail the properties of the mapping
  $t_{\mathrm{out}} = t_{\mathrm{out}}(t_{\mathrm{inn}})$. More
  specifically, we have found that the derivative $dt_{\mathrm{out}} /
  dt_{\mathrm{inn}}$ is not constant and starts as being larger than
  unity (indicating that initially the coordinate time at $\scri^+$
  runs faster than the time at ${\cal H}^+$), but then oscillates
  around unity at late times. This is consistent with the fact that as
  stationarity is approached, the evolution vector $t^{a}$ adapts to
  the timelike Killing vector.}. More information about the mapping 
  between $t_{\mathrm{out}}$  and $t_{\mathrm{in}} $ is found in Appendix \ref{a:mapping}.

\subsubsection{A critical assessment of the correlation}

\begin{figure}[t]
\begin{center}
\includegraphics[angle=0, width=8.0cm,clip=true]{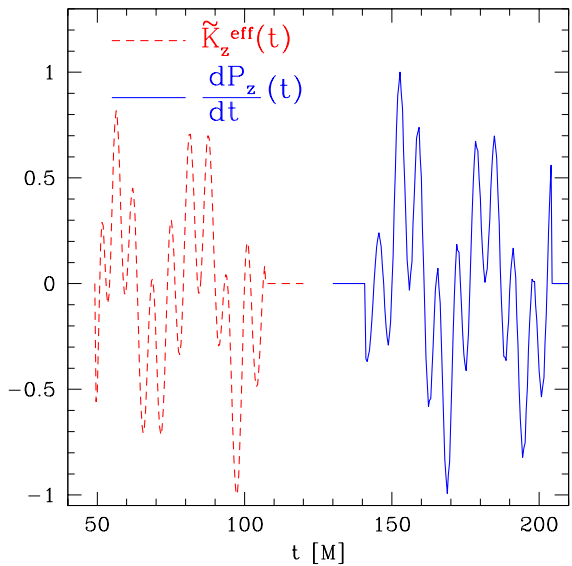}
\end{center}
\caption{Quantities on ${\cal H}^+$ (red dashed line) and $\scri^+$
  (blue solid line) as shown in Fig.~\ref{fig:flux_time} but without
  an overall exponential decay in time. The close similarity in the
  signals is confirmed by the very large correlation which is ${\cal M}
    =0.87$, obtained without a time mapping.}
\label{fig:flux_without_expdecay}
\end{figure}

Of course, it is reasonable to question if finding such a high
correlation is just a bias in our methodology. Certainly, our strategy
of identification of maxima and minima in the signals enhances the
correlations when time-stretching issues are involved. However, it
does not guarantee by itself the high (positive) values found for
${\cal M}$. More specifically, once a first couple of maxima (or
minima) are identified in the two signals, the subsequent couples of
maxima and minima constructed from the data in each signal are
automatically fixed. As a consequence, high correlations are possible
only if the sequence of signs in the extrema of the two signals is
exactly the same. In addition to consideration above, one may also
argue that the high correlation found is just the result of the very
rapid decay of the signals, which makes the first couple of maxima and
minima play a dominant role in the estimate, possibly shadowing the
role of the smaller peaks appearing at later times. To address this
point and weight equally all parts of the signals, we model them as
exponentially decaying oscillating functions, \ie
$h^{\kappa}_{\mathrm{inn}}(t) \equiv e^{\kappa_{\mathrm{inn}} t}
h_{\mathrm{inn}}(t)$ and $h^{\kappa}_{\mathrm{out}}(t)\equiv
e^{\kappa_{\mathrm{out}} t}h_{\mathrm{out}}(t)$. This is applied to
the signals without the time correction provided by the map $t_{\rm
  out} = t_{\rm out}(t_{\rm inn})$ , finding
\begin{equation}
\label{e:kappa_decay}
M \kappa_{\mathrm{inn}}= 0.179 \pm
0.005 \,, \qquad  M \kappa_{\mathrm{out}}= 0.181 \pm 0.006 \ ,
\end{equation}
through a least-square fitting. The resulting functions
$h^{\kappa}_{\mathrm{inn}}(t)$ and $h^{\kappa}_{\mathrm{out}}(t)$
once the exponential decay has been subtracted are shown in
Fig.~\ref{fig:flux_without_expdecay}. Once again it is apparent that
the two time series are very similar and indeed the matching computed
even without introducing any time mapping is ${\cal
  M}(h^{\kappa}_{\mathrm{inn}}, h^{\kappa}_{\mathrm{out}})= 0.87$ and
thus remarkably high.

\begin{figure}[t]
\begin{center}
\includegraphics[angle=0, width=8.0cm,clip=true]{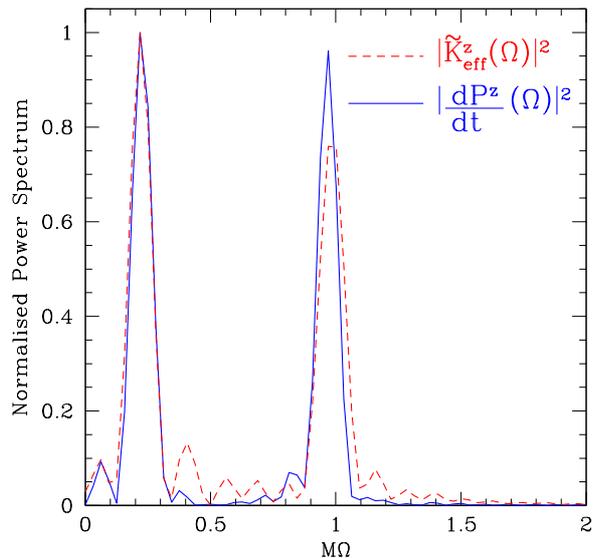}
\end{center}
\caption{Normalized power spectrum for $K_z^{\rm eff}(\Omega)$ (red
  dashed line) and $(dP_z/dt)(\Omega)$ (blue solid line) as measured
  at ${\cal H}^+$ and $\scri^+$, respectively (\cf
  Figure~\ref{fig:flux_without_expdecay}).  Both spectra are dominated
  by two frequencies: $\Omega^{\mathrm{inn}}_1=0.22 \pm 0.04$ and
  $\Omega^{\mathrm{inn}}_2=0.98 \pm 0.05$ (red dashed line) and
  $\Omega^{\mathrm{out}}_1=0.22 \pm 0.04$ and
  $\Omega^{\mathrm{out}}_2=0.97 \pm 0.04$ (blue solid line) which are
  linear combinations of the quasinormal ringing modes of the merged
  BH (\cf Table \ref{t:frequencies_decays}).}
\label{fig:power_spect}
\end{figure}

The main reason behind the good correlation found also for the undamped
signals is that the next-to-leading-order term, \ie term $\tilde{\cal
  N}_3\tilde{\cal N}_4$ and the corresponding $\ell=3$ and $\ell=4$
coupling in Eq.~(\ref{e:Bondi_momentum_time_derivative_m=0}), are much
smaller than the leading-order term ${\cal N}_2 {\cal N}_3$. Indeed,
we have found that it is possible to express to a very good
approximation $\tilde{K}^{\mathrm{eff}}_z \sim \tilde{\cal N}_2
\tilde{\cal N}_3$ and $(dP_z^{\cal B}/dt) \sim {\cal N}_2 {\cal
  N}_3$. This is confirmed by the corresponding power spectra, which
are shown in Fig.~\ref{fig:power_spect} and are dominated in both
cases by two frequencies: $\Omega^{\mathrm{inn}}_1=0.22 \pm 0.04,\,
\Omega^{\mathrm{inn}}_2=0.98 \pm 0.05$ for the signal
$h_{\mathrm{inn}}(t)$, and $\Omega^{\mathrm{out}}_1=0.22 \pm 0.04,\,
\Omega^{\mathrm{out}}_2=0.97 \pm 0.04$ for the signal
$h_{\mathrm{out}}(t)$.

These frequencies are closely related to the quasinormal modes of the
merged BH, interfering to lead to a beating signal. To see this, we
model each function $\tilde{\cal N}_\ell$ and ${\cal N}_\ell$ as an
exponentially damped sinusoid, \ie $\tilde{\cal N}_\ell\sim
e^{-\kappa^{\tilde{\cal N}}_\ell t} \mathrm{\sin}(\Omega^{\tilde{\cal
    N}}_\ell t + \varphi^{\tilde{\cal N}})$ and ${\cal N}_\ell\sim
e^{-\kappa^{{\cal N}}_\ell t} \mathrm{\sin}(\Omega^{{\cal N}}_\ell t +
\varphi^{\cal N})$.  Then, under the approximation
$\tilde{K}^{\mathrm{eff}}_z \sim \tilde{\cal N}_2 \tilde{\cal N}_3$
and $(dP_z^{\cal B}/dt) \sim {\cal N}_2 {\cal N}_3$ it follows
\begin{equation}
\label{e:QNM_horizon} 
\Omega^{\tilde{\cal N}}_{\ell=2} = \frac{\Omega^{\mathrm{inn}}_2-
\Omega^{\mathrm{inn}}_1}{2}\,, \quad 
\Omega^{\tilde{\cal N}}_{\ell=3} =
\frac{\Omega^{\mathrm{inn}}_2+\Omega^{\mathrm{inn}}_1}{2}\,,
\end{equation}
at ${\cal H}^+$, whereas 
\begin{equation}
\label{e:QNM_scri}
\Omega^{\cal N}_{\ell=2} = \frac{\Omega^{\mathrm{out}}_2-\Omega^{\mathrm{out}}_1}{2}\,, \quad 
\Omega^{\cal N}_{\ell=3} = \frac{\Omega^{\mathrm{out}}_2+\Omega^{\mathrm{out}}_1}{2}\,,
\end{equation}
at $\scri^+$, consistent with the ``beating'' behavior shown by the
power spectra in Fig.~\ref{fig:power_spect}.  Similarly, the decay
time scales are then given by
\begin{equation}
\label{e:QNM_scri_kappa}
\kappa_{\mathrm{inn}} = \kappa^{\tilde{\cal N}}_{\ell=2} +
\kappa^{\tilde{\cal N}}_{\ell=3} \ \ , \ \ \kappa_{\mathrm{out}} =
\kappa^{{\cal N}}_{\ell=2}+\kappa^{{\cal N}}_{\ell=3}. 
\end{equation}
These frequencies and time scales match very well the real
$(\omega^{\rm R}_\ell)$ and imaginary $(\omega^{\rm I}_\ell)$ parts of
the fundamental $(n=0)$ quasi-normal-modes (QNM) eigenfrequencies of a
Schwarzschild BH~\cite{Berti:2009kk}. A detailed comparison is
presented in Table~\ref{t:frequencies_decays}, whose first six columns
report the properties of the signals $h_{\mathrm{inn}}(t)$ and
$h_{\mathrm{out}}(t)$ in their constituent frequencies $\Omega^{\cal
  N}_{\ell=2,3}$  and $\Omega^{\tilde{\cal  N}}_{\ell=2,3}$ defined in
Eqs.~\eqref{e:QNM_horizon}--\eqref{e:QNM_scri}, and compare them with
the corresponding real parts of the eigenfrequencies of a
Schwarzschild BH, $\omega^{\rm R}_{\ell=2,3}$. The close match in the
oscillatory part is accompanied also by a very good correspondence in
the decaying part of the signal. Defining, in fact, the overall decay
time in terms of the imaginary parts of the QNM eigenfrequencies, \ie
as $\kappa_{\mathrm{decay}}\equiv\omega_{\rm I}^{\ell=2}+\omega_{\rm
  I}^{\ell=3}$, it is easy to realize from the last three columns in
Table~\ref{t:frequencies_decays}, that this decay time is indeed very
close to the one associated to the signal at the two screens [\cf Equations~\eqref{e:QNM_scri}
and~\eqref{e:QNM_scri_kappa}].

This role of QNMs is not entirely surprising for a measure at
$\scri^+$, but it is far less obvious to see it imprinted also for a
quantity measured at ${\cal H}^+$. This indicates that the bulk
spacetime dynamics responsible for the recoil physics is a relatively
mild one, so that a QNM ringdown behavior dominates the dynamics of
the deformed single AH and imprints the properties of the radiated
linear momentum.  It is interesting that a purely (quasi-)local study
of the AH geometric properties permits us to read the behavior of
quantities which are intrinsically defined at infinity, thus
confirming the main thesis in Ref.~\cite{Rezzolla:2010df}.

\begin{table*}[ht]
\caption{The first six columns offer a comparison between the
  properties of the signals $h_{\mathrm{inn}}(t)$ and
  $h_{\mathrm{out}}(t)$ in their constituent frequencies $\Omega^{\cal
    N}_{\ell=2,3}$ defined in
  Eqs.~\eqref{e:QNM_horizon}--\eqref{e:QNM_scri}, with the
  corresponding real parts of the eigenfrequencies of a Schwarzschild
  BH, $\omega^{\rm R}_{\ell=2,3}$. The last three columns show instead
  a comparison between the damping times $\kappa_{\mathrm{inn,out}}$
  defined in~Eq.~\eqref{e:QNM_scri_kappa}, with the corresponding
  decay time $\kappa_{\mathrm{decay}}$ computed from the imaginary
  parts of the eigenfrequencies. In all cases, the close match is
  remarkable and not at all obvious for quantities measured at ${\cal
    H}^+$.}
\label{t:frequencies_decays}
\begin{center}
\begin{ruledtabular}
\footnotesize{
\begin{tabular}[t]{|ccc|ccc|ccc|}
$M\Omega^{\tilde{\cal N}}_{\ell=2}$ & $M\Omega^{\cal N}_{\ell=2}$ & $M\omega^{\rm R}_{\ell=2}$  & 
$M\Omega^{\tilde{\cal N}}_{\ell=3}$ & $M\Omega^{\cal N}_{\ell=3}$ & $M\omega^{\rm R}_{\ell=3}$  & 
$M\kappa_{\mathrm{inn}}$   & $M\kappa_{\mathrm{out}}$          & $M\kappa_{\mathrm{decay}}$ \\
&&&&&&&&\\
\hline
&&&&&&&&\\
$0.38 \pm 0.04$         & $0.37 \pm 0.04$               & $0.37367$ & 
$0.60 \pm 0.04$         & $0.59 \pm 0.04$               & $0.59944$ & 
$0.181 \pm 0.006$       & $0.179 \pm 0.005$             & $0.18166$ \\
\end{tabular}
}
\end{ruledtabular}
\end{center}
\end{table*}

\subsubsection{Antikicks and the Slowness Parameter}

\begin{figure*}[t]
\begin{center}
\includegraphics[width=5.9cm,clip=true]{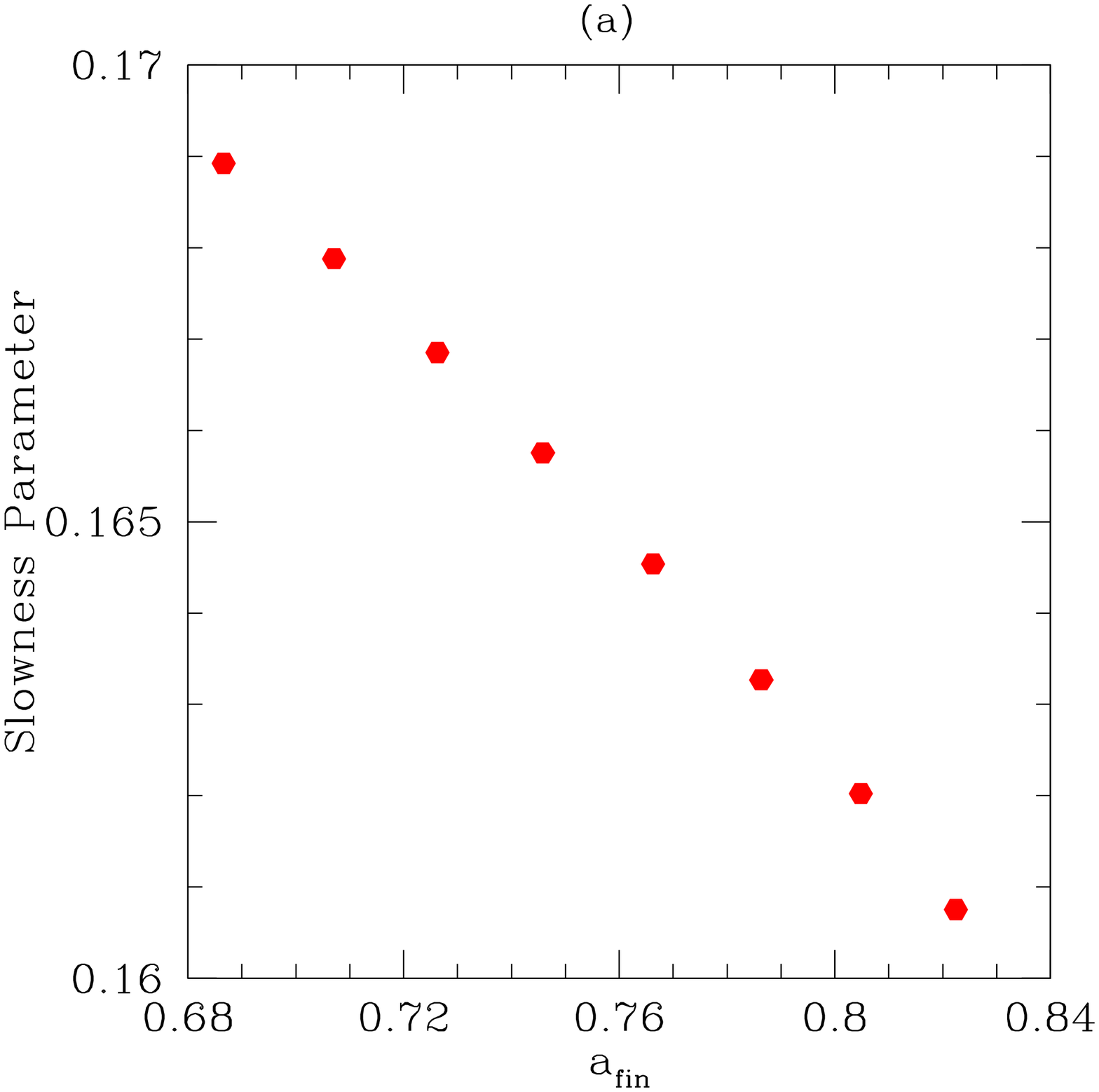}
\includegraphics[width=5.9cm,clip=true]{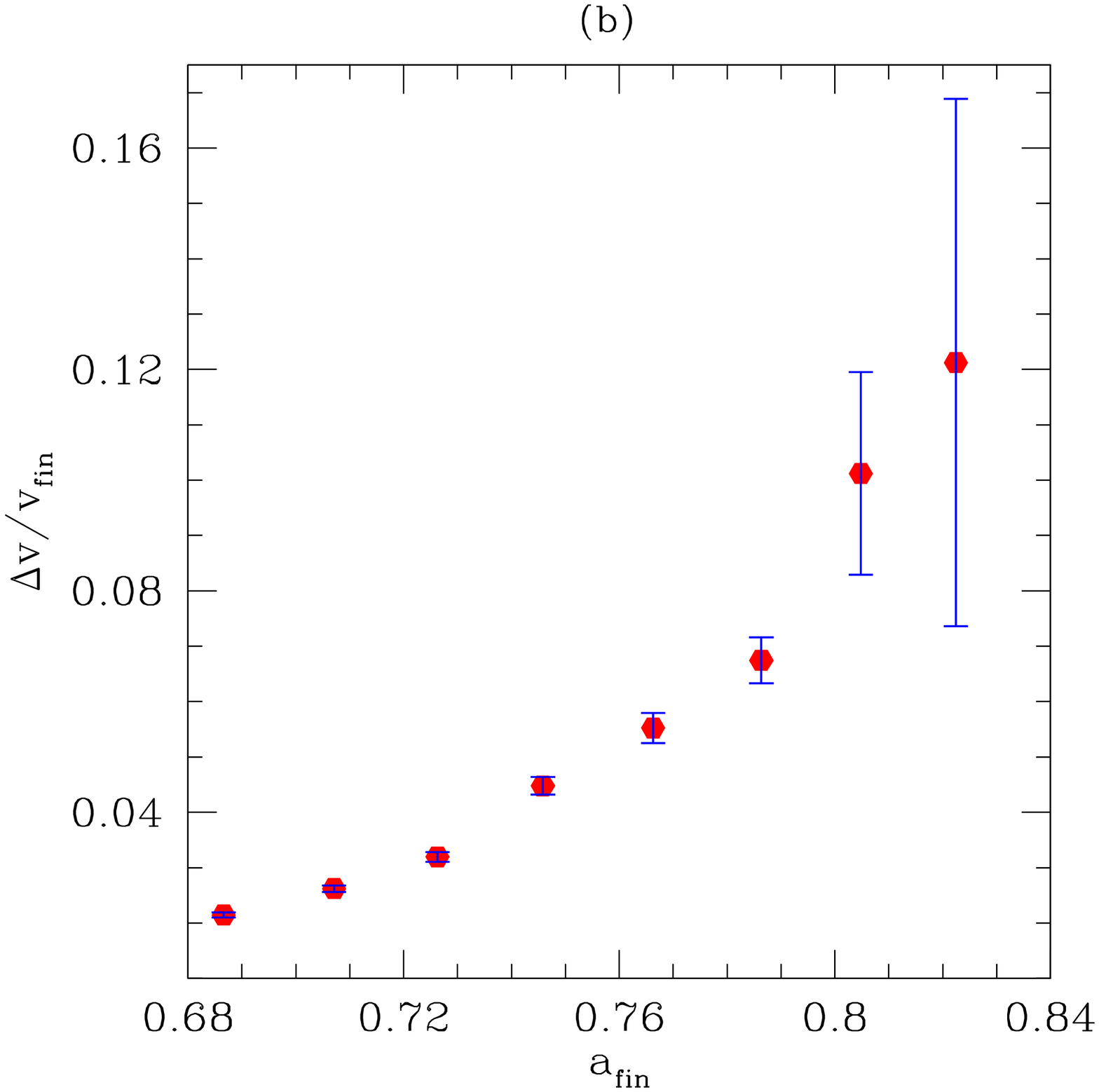}
\includegraphics[width=5.9cm,clip=true]{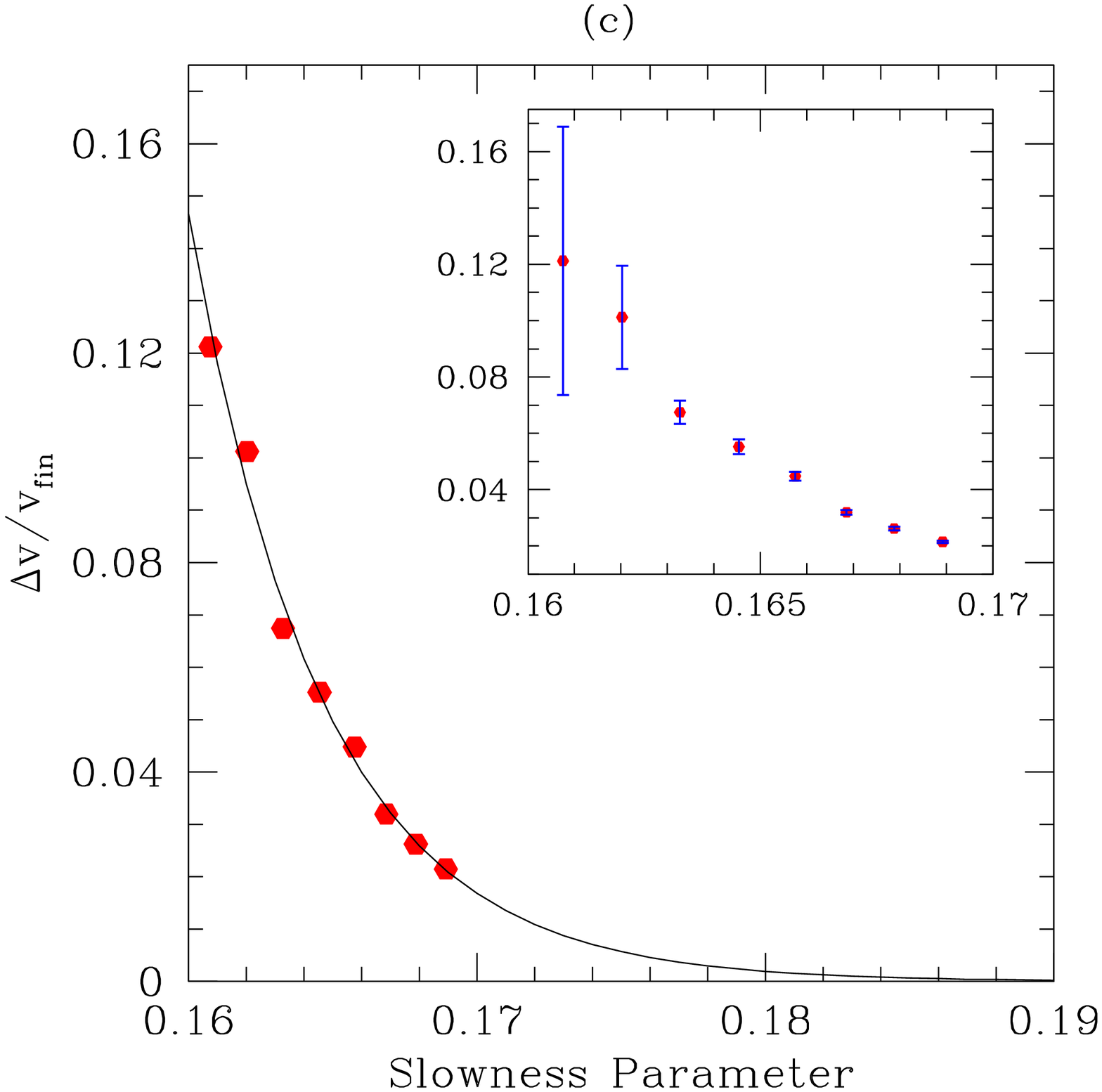}
\end{center}
\caption{Predictability of the antikick in terms of the slowness
  parameter and the antikick. The left panel shows the correlation
  between the slowness parameter $P=T/\tau$ as computed in
  Eq.~\eqref{e:T_tau_P} and the dimensionless spin of a BH
  $a_{\textrm{fin}}$. The mid panel shows instead the good correlation
  between the relative antikick velocity $\Delta v/v_{\textrm{fin}} =
  v_{\textrm{k}}/v_{\textrm{fin}}$ and the dimensionless final spin,
  using the data taken from Ref.~\cite{Pollney:2007ss:shortal}
  (indicated with error bars are the estimated numerical
  errors). Finally, the right panel combines the first two and shows
  the correlation between the antikick velocity and the slowness
  parameter. In particular, the good exponential fitting shows the
  consistency with a vanishing antikick for a slowness parameter 
  $P\sim 1$, as discussed in \cite{Price:2011fm}.}
\label{fig:SlownessParameter1}
\end{figure*}

As a concluding remark for this section, we make use of our results,
and, in particular, on the spectral and decaying properties of our
measures on the screens, to make contact with the analysis carried out
in~\cite{Price:2011fm}. More specifically, we can define a
characteristic decay time $\tau \equiv (2\pi)/\kappa^{\rm inn/ out}$
and an oscillation-characteristic time $T \equiv
2\pi/\Omega^{\mathrm{inn/out}}_2$, from which to build our equivalent
of the {\em ``slowness parameter''} $P\equiv T/\tau$ introduced
in~\cite{Price:2011fm}. The specific case discussed above then yields
$\tau\simeq 34.9 M$, $T\simeq 6.4 M$ and thus $P \simeq 0.18$. As
detailed in Ref.~\cite{Price:2011fm}, small antikicks should happen
when the two timescales are comparable, thus corresponding to an
oscillation which is over-damped. This expectation is indeed confirmed
by the recoil velocity shown in Fig.~\ref{fig:BBH_v_kick}, where the
relative antikick is about $\sim 30\%$ and thus compatible with the
slowness parameter that we have associated to our process [see also
the discussion below on the application of Eqs.~\eqref{e:T_tau}
and~\eqref{e:T_tau_P} in Fig.~\ref{fig:SlownessParameter1}].  This
qualitative agreement with the phenomenological approach discussed in
Ref.~\cite{Price:2011fm} is very natural. While we here concentrate on
modeling the local curvature properties at the horizon,
Ref.~\cite{Price:2011fm} concentrates on the spectral features of the
signal at large distances. Since we have demonstrated that the two are
highly correlated, it does not come as a surprise that the two
approaches are compatible. Looking at the local horizon's properties
has however the added value that it provides a precise framework in
which to predict not only the strength of the antikick, but also its
directionality. Furthermore, such an approach permits an
interpretation of BH dynamics in terms of viscous hydrodynamics, as we
will discuss in detail in paper II. In particular, we shall show there
that the horizon-viscous analogy naturally leads to a geometric
prescription for an (instantaneous) slowness parameter $P$, in terms
of timescales $\tau$ and $T$ respectively related to bulk and shear
viscosities.

The logic developed above for the calculation of the slowness
parameter can be brought a step further by assuming that the final BH
produced by the merger of a binary system in quasicircular orbit can
be described at the lowest order by an oscillation and decay times
\begin{eqnarray}
\label{e:T_tau}
\tau &\equiv& \frac{2\pi}{\omega_{\ell=2}^{\textrm{I}} +\omega_{\ell=3}^{\textrm{I}} } \,, \qquad
T \equiv \frac{2\pi}{\omega_{\ell=2}^{\textrm{R}} +\omega_{\ell=3}^{\textrm{R}} } \,,
\end{eqnarray}
to which corresponds a slowness parameter defined as 
\begin{eqnarray}
\label{e:T_tau_P}
P&\equiv&\frac{T}{\tau}=\frac{\omega_{\ell=2}^{\textrm{I}} +\omega_{\ell=3}^{\textrm{I}}  }
 {\omega_{\ell=2}^{\textrm{R}} +\omega_{\ell=3}^{\textrm{R}}}\,.
\end{eqnarray}
Using the semianalytic expressions derived for estimating the spin of
the final BH, \eg~\cite{Rezzolla-etal-2007, Rezzolla-etal-2007b,
  Barausse:2009uz}, it is possible to predict the values of $\tau$ and
$T$ for any binary whose initial spins and masses are known, and thus
predict qualitatively through $P$ the strength of the antikick which
will be produced in any of these configurations. We have tested this
idea by considering the data presented in
Ref.~\cite{Pollney:2007ss:shortal} both for the kick/antikick
velocities and for the final spin of the merged BH. This conjecture
about the predictability of the antikick in terms of the slowness
parameter is indeed supported by the example data collected in
Fig.~\ref{fig:SlownessParameter1}. More specifically, the left panel
in Fig.~\ref{fig:SlownessParameter1} shows the correlation between the
slowness parameter $P=T/\tau$ as computed in Eq.~\eqref{e:T_tau_P} and
the dimensionless spin of a BH
$\tilde{a}_{\textrm{fin}}=J_{\textrm{fin}}/M_{\textrm{fin}}^2$
produced, for example, in the merger of a binary system (expressions
to estimate the QNM eigenfrequencies for rotating BHs can be found in
a number of works which are collected in the
review~\cite{Berti:2009kk}). The middle panel shows instead the good
correlation between the relative antikick velocity $\Delta
v/v_{\textrm{fin}} \equiv (v_{\textrm{max}} - v_{\textrm{fin}}) /
v_{\textrm{fin}} = v_{\textrm{k}}/v_{\textrm{fin}}$ and the
dimensionless final spin as computed from the data taken from
Ref.~\cite{Pollney:2007ss:shortal} (indicated with error bars are the
estimated numerical errors). Finally, the right panel combines the
first two and shows the searched correlation between the antikick
velocity and the slowness parameter. It also shows with a solid line
an exponential fit, which suggests a vanishing antikick for a slowness
parameter $P\sim 1$. All in all, this figure confirms also for the
case of binaries in quasicircular orbits the
suggestion~\cite{Price:2011fm} that the smaller the slowness parameter
$P$ gets, the larger is the expected value of the antikick.  Large
antikicks should then be expected for $P\ll 1$~\cite{Price:2011fm}.
Furthermore, it highlights that it is indeed possible to predict
qualitatively the antikick merely on the basis of the initial
properties of the BHs when the binary is still widely separated.

\section{Conclusions}
\label{s:conclusions}

We have demonstrated that qualitative aspects of the post-merger
recoil dynamics at infinity can be understood in terms of the
evolution of the geometry of the common horizon of the resulting black
hole. This extends to binary black-hole spacetimes the conclusions
presented in Ref.~\cite{Rezzolla:2010df} based on Robinson-Trautman
spacetimes. More importantly, we have shown that suitably built
quantities defined on inner and outer world tubes (represented either
by dynamical horizons or by timelike boundaries) can act as test
  screens responding to the spacetime geometry in the bulk, thus
opening the way to a cross-correlation approach to probe the dynamics of
spacetime.

The extension presented here is nontrivial and it involves the
construction of a phenomenological vector
$\tilde{K}^{\mathrm{eff}}_i(t)$ from the Ricci curvature scalar
${}^2\!R$ on the dynamical-horizon sections, which then captures the
global properties of the flux of Bondi linear momentum
$(dP_i^{\mathrm{B}}/dt)(t)$ at infinity, namely, (proportional to) the
acceleration of the BH. At the same time, the proposed approach
involves the development of a cross-correlation methodology which is
able to compensate for the in-built gauge character of the time
evolution on the two surfaces. A proper mapping between the times on
the two surfaces is needed and its gauge nature highlights that the
physical information encoded in the surface quantities is not in its
{\em local} (arbitrary) time dependence, but rather in the {\em
  global} structure of successive maxima and minima.

By analyzing Robinson-Trautman spacetimes, Ref.~\cite{Rezzolla:2010df}
proposed that when a single horizon is formed during the merger of two
BHs, the observed decelerations/accelerations of the newly formed BH
can be understood in terms of the dissipation of an anisotropic
distribution of the Ricci scalar curvature on the horizon. The results
presented here confirm this picture, although through quantities which
are suited to BH spacetimes. Being computed on the horizon, these
quantities reflect the properties of the BH and, in particular, its
exponentially damped ringing. The interplay between oscillation and
decay timescales associated with this
process, which are inevitably imprinted in our geometric variables,
explain the late qualitative features of the recoil dynamics, in particular,
the antikick, in natural connection with the approach discussed in
Ref.~\cite{Price:2011fm}, where the antikick is explained in terms of
the spectral features of the signal at large distances. Because we
have shown that the latter is closely correlated with the signal at
the horizon, we can adopt the same slowness parameter introduced in
Ref.~\cite{Price:2011fm} to predict qualitatively the magnitude of the
antikick from the merger of BH binaries with spin aligned to the
orbital angular momentum, finding a very good agreement with the
numerical data.

As a final remark we note that looking at the horizon's properties has
the added value that it provides a precise framework in which to
predict not only the strength of the antikick, but also its
directionality. Furthermore, as we discuss in detail in paper II, our
geometric (cross-correlation) framework presents a number of close
connections with (and potential implications on) the literature
developing around the use of horizons to study the dynamics of BHs, as
well as with the interpretations of such dynamics in terms of a
viscous hydrodynamics analogy.  Much of the machinery developed using
dynamical trapping horizons as inner screens can be extended also when
a common horizon is not formed (as in the calculations reported in
Ref.~\cite{Sperhake:2010uv}). While in such cases the identification
of an appropriate hypersurface for the inner screen can be
considerably more difficult, once this is found its geometrical
properties can be used along the lines of the cross-correlation
approach discussed here for dynamical horizons.


\begin{acknowledgments} 
It is a pleasure to thank A. Saa, M. Koppitz, B. Krishnan, F. Ohme,
H. Oliveira, B. Schutz, I. Soares and A. Tonita for useful
discussions.  This work was supported in part by the DAAD and the DFG
grant SFB/Transregio~7. JLJ acknowledges support from the Alexander
von Humboldt Foundation, the Spanish MICINN (FIS2008-06078-C03-01) and
the Junta de Andaluc\'ia (FQM2288/219). The computations were
performed on the Datura cluster at the AEI and on the Teragrid network
(alloca- tion TG-MCA02N014).
\end{acknowledgments} 

\appendix
\section{Correlation and matching of time series} 
\label{a:correlation}

The correlation function ${\cal C}(h_1,h_2;\tau)$ introduced in
Sec.~\ref{s:correlation} provides information in the time domain about
the comparison of temporal series $h_1(t)$ and $h_2(t)$. Its Fourier
transform defines the {\em cross-spectrum}${\cal C}(h_1,h_2;f)$  
of $h_1$ and $h_2$, providing the corresponding analysis in the
frequency domain. It has the form
\begin{equation}
\label{e:cross_spectrum}
{\cal C}(h_1,h_2;f) = \tilde{h}_1(f)\tilde{h}_2^*(f) \,,
\end{equation}
where Fourier transform conventions are
\begin{equation}
\label{e:Fourier_transform}
\tilde{h}_i(f)=\int_{-\infty}^\infty h_i(t) e^{i2\pi f t}dt \ \,, \ \ 
h_i(t)=\int_{-\infty}^\infty \tilde{h}_i(f) e^{-i2\pi f t}df \,.
\end{equation}
Choosing a measure $(S_n(|f|))^{-1}df$, a natural scalar product
between functions $h_1(t)$ and $h_2(t)$ (or $\tilde{h}_1(f)$ and
$\tilde{h}_2(f)$) is introduced as
\begin{equation}
\label{e:scalar_product}
\langle h_1,h_2\rangle = \int_{-\infty}^\infty  
\frac{\tilde{h}_1(f)\tilde{h}_2^*(f)}{S_n(|f|)}df.
\end{equation}

In GW data analysis $S_n(|f|)$, the {\em noise power-spectral
  density}, is associated with the spectral sensitivity of the
instrument. In our case we have no {\em a priori} knowledge about
$S_n(|f|)$, and we choose $S_n(f)=1$.  The scalar product
$\langle\cdot,\cdot\rangle$ introduces the natural projection between
$\tilde{h}_1(f)$ and $\tilde{h}_2(f)$. Their normalized scalar product
defines the {\em overlap}
\begin{equation}
\label{e:overlap}
{\cal O}[h_1,h_2] \equiv \frac{\langle h_1,h_2\rangle}
{\sqrt{\langle h_1, h_1\rangle\langle h_2,h_2\rangle}}\,.
\end{equation}
Fixing one of the functions, say $h_1(t)$, we can  
consider its overlap with the function resulting by shifting $h_2(t)$ in time by
a time lag $\tau$, \ie $h_2(t+\tau)$. In the frequency domain
this amounts to calculate the overlap between $\tilde{h}_1(t)$ and 
$\tilde{h}_2(f,\tau)=\tilde{h}_2(f)e^{-i2\pi f \tau}$. Maximizing
over $\tau$ provides the {\em best match} estimator 
\begin{eqnarray}
\label{e:best_match}
{\cal M}(h_1,h_2)&\equiv& 
\max\limits_{\tau} \{ {\cal O}[h_1,h_2(\tau)]\} =
 \max\limits_{\tau}\frac{\langle h_1,h_2(\tau)\rangle}
{\sqrt{\langle h_1, h_1\rangle \langle h_2,h_2\rangle}}
\nn \\
&=&
\max\limits_{\tau}\frac{\int_{-\infty}^\infty  \tilde{h}_1(f)
\tilde{h}_2^*(f)e^{-i2\pi f \tau}df}
{\left(\int_{-\infty}^\infty  |\tilde{h}_1(f)|^2df
\int_{-\infty}^\infty  |\tilde{h}_2(f)|^2 df\right)^{\frac{1}{2}}}\,. \notag \\
\end{eqnarray}
We note that, according to expression (\ref{e:cross_spectrum}), the
numerator in the second line of Eq.~(\ref{e:best_match}) is the
inverse Fourier transform of the cross-spectrum function ${\cal
  C}(h_1,h_2)(f)$. Therefore the numerator in the expression for
${\cal M}(h_1,h_2)$ is just the correlation function ${\cal
  C}(h_1,h_2)(\tau)$.  Regarding the denominator, we use Parseval's
identity
\begin{equation}
\label{e:Parseval}
 \int_{-\infty}^\infty  |\tilde{h}_i(f)|^2 df = \int_{-\infty}^\infty  |h_i(t)|^2 dt,
\end{equation}
and the expression for the autocorrelation of functions $h_i(t)$
\begin{equation}
\label{e:auto_correlation}
\int_{-\infty}^\infty  |h_i(t)|^2 dt = {\cal C}(h_i,h_i;\tau=0) \ \,.
\end{equation}
We then recover expression (\ref{e:correl_estimator}) for  ${\cal M}(h_1,h_2)$
in terms of the correlation function  ${\cal C}(h_1,h_2;\tau)$.

\section{Mapping time series on the screens} 
\label{a:mapping}

As discussed in Sec. \ref{s:correlation}, a built-in gauge mapping between
sections of $\scri^+$ and the horizon ${\cal H}^+$ defined by the
spacetime slicing leads to a
stretching of the time coordinate between the two screens. 

\begin{figure*}
\begin{center}
\includegraphics[width=5.8cm]{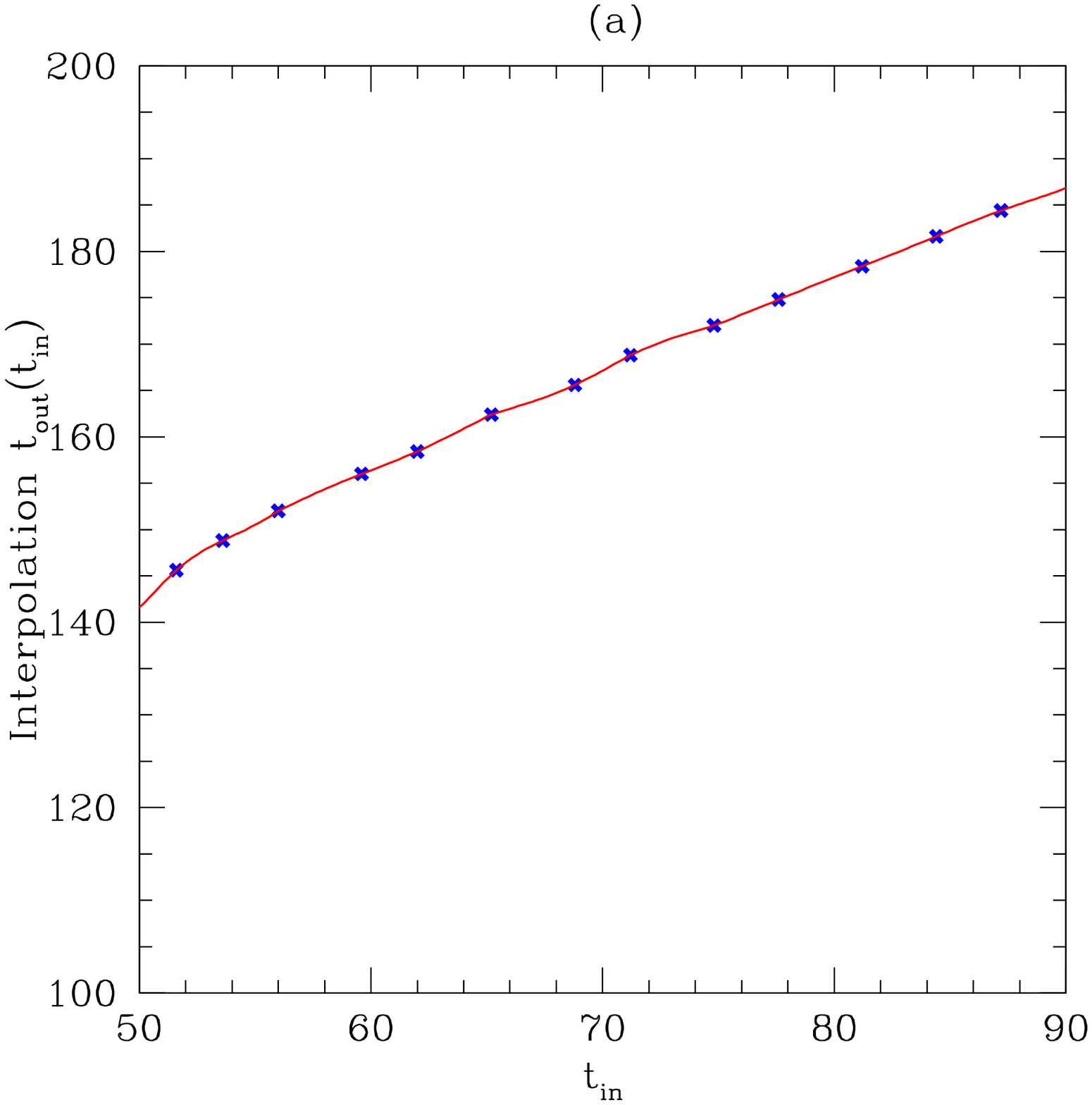}
\hskip 0.1cm
\includegraphics[width=5.8cm]{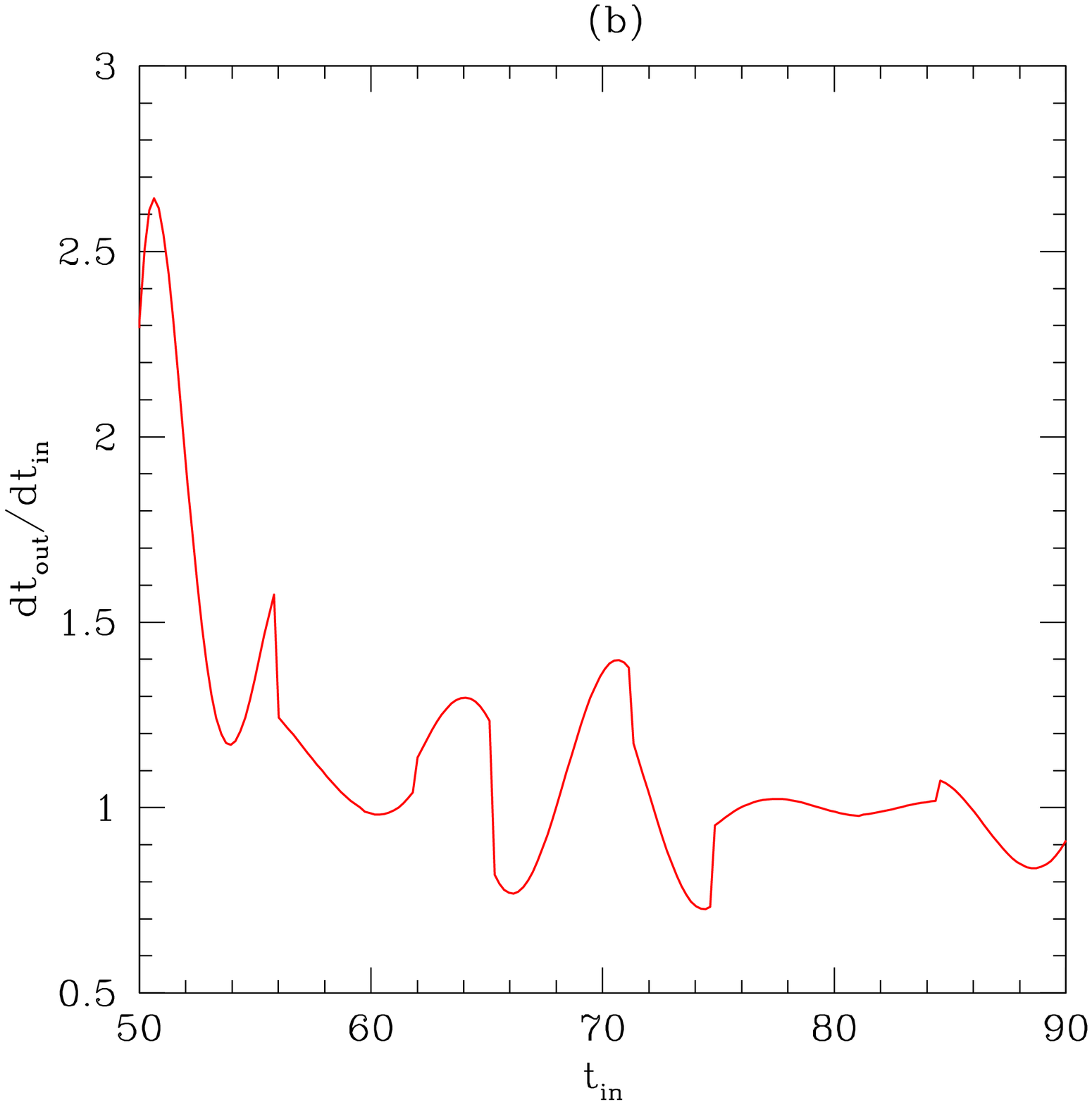}
\hskip 0.1cm
\includegraphics[width=5.8cm]{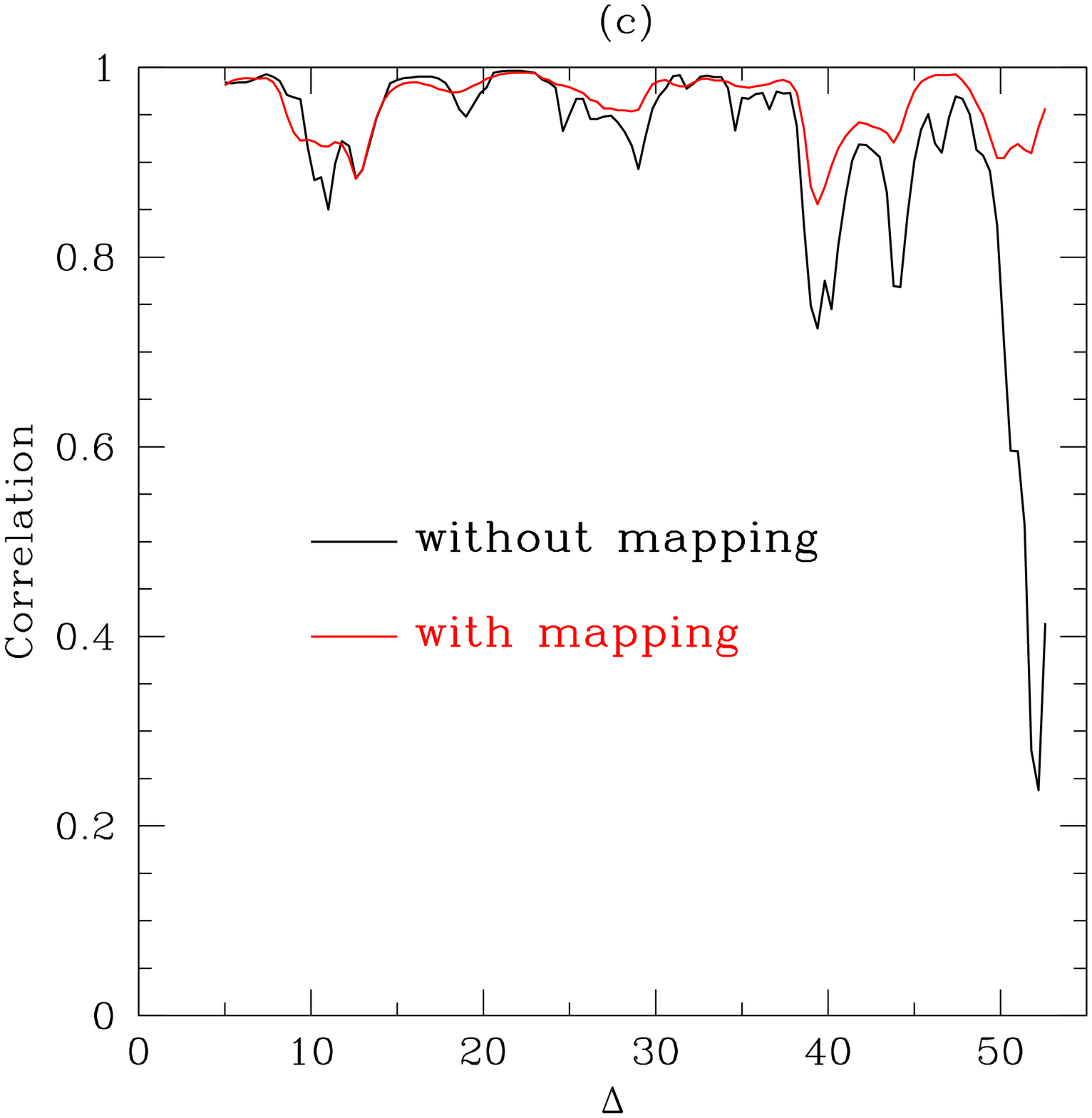}
\end{center}
\caption{Mapping between the coordinate time $t_{\rm in}$ and $t_{\rm out}$ measured at the inner/outer screens. The first panel shows the interpolation of the function $t_{\rm out}=t_{\rm out}(t_{\rm in})$ constructed from the comparison between the sequence of maxima and minima in the signals $h_{\rm inn}(t)$ and $h_{\rm out}(t)$. The middle panel depicts the interpolation's derivative $dt_{\rm out}/dt_{\rm in}$. In particular, it shows that initially the coordinate time at $\scri^+$ runs faster than the time at ${\cal H}^+$ and than oscillates around unity at late times. This behavior is consistent with the approach to stationarity. Finally, the right panel presents the correlation number ${\cal M}$ as a function of time intervals $\Delta$. The red curve shows the correlation without the using the mapping $t_{\rm out}=t_{\rm out}(t_{\rm in})$  to correct the time stretching between the signals on the two screens, whereas the black curve takes the effect into account and gives ${\cal M} \geqslant 0.9$. This correction is crucial to disentangle coordinate
from real effects at early times. 
}
\label{fig:mapping}
\end{figure*}

A comparison based on sequences of maxima and minima 
in the signals $h_{\mathrm{inn}}(t)$
and $h_{\mathrm{out}}(t)$ allows us to construct the mapping 
$t_{\mathrm{out}}(t_{\mathrm{inn}})$, which is here depicted 
in Fig.\ref{fig:mapping}a.  Also shown in Fig.\ref{fig:mapping}b is the derivative of this interpolation, 
crucial to assess the relative rate of the considered coordinate times. 
In particular, it addresses the behavior discussed in footnote (\ref{foot:mapping}). 

Finally, Fig.\ref{fig:mapping}c presents the correlation number ${\cal M}$
 between the two signals as a function of time intervals $\Delta$. The construction
 of the intervals $\Delta$ is based on the sequences of maxima and minima identified 
 in the two signals. In this way, we fix the final time in both series as $t^{\rm final}_{\rm inn} = 96.8 M$ and $t^{\rm final}_{\rm out} = 194.4 M$ and then we establish windows $\Delta = t^{\rm final} - t^{\rm initial}$, starting from $t^{\rm initial}_{\rm inn} = 49.2 M$ and $t^{\rm initial}_{\rm out} = 140.4 M$. 
 For the red curve, the correlation is evaluated without 
 using the mapping $t_{\rm out} = t_{\rm out}(t_{\rm inn})$ to correct the stretching, while the black curve takes the effect into account. This latter figure shows that the correction through
the mapping $t_{\rm out} = t_{\rm out}(t_{\rm inn})$ is crucial to disentangle coordinate
from real effects at early times. 

\bibliographystyle{apsrev4-1-noeprint.bst}
\bibliography{aeireferences}

\end{document}